\documentclass[12pt]{article}
\usepackage{amsbsy}
\usepackage{amsfonts}
\usepackage{amssymb}

\usepackage{bm}
\usepackage{leftidx}
\usepackage{enumerate}
\usepackage{graphics}
\usepackage{epsfig}
\usepackage{color}
\usepackage{nicefrac}
\usepackage{mathrsfs}
\usepackage{bbm}
\usepackage{amsmath}
\usepackage{amsthm}

\definecolor{gruen}{rgb}{0,0.625,0}  

\newlength{\TZ}
\newcommand{\BEQ}{\begin{equation}}     
\newcommand{\BEA}{\begin{eqnarray}}
\newcommand{\BD}{\begin{displaymath}}
\newcommand{\EEQ}{\end{equation}}       
\newcommand{\EEA}{\end{eqnarray}}
\newcommand{\ED}{\end{displaymath}}
\newcommand{\bb}{\begin{eqnarray}}
\newcommand{\ee}{\end{eqnarray}}

\newcommand{\e}{{\rm e}}

\newcommand{\eps}{\varepsilon}          
\newcommand{\vep}{\varepsilon}          
\newcommand{\vph}{\varphi}              
\newcommand{\D}{{\rm d}}                
\newcommand{\ud}{\,\mathrm{d}}
\newcommand{\II}{{\rm i}}               
 
\newcommand{\demi}{\frac{1}{2}}         
\newcommand{\tr}{{\rm tr\,}}            
\newcommand{\wht}[1]{\widehat{#1}}      
\newcommand{\lap}[1]{\overline{#1}}     

\newcommand{\dicht}{\rho}
\newcommand{\up}[1]{b^{\dagger}_{\vec{#1}}}
\newcommand{\down}[1]{b_{\vec{#1}}}
\renewcommand{\vec}[1]{\boldsymbol{#1}} 
\newcommand{\Tr}[1]{\operatorname{tr}\left( #1 \right)} 
\newcommand{\q}[1]{q_{#1}}
\newcommand{\p}[1]{\pi_{#1}}

\newcommand{\cleanint}[2]{\int \limits_{#1}^{#2}}                                        

\catcode`\@=11
\def\numberbysection{\@addtoreset{equation}{section}
        \def\theequation{\thesection.\arabic{equation}}}
\numberbysection

\graphicspath{{./graphics/}}           

\begin{document}
\begin{titlepage}

\vskip 1.5 cm
\begin{center}
{\Large \bf Lindblad dynamics of the quantum spherical model}
\end{center}

\vskip 2.0 cm
\centerline{{\bf Sascha Wald}$^{a,b,}$\footnote{swald@sissa.it}, {\bf Gabriel T. Landi}$^{c}$  
and {\bf Malte Henkel}$^{a,d,e}$\footnote{address after 1${\rm st}$ of January 2018: Laboratoire de Physique et Chimie Th\'eoriques (CNRS UMR),
Universit\'e de Lorraine Nancy, B.P. 70239, F - 54506 Vand{\oe}uvre-l\`es-Nancy Cedex, France}}
\vskip 0.5 cm
\begin{center}
$^a$ Groupe de Physique Statistique, 
D\'epartement de Physique de la Mati\`ere et des Mat\'eriaux, 
Institut Jean Lamour (CNRS UMR 7198),  Universit\'e de Lorraine Nancy, 
B.P. 70239, \\ F -- 54506 Vand{\oe}uvre l\`es Nancy Cedex, France
\\ \vspace{.5cm}
$^b$ SISSA - International School for Advanced Studies, via Bonomea 265, I -- 34136 Trieste, Italy
\\ \vspace{.5cm}
$^c$ Instituto de F\'{\i}sica, Universidade de S\~ao Paulo, Caixa Postal 66318, \\ 05314-970 S\~ao
Paulo, S\~ao Paulo, Brazil
\\ \vspace{.5cm}
$^d$ Rechnergest\"utzte Physik der Werkstoffe, Institut f\"ur Baustoffe (IfB), \\ 
ETH Z\"urich, Stefano-Franscini-Platz 3, CH - 8093 Z\"urich, Switzerland
\\ \vspace{.5cm}
$^e$ Centro de F\'{i}sica Te\'{o}rica e Computacional, Universidade de Lisboa, \\P--1749-016 Lisboa, Portugal
\end{center}

\begin{abstract}
The purely relaxational non-equilibrium dynamics of the quantum spherical model as described through a Lindblad equation is analysed. It is
shown that the phenomenological requirements of reproducing the exact quantum equilibrium state as stationary solution and
the associated classical Langevin equation in the classical limit $g\to 0$ fix the form of the Lindblad dissipators, up to an
overall time-scale. In the semi-classical limit, the models' behaviour becomes effectively the one of the classical analogue, with
a dynamical exponent $z=2$ indicating diffusive transport, 
and an effective temperature $T_{\rm eff}$, renormalised by the quantum coupling $g$. A different behaviour
is found for a quantum quench, at zero temperature, deep into the ordered phase $g\ll g_c(d)$, for $d>1$ dimensions. 
Only for $d=2$ dimensions, a simple scaling behaviour holds true, with a dynamical exponent $z=1$ indicating ballistic transport, 
while for dimensions $d\ne 2$, logarithmic corrections to scaling arise. The spin-spin correlator, 
the growing length scale and the time-dependent susceptibility show the existence of several logarithmically different
length scales. 
\end{abstract}

\vfill
PACS numbers: 05.30.-d, 05.30.Jp, 05.30.Rt, 64.60.De, 64.60.Ht, 64.70.qj \\~\\

\end{titlepage}

\setcounter{footnote}{0}

\section{Introduction}

The statistical mechanics of non-equilibrium open system continues to pose many challenges, 
related to the absence of a unified framework for their formulation. 
Here, we shall be concerned with non-equilibrium relaxations of open {\em quantum} systems. 
In the vicinity of equilibrium, linear-response theories such as the
Kubo formula or the Landauer-B\"uttiker formalism may be used \cite{Kub57,Jeon95,Mah}. 
But such approaches cannot describe the system's behaviour far from equilibrium, for instance after a quench from one physical phase into
another. Studies on the physical ageing of glassy and non-glassy systems after such quenches have led 
to a precise understanding of the associated phenomena and
in particular have made it clear that the competition between several distinct, but equivalent, 
equilibrium states may prevent the system to relax to an equilibrium state at all, even if
the microscopic dynamics does satisfy detailed balance \cite{Cugl03,Mar05,Maz06,Henk10,Taeu14}. 

Often-used phenomenological approaches to classical dissipative systems include master equations 
for the probability distributions or Langevin equations for the observables. 
\textcolor{black}{Various types of critical dynamics have been identified \cite{Hohe77,Maz06,Cala05,Taeu14}. 
Here, we shall concentrate on purely relaxational dynamics, often referred to as {\em model-A dynamics}.}
A major distinction of {\em quantum systems} with respect to classical ones is the presence of a conjugate momentum 
$p_n$ for each classical observable $s_n$, both to be considered
as operators, such that canonical commutation relations $[s_n, p_m] = \II\hbar \delta_{n,m}$ 
hold true. From the point of view of a phenomenological classical description, 
this raises the requirement to re-formulate the dynamics in such a way that these prescribed conservation laws should be obeyed. 
Therefore, simplistic approaches such as 
phenomenological Kramers equations, for the observables $s_n$ and the momenta $p_m$, 
supplemented by phenomenological damping terms, are inadequate, since they lead to
the violation of the canonical commutation relations,
on time-scales of the order of the inverse damping rate, such that an effectively classical dynamics remains \cite{Carm99}. 

The open-system dynamics of a quantum system is most ideally studied using the concept of dynamical 
semi-groups and completely positive trace-preserving (CPTP) dynamics. 
In the Heisenberg picture, this may be implemented using the tools of quantum Langevin equations \cite{Gar04}. 
Conversely, in the Schr\"odinger picture, that is most readily accomplished using Lindblad master equations \cite{Lind76,Breu02,Engl02}.

Formally, the Lindblad equation preserves the trace, the hermiticity and the positivity of the reduced density matrix $\rho$. 
On the other hand, it is not considered straightforward
to write down explicit expressions for the Lindblad dissipators for generic many-body systems, 
although well-established formalisms exist for few-body systems, see e.g. 
\cite{Breu02,Atta06,Atta07,Weiss12,Scha14,Wald16}. 
Finally, if such expressions have been obtained, actually solving a Lindblad equation is still far from obvious. 
Some results exist for one- or two-body problems, see \cite{Breu02,Weiss12,Scha14}. 
For {\it fermionic} many-body {\it chains}, exact solutions have been found by establishing relationships
with $1D$ quantum integrability, see \cite{Pros10,Pros11,Kare13} and \cite{Pros15} for a recent review. 
Indeed, integrable models are relevant for the understanding of a large
range of experiments, see \cite{Batc16} for a recent review. But by their very mathematical nature, 
such techniques are limited to one-dimensional systems. 

In order to provide insight beyond purely numerical studies, a versatile and non-trivial exactly solvable model is sought. 
In equilibrium statistical mechanics, the so-called {\em spherical model}
of a ferromagnet (see section~2 for the precise definition) \cite{Berl52,Lewi52} has since a long time served for such purposes. 
In the classical formulation with ferromagnetic nearest-neighbour interactions, 
it undergoes a continuous phase transition at a critical temperature $T_c>0$ for spatial dimensions  $d>2$
($d$ can be treated as a continuous parameter).  
For $2<d<4$ dimensions, the critical exponents are distinct from those of mean-field theory. 
The standard formulation in terms of classical spins has the drawback that the third fundamental
theorem of thermodynamics is not obeyed, since the specific heat $c_h=1$ for temperatures $T<T_c$ \cite{Berl52}. 
This can be cured however, by adjoining  to each spin variable $s_n$ 
a canonically conjugate momentum $p_n$ and adding a kinetic energy term, with a quantum coupling $g$, to the Hamiltonian $H$, 
thus arriving at the {\it quantum spherical model} ({\sc qsm}) \cite{Ober72}. Then the specific heat $c_h$ vanishes
indeed as $T\to 0$, as it should be \cite{Niew95,Vojt96,Oliv06,Sach11}. The model's properties near the critical temperature 
$T_c(g)>0$ are the same as in the classical spherical model.
However, at temperature $T=0$, there is for $d>1$ dimensions a quantum critical point, at some $g=g_c>0$, 
which is in the same universality class as the classical model 
in $d+1$ dimensions \cite{Kogu79,Henk84a,Niew95,Vojt96,Bran00,Oliv06,Sach11,Sred79,Dutt15,Wald15}. The formulation of the spherical model contains  
the so-called `spherical constraint'. The exact  solution of the model reduces to
establishing the constraint equation for the associated Lagrange multiplier which at equilibrium must be found from the solution 
of a transcendent equation. Turning to the dynamics, the kinetics of the classical spherical model can be described in terms of a 
Langevin equation, such that the spherical constraint reduces to a Volterra integral
equation for the now time-dependent Lagrange multiplier \cite{Ronc78,Godr00b}. 
Many aspects of the non-equilibrium dynamics of the model have been analysed in great detail, including extensions to the
spherical spin glass and to the growth of interfaces \cite{Ronc78,Coni94,Cugl95,Godr00b,Fusc02,Pico02,Cala05,Fort12,Taeu14,Henk15,Dura17}. 

Here, we shall explore aspects of the non-equilibrium {\em quantum} dynamics of the {\sc qsm}. 
In order to construct the Lindblad dissipators, we shall require that
these are chosen such that (i) the correct quantum equilibrium state emerges as the stationary state of the dynamics and 
(ii) in the classical limit $g\to 0$, the correct classical
Langevin dynamics should be recovered.\footnote{\textcolor{black}{This classical
dynamics is in the universality class of the $\mbox{\rm O}(n)$-model in the $n\to\infty$ limit with purely relaxational 
model-A dynamics \cite{Hohe77,Maz06,Henk10,Taeu14}.}} 
As we shall see, this fixes the form of the Lindblad dissipators, 
up to the choice of an overall time scale. To do so, we recall in section~2 the definition
of the quantum spherical model and the main properties of its equilibrium phase diagram. 
In section~3, the Lindblad dissipators will be constructed in two different ways. 
First, we shall follow the traditional route of system-plus-reservoir methods \cite{Breu02,Scha14}. 
Inspired by recent constructions of free bosonic quantum systems \cite{Guim16,Sant16}, 
we shall give an explicit description for the phonons which make up the reservoir. 
We also discuss how this construction must be amended to take the spherical constraint into account. 
In section~4, we derive the associated equations of motion for the observables. 
Independently of any specific model for the reservoir, we shall show how a comparison with the
classical limit $g\to 0$ (whenever available) determines the form of the Lindblad dissipators.
This also clarifies further the interpretation of the phonon reservoir model. 
The formal closed-form solution for spin- and momentum-correlators will be derived. 
The most difficult part of any spherical-model calculation is the solution of the spherical constraint, 
which becomes in our case a highly non-trivial integro-differential equation. Since a full solution of this equation is very difficult, 
we shall focus on two special cases. 
First, in section~5, we analyse the semi-classical limit, which can be used to describe the leading quantum correction to 
the order-disorder phase transition at temperature $T=T_c(g)$. 
By construction, the Lindblad equation does preserve quantum coherence. Still, we find from the explicitly computed 
spin-spin correlator that to leading order in $g$, 
the dynamical critical behaviour, for temperatures $T>T_c(g)$, $T=T_c(g)$ or $T<T_c(g)$ is exactly the one of the classical 
\textcolor{black}{(purely relaxational model-A)} dynamics, 
where quantum effects only manifest themselves through the appearance of a new effective temperature $T\mapsto T_{\rm eff}(g)$. 
For quenches to $T\leq T_c(g)$, dynamical scaling holds with a dynamical exponent $z=2$, which indicates diffusive motion of the basic
degrees of freedom. 
Having thus confirmed the consistency of the Lindblad formalism applied to the quantum spherical model, we analyse in section~6
what happens for a quantum quench deeply into the ordered phase, through an exact analysis 
of the leading long-time and large-distance behaviour of the spin-spin and
momentum-momentum correlators. 
We find a very rich behaviour which subtly depends on  the spatial dimension $d$. Indeed, for $d=2$, simple dynamical scaling holds true, while
for $d>2$, several logarithmically different time-dependent length scales appear, which implies a multi-scaling phenomenology. 
This leading behaviour is independent of the damping $\gamma$ and the limit $\gamma\to 0$ of closed quantum systems can be taken. 
For $d<2$, logarithmic corrections to scaling appear as well, but are of a different nature since the model's behaviour now depends on $\gamma$.
The dynamical exponent is always $z=1$, up to eventual logarithmic corrections, indicative of ballistic motion.  

The technical details of the calculations are covered in several appendices. Appendix~A describes how to derive the equilibrium form of the
quantum spherical constraint, appendix~B gives the analysis of the effective Volterra equation in the semi-classical limit. Appendices~C and~D give
the necessary mathematical details for reducing asymptotically the spherical constraint to a transcendental equation involving Humbert functions, 
for the case of deep quantum quenches. 
This equation is solved asymptotically in appendices~E and~F. 
The scaling of the two-point correlator is analysed in appendix~G.

\section{\label{sec:model}Quantum spherical model: equilibrium}

The {\bf S}pin-{\bf A}nisotropic {\bf Q}uantum {\bf S}pherical {\bf M}odel ({\sc saqsm}) \cite{Wald15} is defined  by a set of `spin operators' 
$s_{\vec{n}}=s_{\vec{n}}^{\dagger}$, attached to the sites $\vec{n}$ of a 
$d$-dimensional hyper-cubic lattice $\mathcal{L}\subset \mathbb{Z}^d$ with $\mathcal{N}=N^d$ sites.
For each spin variable we define the corresponding conjugated momentum  
$p_{\vec{n}}=p_{\vec{n}}^{\dagger}$ \cite{Ober72},  which satisfies the canonical commutation relations
\BEQ
[s_{\vec{n}}, p_{\vec{m}}] = \II\, \delta_{\vec{n},\vec{m}}
\EEQ
Throughout, we shall use units such that $\hbar = 1$. 
For nearest-neighbour interactions, and with periodic boundary conditions, the Hamiltonian is 
\BEQ
\label{eq:H}
H = \sum\limits_{\vec{n}\in \mathcal{L}}\left[ \frac{g}{2}\bigg( p_{\vec{n}}^2 
-\frac{1-\lambda}{2\mathcal{S}}\sum_{\left<\vec{n},\vec{m}\right>}  p_{\vec{n}} p_{\vec{m}}\bigg) 
+  \mathcal{S} \bigg( s_{\vec{n}}^2 -\frac{1+\lambda}{2\mathcal{S}}\sum_{\left<{\vec{n}},{\vec{m}}\right>}  s_{\vec{n}}s_{\vec{m}} \bigg) \right]    
\EEQ
%
%
\begin{figure}[t]
\begin{center}
\includegraphics[width=0.6\textwidth]{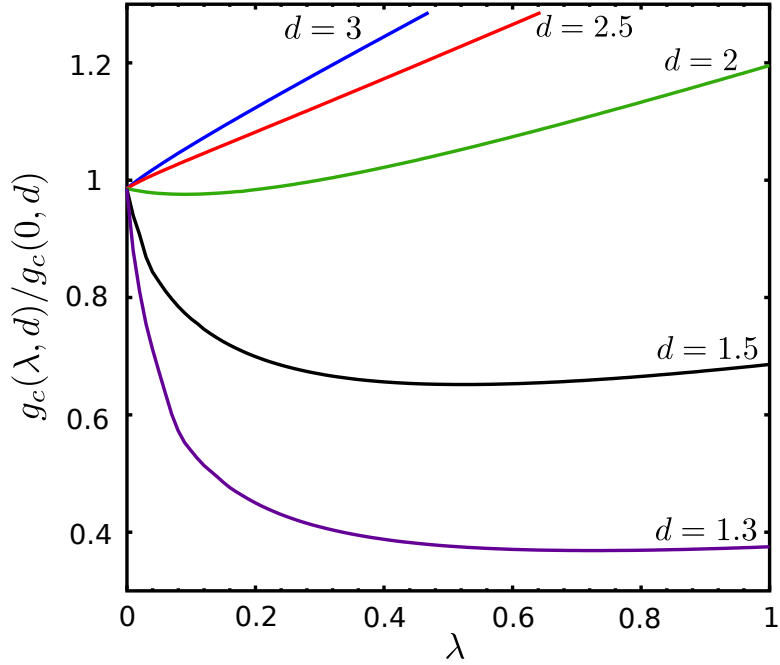}
\caption{\label{fig:GGPhasen}
Equilibrium quantum phase diagram of the {\sc saqsm} at $T=0$, for dimensions $d=[1.3,1.5,2,2.5,3]$, from bottom to top. 
We show for each dimension the critical line $g_c(\lambda)$ below which the
system is a quantum ferromagnet. Above these lines order is destroyed by quantum fluctuations. After \cite{Wald15}.}
\end{center}
\end{figure}
%
\noindent
where $\left< \vec{n},{\vec{m}} \right> $ are pairs of nearest-neighbour sites ${\vec{m}}$ and ${\vec{n}}$.
The parameter $\lambda$ describes the spin-anisotropy in the interactions 
(this can be seen explicitly by going over to bosonic degrees of freedom \cite{Wald15}) and the
usually studied {\sc qsm} \cite{Henk84a,Vojt96,Oliv06} is the special case $\lambda=1$. 
The parameter $g$ is the quantum coupling, such that for $\lambda=1$ and $g\to 0$, 
the spin operators become real numbers $s_{\vec{n}}\in\mathbb{R}$ and 
one recovers the classical spherical model \cite{Berl52}. Finally, 
the {\it spherical parameter} $\mathcal{S}$ is a Lagrange multiplier, to be chosen self-consistently 
in order to satisfy the so-called mean \textit{spherical constraint} \cite{Lewi52}
\BEQ\label{eq:constraint1}
\sum\limits_{{\vec{n}}\in \mathcal{L}} \left\langle s_{\vec{n}}^2 \right\rangle = \mathcal{N} 
\EEQ
The quantum Hamiltonian is invariant under the duality transformation $\mathscr{D}$ given by
\BEQ \label{eq:dual}
\lambda \leftrightarrow -\lambda \;\; , \;\; s_{\vec{n}} \leftrightarrow \sqrt{\frac{g}{2\mathcal{S}}\,}\: p_{\vec{n}} 
\EEQ
We shall strive to find a Lindblad dissipator which will preserve this symmetry. 
The equilibrium phases at temperature $T=0$, and the dimension-dependent transition lines
$g_c(\lambda,d)$ are shown in figure~\ref{fig:GGPhasen}. A re-entrant
phase transition is seen for  $1<d\lesssim2.065$ when  $\lambda$ is small enough, without a 
known counterpart in the fermionic analogues of the {\sc saqsm} \cite{Wald15}. 
This illustrates the non-trivial nature of the ground state of $H$.

The Hamiltonian~(\ref{eq:H}) is readily diagonalised by first going over to to Fourier space. 
We define the (non-hermitian) operators $q_{\vec{k}}=q_{-\vec{k}}^{\dagger}$ 
and $\pi_{\vec{k}}=\pi_{-\vec{k}}^{\dagger}$, along with the inverse transformations 
\begin{subequations} \label{eq:TransFourier}
\begin{align} \label{eq:fourier}
q_{\vec{k}} &:= \mathcal{N}^{-1/2}\sum_{n \in \mathcal{L}} s_{\vec n}\:\e^{-\II \vec{n}\cdot  \vec{k} }\;\; , \;\; 
\pi_{\vec{k}} := \mathcal{N}^{-1/2}\sum_{\vec{n} \in\mathcal{L}}p_{\vec{n}}\:\e^{\II \vec{n} \cdot \vec{k} }\\
\label{eq:fourier_inv}
s_{\vec{n}} &= \mathcal{N}^{-1/2}\sum_{\vec{k} \in \mathcal{B}} q_{\vec{k}}\:\e^{\II \vec{n} \cdot  \vec{k} } \hspace{0.24truecm}\;\; , \;\; 
p_{\vec{n}} = \mathcal{N}^{-1/2}\sum_{\vec{k} \in \mathcal{B}}^{} \pi_{\vec{k}}\: \e^{-\II \vec{n} \cdot \vec{k} } \ .
\end{align}
\end{subequations}
where the momentum $\vec{k}$ lies in the first \textit{Brillouin zone} 
$\mathcal{B}:=\{\vec{k}=(k_1 \ldots k_d)|k_i\in\{-\frac{\pi}{N} \ldots \frac{\pi}{N}\}$.
These operators obey the canonical commutation relations 
\BEQ	\label{eq:comm}
\left[\q{\vec{k}},\p{\vec{k}'}\right] = \II\,\delta_{\vec{k},\vec{k}'}
\EEQ
and the transformation (\ref{eq:fourier}) casts the Hamiltonian~(\ref{eq:H}) into the form
\BEQ \label{eq:ham}
H=\sum_{\vec{k} \in \mathcal{B}}
\left[\frac{g}{2\mathcal{S}}\Lambda_{-;\vec{k}}^2\,{\pi}_{\vec{k}}\pi_{-\vec{k}}+\Lambda_{+;\vec{k}}^2\,{q}_{\vec{k}}{q}_{-\vec{k}} \right]
\EEQ
where
\begin{equation}\label{eq:LambdaDef}
\Lambda_{\pm;\vec{k}} := \sqrt{\mathcal{S}+\frac{1\pm\lambda}{4}(\omega_{\vec{k}}-2d)\,} 
\hspace{1cm} \text{with} \hspace{1cm} \omega_{\vec{k}} := 2\sum_{j=1}^d (1-\cos k_j)
\end{equation}
In the same manner, the spherical constraint~(\ref{eq:constraint1}) is transformed as
\BEQ\label{eq:constraint2}
\sum_{\vec{k}\in\mathcal{B}} \left<q_{\vec{k}} q_{-\vec{k}} \right> = \mathcal{N} \ .
\EEQ
The Hamiltonian (\ref{eq:ham}) is now diagonalised by introducing the \textit{bosonic ladder operators}
\begin{subequations} 
\begin{align} \label{eq:map}
\q{k} &= \alpha_{\vec{k}}\frac{\down{k} + b_{-\vec{k}}^\dagger}{\sqrt{2}} \hspace{1.2truecm}\;\; , \;\; 
\pi_{\vec{k}} = \frac{\II}{\alpha_{\vec{k}}} \frac{\up{k} - b_{-\vec{k}}}{\sqrt{2}}, \\ 
\label{eq:inverse-map} 
\down{k} &=\frac{\alpha_{\vec{k}}}{\sqrt{2}} \left( \frac{q_{\vec{k}}}{\alpha_{\vec{k}}^2}  + \II \pi_{-\vec{k}}\right) \;\; , \;\; 
\up{k} = \frac{\alpha_{\vec{k}}}{\sqrt{2}} \left( \frac{q_{-\vec{k}}}{\alpha_{\vec{k}}^2}  - \II \pi_{\vec{k}}\right)
\end{align}
\end{subequations}
where
\BEQ\label{eq:alpha}
\alpha_{\vec{k}} = \left( \frac{g}{2\mathcal{S}}\right)^{{1}/{4}}\sqrt{\frac{\Lambda_{-;\vec{k}}}{\Lambda_{+;\vec{k}}}}
\EEQ
The operators $\down{k}$ and $\up{k}$ obey the usual Weyl-Heisenberg algebra  $\big[\down{k},\up{k'} \big] = \delta_{\vec{k},\vec{k}'} $.  
The Hamiltonian in eq.~(\ref{eq:ham})  then becomes 
\BEQ\label{eq:H_diag}
H = \sum_{\vec{k}\in \mathcal{B}}E_{\vec{k}} \left( \up{k} \down{k} + \demi\right) \;\; , \;\; 
E_{\vec{k}} = \sqrt{2\frac{g}{\mathcal{S}}\,}\, \Lambda_{+;\vec{k}} \Lambda_{-;\vec{k}} \ .
\EEQ
\subsubsection*{The isotropic case}
For technical simplicity, we shall focus on the Lindblad equation in the isotropic case $\lambda=1$. 
Then, eq.~(\ref{eq:LambdaDef}) reduces to $\Lambda_{-;\vec{k}} = \sqrt{\mathcal{S}}$ and 
\BEQ\label{eq:LambdaLambdaLambda}
\Lambda_{+;\vec{k}} =: \Lambda_{\vec{k}} =\sqrt{ \mathcal{S} - d + \omega_{\vec{k}}/2\,}
\EEQ
The energy $E_{\vec{k}}$ in eq.~(\ref{eq:H_diag}) and the parameter $\alpha_k$ in eq.~(\ref{eq:alpha})  simplify to 
\begin{equation}
E_{\vec{k}} = \sqrt{2 g\,} \cdot \Lambda_{\vec{k}} = \sqrt{g\,}\sqrt{2(\mathcal{S}-d) + \omega_{\vec{k}}\,}
\;\; , \;\; 
\alpha_{\vec{k}} = \left(\frac{g}{2}\right)^{1/4} \frac{1}{\sqrt{\Lambda_{\vec{k}}\,}\, }
\end{equation}
In the long-wavelength limit we may expand the cosines and write 
\begin{equation} \label{2.17}
E_k \simeq \sqrt{2  g\,}\bigg[\sqrt{\mathcal{S}-d\,}\, + \demi\frac{k^2}{\sqrt{\mathcal{S}-d\,}\,}\bigg]
\end{equation}
where $k^2 = k_1^2 + \ldots + k_d^2$. 
The last term in (\ref{2.17}) represents a non-relativistic massive dispersion relation, whereas the first term represents a chemical potential term.
Clearly, thermodynamic stability is realised if $\mathcal{S}\geq d$. Furthermore, 
the zero-momentum energy gaps vanish for $\mathcal{S}= d$. In complete analogy with
the equilibrium spherical model, classical \cite{Berl52} or quantum \cite{Henk84a}, this last condition defines the critical point.

\section{Construction of the Lindblad master equation}
\label{sec:micro}

Now, we discuss how to describe the dynamics of the {\sc qsm}
in contact with a heat bath. We shall explicitly admit the Markov property in the dynamics and
the \textit{weak-coupling limit} of the coupling between the system and the bath. 
It is well-established that under these hypotheses, the most general description of 
the quantum dynamics of a system interacting with a reservoir is a non-unitary time-evolution 
of the reduced density matrix $\dicht$, via the \textit{Lindblad equation} \cite{Breu02,Scha14}
\BEQ \label{Lind1}
\partial_t \dicht = -\II \left[ H, \dicht \right] +\mathcal{D}(\dicht)
\EEQ
Herein, the \textit{dissipator} $\mathcal{D}(\dicht)$ describes the relaxation towards equilibrium. 
In the case of a single harmonic oscillator, interacting
with a thermal bath, made of a phonon or photon gas, at the fixed temperature $T$ \cite{Lind76,Breu02,Scha14}
\BEQ \label{Lind2}
\mathcal{D}(\dicht) = \gamma(E) \bigg((\bar{n}+1) \bigg[b \rho b^\dagger - \frac{1}{2} \{b^\dagger b, \rho\} \bigg] 
+ \bar{n} \bigg[b^\dagger \rho b - \frac{1}{2} \{b b^\dagger, \rho\} \bigg] \bigg)
\EEQ
where $E$ is the energy of the oscillator, $b$ and $b^\dagger$ are the bosonic ladder operators of the system, $\{A,B\}=AB+BA$ is the anti-commutator, 
$\gamma(E)$ is the \textit{damping parameter} which also depends substantially on the bath and 
\BEQ\label{eq:BE}
\bar{n} = \bar{n}(E) = \left(\e^{E/T}-1 \right)^{-1}
\EEQ 
is the Bose-Einstein occupation number at bath temperature $T$. 
This quantum master equation (\ref{Lind1},\ref{Lind2}) preserves essential properties of the density matrix 
$\dicht$, namely \textit{trace}, \textit{complete positivity} and \textit{hermiticity} \cite{Lind76}.
In addition, the Schr\"odinger picture is used for the bosonic operators $b$, $b^{\dagger}$. 
Hence the commutator $\big[b , b^\dagger \big]= 1$ is time-independent and its conservation
is an intrinsic property of the formalism \cite{Carm99}. 

For our many-body problem, further consistency requirements are necessary: 
\begin{enumerate}
\item the quantum equilibrium state must be a stationary state of eqs.~(\ref{Lind1},\ref{Lind2}),
\item this should imply that in the $g\to 0$ limit, the classical equilibrium state must be a stationary state,
\item the classical Langevin dynamics must follows in the limit $g\to0$, for all times.
\end{enumerate}
It turns out that these requirements can all be met, in an essentially unique way. 
The final Lindblad equation of the {\sc saqsm} will come out to read 
\BEA \nonumber
\partial_t\dicht = -\II\big[ H,\dicht \big]  &+&  \gamma_0 \sum_{{\vec{k}}\in\mathcal{B}}\left[\left(\frac{1+\lambda}{2}\right)^2\Lambda_{-;{\vec{k}}}^2+
\left(\frac{1-\lambda}{2}\right)^2 \Lambda_{+;{\vec{k}}}^2\right]\frac{\Lambda_{+;{\vec{k}}}^2\Lambda_{-;{\vec{k}}}^2}{\mathcal{S}^2}\times\\ 
& &\times \bigg[\left(\bar{n}_{\vec{k}}+1\right) \bigg(b_{\vec{k}} \rho b_{\vec{k}}^\dagger -\frac{1}{2} \{b_{\vec{k}}^\dagger b_{\vec{k}}, \rho\} \bigg)
+\ \bar{n}_{\vec{k}} \bigg(b_{\vec{k}}^\dagger \rho b_{\vec{k}} - \frac{1}{2} \{b_{\vec{k}} b_{\vec{k}}^\dagger, \rho\} \bigg)\bigg] ~~~~
\label{Lind3}
\EEA
Herein, the only free parameter is the constant $\gamma_0$  which sets the time-scale. Clearly, the dissipator does depend 
on the spherical parameter $\mathcal{S}$. The derivation of (\ref{Lind3}) is made first for a free bosonic system, 
without taking the spherical constraint into account. 
At the end, through the spherical constraint which must hold at all times, $\mathcal{S}=\mathcal{S}(t)$ becomes time-dependent. 
This will turn out to make the solution of the spherical constraint considerably more complicated than 
at equilibrium (and also with respect to the classical dynamics).

\noindent 
Two different ways of deriving (\ref{Lind3}) will be presented:  
\begin{enumerate}[(i)]
\item One may consider explicitly the system-reservoir coupling and go through the standard route, with  the usual approximations \cite{Breu02}. 
The bath properties are taken into account through the explicit time-dependent phonon (or photon) correlators. 
This gives a formal derivation of the Lindblad equation and will be carried out in the remainder of this section. 
\item For the purpose of model-building, an alternative and more phenomenological approach might be useful. 
As we shall show in section~4, one may start from a generic form of the
dissipator, essentially a sum of terms of the form (\ref{Lind2}) for each mode, and with yet unspecified damping constants $\gamma_{\vec{k}}$.
We then derive quantum equations of motion for certain observables. 
Comparison of these equations of motion with the known classical $g\to 0$ limit (if available) then fixes the $\gamma_{\vec{k}}$. 
\end{enumerate}
At the end, both procedures lead to the same Lindblad equation (\ref{Lind3}).

\subsection{General structure of the system-bath coupling}

Now, largely following \cite{Breu02}, but with the few adaptations required for the {\sc qsm}, we introduce the open-system dynamics. 

For clarity, we begin treating just a single spin, say $s_{\vec{n}}$, coupled to the bath. 
The coupling of several spins is readily obtained at the end. 
As usual, the bath will be modelled by an infinite number of bosonic `phonon' degrees of freedom, with the bath hamiltonian 
\BEQ
H_B = \sum_\ell \Omega_\ell \eta_\ell^\dagger \eta_\ell 
\EEQ
with the bosonic operators ${\eta}_\ell$ and their corresponding frequencies $\Omega_\ell$.
The system-bath interaction Hamiltonian is assumed to take the form 
\BEQ\label{HI}
H_{I} = \sum\limits_\ell f_{\ell}\  A_{\vec{n}} \otimes (\eta_\ell + \eta_\ell^\dagger)
\EEQ
where $f_{\ell}\in\mathbb{R}$ are coupling constants and $A_{\vec{n}}$ is a hermitian system operator. 
There is a certain freedom in the choice of $A_{\vec{n}}$. Here, rather than a simplistic coupling to only the 
spin operator $s_{\vec{n}} =s_{\vec{n}}^\dagger$ or only to the
momentum operator $p_{\vec{n}} = p_{\vec{n}}^\dagger$, we prefer a coupling which preserves the invariance under the 
duality transformation $\mathscr{D}$, see eq~(\ref{eq:dual}). 
The most general linear operator compatible with duality is
\BEQ
A_{\vec{n}} = \frac{1+\lambda}{2} \frac{s_{\vec{n}}}{\sqrt{g\,}\,} + \frac{1-\lambda}{2}\frac{p_{\vec{n}}}{\sqrt{2\mathcal{S}\,}\,}  \ .
\EEQ
In the weak-coupling limit, the action of the bath is described approximately 
by a Lindblad equation for the reduced density matrix $\dicht$ of the system 
\BEQ\label{eq:lindblad}
\partial_t\rho = -\II [H,\rho] + \mathcal{D}_n(\rho)
\EEQ
where the first term describes the unitary evolution and
$\mathcal{D}_n(\rho)$ is the Lindblad dissipator corresponding to the interaction~(\ref{HI}).
The expression for $\mathcal{D}_n(\rho)$ is most commonly derived using the method of eigenoperators \cite{Breu02}.

To make this presentation self-contained, we rapidly recall the main steps before applying it to the {\sc saqsm}. 
Consider a Hamiltonian $H$ with energy levels $\epsilon$ and let $\mathscr{P}_\epsilon$ 
denote the corresponding projection operator onto the subspace of eigenvectors that have energy $\epsilon$. 
Moreover, assume that the system-bath coupling may be described by an interaction Hamiltonian of the form $H_I =  A B$, 
see also (\ref{HI}), 
where $A$ and $B$ are \textit{hermitian} system and bath operators, respectively. 
Define the {\em eigenoperator} $A(\omega)$ corresponding to $A$ via the relation
\BEQ\label{eq:eigenoperator_def}
A(\omega) = \sum\limits_{\epsilon, \epsilon'} \mathscr{P}_\epsilon A \mathscr{P}_{\epsilon'}\; \delta_{\epsilon'-\epsilon , \omega}
\EEQ
where the sum is over all distinct energies $\epsilon$, $\epsilon'$ and  $\delta_{a,b}$ is the Kronecker delta. It can be shown that
\BEQ\label{eq:eigenoperator_comm}
[H,A(\omega)] = - \omega A(\omega) \;\; , \;\; A^\dagger(\omega) = A(-\omega) \ .
\EEQ
The quantities $\omega$ represent all allowed energy differences that may be produced by the action of the operator $A$.

It follows that the Lindblad dissipator corresponding to the interaction $H_I = AB$ reads, 
in the Born-Markov and rotating wave approximations \cite{Breu02} 
\begin{equation}\label{D0}
\mathcal{D}(\rho) = \sum\limits_\omega  \Gamma(\omega) \bigg[A(\omega) \rho A^\dagger(\omega) 
- \frac{1}{2} \{A^\dagger(\omega) A(\omega), \rho\} \bigg]
\end{equation}
where 
\begin{equation}\label{gamma_def}
\Gamma(\omega) = \int_{-\infty}^\infty \!\D t\; \e^{ \II \omega t} \langle B(t) B(0)\rangle
\end{equation}
is the Fourier transform of the bath correlation functions. 
This method therefore allows one to readily write down the dissipator corresponding to a given system-bath interaction. 
However, to do so we must compute the eigenoperator $A(\omega)$ from eq.~(\ref{eq:eigenoperator_def}), 
which requires the full eigenstructure of the Hamiltonian. 
It is also worth noting that this method also produces a Lamb-shift correction to the Hamiltonian. 
However, this correction is usually small and, for simplicity, will be neglected.

\subsection{Evaluation of bath correlation functions}
 
Returning now to our problem, the interaction Hamiltonian~(\ref{HI}) has $A=A_{\vec{n}}$ and $B=\sum_\ell f_{\ell} (\eta_\ell + \eta_\ell^\dagger)$. 
One must compute eq.~(\ref{gamma_def}) for this choice of $B$. If the bath is in thermal equilibrium at a fixed temperature $T$, one has  
\BEQ
\langle B (t) B(0) \rangle = \sum\limits_\ell f_{\ell} f_{\ell} 
\bigg(  e^{-\II \Omega_\ell t} [\bar{n}(\Omega_\ell)+1] +e^{\II \Omega_\ell t} \bar{n}(\Omega_\ell)\bigg)
\EEQ
with the Bose-Einstein distribution $\bar{n}$ defined in eq.~(\ref{eq:BE}).
Inserting this into eq.~(\ref{gamma_def}) leads to  
\BEQ
\Gamma(\omega) = 2\pi \sum\limits_\ell f_{\ell} f_{\ell} \bigg(
\delta_{\omega,\Omega_\ell} [\bar{n}(\Omega_\ell)+1] +
\delta_{\omega,-\Omega_\ell} \bar{n}(\Omega_\ell)\bigg)
\EEQ
If the bath frequencies $\Omega_\ell$ vary continuously in the interval $[0,\infty)$, one may convert the sum to an integral, leading to
\BEQ \label{gamma_cases}
\Gamma(\omega) = \int_0^\infty \!\D \Omega \; \gamma (\Omega) \bigg( 
\delta_{\omega , \Omega} [\bar{n}(\Omega)+1] +
\delta_{\omega ,- \Omega}\bar{n}(\Omega)\bigg)
= \begin{cases} 
\gamma(\omega) [ \bar{n}(\omega) +1 ] & \text{ ,~ if } \omega >0 	\\[0.25cm]
\gamma(|\omega|) \bar{n}(|\omega|)    & \text{ ,~ if } \omega <0
\end{cases} ~~
\EEQ
with the associated \textit{spectral density} 
\BEQ\label{spec_density_1}
\gamma(\Omega)  = 2\pi \sum_\ell f_\ell^2\; \delta(\Omega - \Omega_\ell)
\EEQ
In order to have a definite prediction for the spectral density $\gamma(\omega)$, 
additional physical information about the distribution of bath frequencies is needed. 
In general, one expects that 
\BEQ
\gamma(\Omega) \sim \Omega^\kappa
\EEQ
for some exponent $\kappa$. 
The actual value of $\kappa$ will depend sensibly on the microscopic details of the bath, which in our case we do not know. 
Instead, we shall be guided by the principle that the classical dynamics \cite{Godr00b} should be recovered in an appropriate limit. 
As we shall show below, this turns out to imply the exponent $\kappa = 3$. 

Interestingly, the value of this exponent also follows from another consideration which is common in the context of quantum optics. 
Suppose that our bath bosons have a linear dispersion linear relation (such as, for instance, photons or acoustic phonons). 
Then the index $\ell$ is replaced by the momentum $\bm{k}$ 
and the dispersion relation is written as $\Omega_k = c |\bm{k}|$ where $c$ is the sound/light velocity. 
Transforming the sum in  eq.~(\ref{spec_density_1}) into an integral gives 
\BEQ\label{spec_density_2}
\gamma(\Omega)  \sim f(\Omega)^2 \Omega^2
\EEQ
We now see that we recover the exponent $\kappa = 3$ if we assume that $f_{\bm{k}}$ is proportional to $\sqrt{\Omega_{\bm{k}}}$.
This turns out to be  precisely the dipole approximation minimum coupling \cite{Breu02}. 
Thus, we conclude that we recover the classical Langevin dynamics if we assume a typical electric-field dipole coupling of the spins with the bath bosons. 
In summary, we emerge from this discussion with the result that  
\BEQ\label{spec_density_3}
\gamma(\Omega) = \gamma_0 \Omega^3
\EEQ
where the constant $\gamma_0$ describes the strength of the system-bath coupling.

\subsection{Calculation of the eigenoperators}

To finish the construction of the dissipator~(\ref{D0}) one must find the eigenoperators $A(\omega)$ corresponding to $A = A_{\vec{n}}$. 
First, use eqs.~(\ref{eq:fourier_inv}) and (\ref{eq:map}) to write 
\BEQ \label{3.20}
\hspace{-0.05truecm}A_{\vec{n}} = \frac{(g^{-3}/2\mathcal{S})^{\frac{1}{4}}}{\sqrt{2\mathcal{N}}} 
\sum\limits_{\vec{k} \in \mathcal{B}} e^{\II \vec{n} \cdot \vec{k}} \left(c_{\vec{k}} \down{k} + c_{\vec{k}}^* b_{-\vec{k}}^\dagger\right)
\hspace{.5cm} \text{with} \hspace{.5cm} c_{\vec{k}} = 
\frac{1+\lambda}{2}\sqrt{\frac{\Lambda_{-;\vec{k}}}{\Lambda_{+;\vec{k}}}}+\II\frac{1-\lambda}{2}\sqrt{\frac{\Lambda_{+;\vec{k}}}{\Lambda_{-;\vec{k}}}}
\EEQ
Next we note that, due to the diagonal structure of $H$ in eq.~(\ref{eq:H_diag}), it follows that $[H,b_{\vec{k}}] = - E_{\vec{k}} b_{\vec{k}}$. 
Hence, comparison with eq.~(\ref{eq:eigenoperator_comm}) shows that $b_{\vec{k}}$ 
is an eigenoperator of $H$ with allowed transition frequency $\omega = E_{\vec{k}}$. 
The same is true for $c_{\vec{k}} b_{\vec{k}}$ as well. The full eigenoperator therefore reads 
\BEQ
A_{\vec{n}}(\omega) = \frac{(g^{-3}/2\mathcal{S})^{\frac{1}{4}}}{\sqrt{2\mathcal{N}}} 
\sum\limits_{\vec{k} \in \mathcal{B}} e^{\II \vec{n} \cdot \vec{k}} \left(c_{\vec{k}}\down{k}\delta_{E_{\vec{k}}, \omega} 
+ c_{\vec{k}}^* b_{-\vec{k}}^\dagger \delta_{E_{\vec{k}},-\omega}\right)
\EEQ
The dissipator~(\ref{D0}) corresponding to $A_{\vec{n}}$ being coupled to the bath, will then be
\BEQ\label{D1}
\mathcal{D}_{\vec{n}}(\rho) = \sum\limits_\omega  \Gamma(\omega) \bigg[A_{\vec{n}}(\omega) \rho A_{\vec{n}}^\dagger(\omega) 
- \frac{1}{2} \{ A_{\vec{n}}^\dagger(\omega) A_{\vec{n}}(\omega), \rho\} \bigg]
\EEQ
This expression may be simplified further. To do that, it suffices to look only at the first term
\BEA
\nonumber
& &\sum\limits_\omega  \Gamma(\omega)  A_{\vec{n}}(\omega) \rho A_{\vec{n}}^\dagger(\omega) \\
&=&
\nonumber
\sum\limits_{\omega,\vec{k},\vec{k}'}\frac{\e^{\II \vec{n} \cdot (\vec{k}-\vec{k}')}}{2\ \mathcal{N}}  \Gamma(\omega) \sqrt{\frac{g^{-3}}{2\mathcal{S}}}
\bigg(c_{\vec{k}}\down{k} \delta_{E_{\vec{k}}, \omega} +  c_{\vec{k}}^*b_{-\vec{k}}^\dagger  \delta_{E_{\vec{k}} ,-\omega}\bigg) \rho
\bigg(c_{\vec{k}'}^*\up{k'} \delta_{E_{\vec{k}'}, \omega} +  c_{\vec{k}'}b_{-\vec{k}'}^\dagger \delta_{E_{\vec{k}'} ,-\omega}\bigg) ~~~
\EEA
Since $E_k >0$, see eq.~(\ref{eq:H_diag}), the only terms which will survive the constraints imposed by the 
$\delta$'s are those with $E_{\vec{k}} = E_{\vec{k}'}$. 
Since the energies may be degenerate, this does not necessarily imply that $\vec{k} =\vec{k}'$. But if we carry out the sum over $\omega$ 
and use eq.~(\ref{gamma_cases}), we obtain  
\BEA
\nonumber
& &\sum_{\omega}  \Gamma(\omega)  A_{\vec{n}}(\omega) \rho A_{\vec{n}}^\dagger(\omega) \\
&=& \nonumber 
\sqrt{\frac{g^{-3}}{8\mathcal{S}}}  \sum_{\vec{k},\vec{k'}}  \delta_{E_{\vec{k}},E_{\vec{k'}}} 
\frac{\e^{\II \vec{n}\cdot( \vec{k}-\vec{k'})}}{\mathcal{N}}\gamma(E_{\vec{k}})
\bigg[c_{\vec{k}}c_{\vec{k'}}^*(\bar{n}_{\vec{k}}+1) b_{\vec{k}} \rho b_{\vec{k'}}^\dagger 
+c_{\vec{k}}^*c_{\vec{k'}} \bar{n}_{\vec{k}} b_{\vec{k}}^\dagger \rho b_{\vec{k'}}\bigg] \ .
\EEA
The structure of the other terms in eq~(\ref{D1}) will be similar. Finally, we define 
\BEQ\label{gamma_kq}
\gamma_{\vec{k},\vec{k'}}^{(\vec{n})} =  \sqrt{\frac{g^{-3}}{8\mathcal{S}}}\gamma(E_{\vec{k}})\,  
\frac{\e^{\II \vec{n}\cdot( \vec{k}-\vec{k'})}}{\mathcal{N}}\, c_{\vec{k}}c_{\vec{k'}}^*
\EEQ
The final single-site dissipator~(\ref{D1}) reads  
\BEQ
\label{D2}
\mathcal{D}_{\vec{n}}(\rho) =\sum_{\vec{k},\vec{k'}}   \delta_{E_{\vec{k}}, E_{\vec{k'}}}  \bigg( 
(\bar{n}_{\vec{k}}+1)  \gamma_{\vec{k},\vec{k'}}^{(\vec{n})}\bigg[ \down{k}\rho \up{k'} - \frac{1}{2} \{\up{k'}\down{k}, \rho\}\bigg] 
+ \gamma_{\vec{k'},\vec{k}}^{(\vec{n})} \bar{n}_{\vec{k}}  \bigg[ \up{k} \rho \down{k'} - \frac{1}{2} \{\down{k'} \up{k}, \rho\}\bigg] \bigg)
\EEQ
and describes the action of coupling a single degree of freedom to the heat bath. It couples to all normal modes $b_{\vec{k}}$. 
Furthermore, it is well-known that dissipators of this form will let evolve the system towards the correct thermal Gibbs state $\dicht \sim e^{-H/T}$, 
although only a single spin was coupled to the bath \cite{Breu02}.  

The information which site $\vec{n}$ is coupled to the bath is contained in the factor $\gamma_{\vec{k},\vec{k'}}^{(\vec{n})}$.

\subsection{Effect of coupling the entire system to the bath}

We now extend this to the case where all spins are  coupled to the bath. 
In this case, for each degree of freedom, at site $\vec{n}$, 
we shall have a dissipator $\mathcal{D}_{\vec{n}}(\rho)$ appearing in eq.~(\ref{eq:lindblad}). 
But if we look at eq.~(\ref{D2}) we see that  $n$ only appears in the quantities $\gamma_{\vec{k},\vec{k'}}^{(\vec{n})}$. 
Thus if we sum all dissipators $\mathcal{D}_n(\rho)$ we will get a result with a structure identical to 
eq.~(\ref{D2}), but with $\gamma_{\vec{k},\vec{k'}}^{(\vec{n})}$ replaced by 
\BEQ
\sum_{\vec{n}} \gamma_{\vec{k},\vec{k'}}^{(\vec{n})} 
= \sqrt{\frac{1}{8 g^3 \mathcal{S}}\,}\:\gamma(E_{\vec{k}})\, |c_{\vec{k}}|^2\, \delta_{\vec{k},\vec{k'}} 
=: \delta_{\vec{k},\vec{k'}}\: \gamma_{\vec{k}}
\EEQ
where, using also eq.~(\ref{spec_density_3}) along with (\ref{eq:H_diag},\ref{3.20}), we find
\BEQ\label{spec_density_4}
\gamma_{\vec{k}} = \gamma_0\left[ \left(\frac{1+\lambda}{2}\right)^2 \Lambda_{-;\vec{k}}^2 
+ \left(\frac{1-\lambda}{2}\right)^2 \Lambda_{+;\vec{k}}^2\right]\frac{\Lambda_{+;\vec{k}}^2\Lambda_{-;\vec{k}}^2}{\mathcal{S}^2}
\EEQ
Specific calculations will only be carried out for the isotropic case $\lambda=1$ for which (\ref{spec_density_4}) simplifies to 
\begin{subequations} \label{Lind4}
\BEQ\label{spec_density_5}
\gamma_{\vec{k}} = \gamma_0  \Lambda_{\vec{k}}^2 
\EEQ
see also eq.~(\ref{eq:LambdaLambdaLambda}). The final dissipator, after having summed over $\vec{n}$, reads
\BEQ
\mathcal{D}(\rho) = \sum_{\vec{k} \in \mathcal{B}}\gamma_{\vec{k}}	 \bigg(
  (\bar{n}_{\vec{k}}+1)  \bigg[ b_{\vec{k}} \rho b_{\vec{k}}^\dagger - \frac{1}{2} \{b_{\vec{k}}^\dagger b_{\vec{k}}, \rho\}\bigg] 
+ \bar{n}_{\vec{k}}  \bigg[ b_{\vec{k}}^\dagger \rho b_{\vec{k}} - \frac{1}{2} \{b_{\vec{k}} b_{\vec{k}}^\dagger, \rho\}\bigg] \bigg)
\EEQ
\end{subequations}
This is our final result (\ref{Lind3}) for the microscopic derivation of the Lindblad dissipator. 

Recall that this dissipator satisfies detailed balance, as shown in \cite{Breu02}. 
Therefore, {\it modulo} an ergodicity assumption, the Lindblad equation
(\ref{Lind3}) will thermalise the system, irrespective of the initial condition, 
to the unique steady-state with reduced density matrix $\dicht \sim e^{-H/T}$. 

This entire discussion did not take into account the spherical constraint (\ref{eq:constraint2}). 
If one uses it in an {\it ad hoc} fashion, one would consider
$\mathcal{S}=\mathcal{S}(t)$ as time-dependent. Then either the couplings to the bath or the bath properties themselves, described by
$\gamma_{\vec{k}}$, $\bar{n}_{\vec{k}}$ and the operators $b_{\vec{k}}$ must be considered time-dependent. 
Pragmatically, one considers an effectively time-dependent
dissipator $\mathcal{D}_t$ which will always have as its target state the instantaneous Hamiltonian $H=H(t)$ 
of the system, such that formally $\mathcal{D}_t(e^{-\beta H(t)}) = 0$. 
Physically, that means that the time-dependent changes in $H$ should be slow enough, 
which in turn should be the case if the changes in $\mathcal{S}(t)$ should be more slow
than the typical bath correlation times. Since the eventual applications we are interested in concern 
the slow power-law relaxations after a quantum quench into the
two-phase coexistence regime with formally infinite relaxation times, 
we expect that these kinds of physical requirements should be satisfied. 

More systematically, one should not have imposed a spherical constraint, 
but rather have considered a second bath in order to implement it, at least on average. 
Since we expect that for sufficiently long times, 
the effective equations of motion should become the same as those we are going to study in the next section, 
we have not carried out this explicitly. At the present time, 
we consider it more urgent to arrive at some understanding of the qualitative consequences of the
equations of motion on the long-time behaviour of the non-equilibrium correlators.

\section{Dynamical equations for observables}
\label{sec:dyn}

In this section, we shall examine the dynamical equations governing the evolution of certain important observables under the influence of the heat bath. 
For any observable $\mathcal{O}$, the time-dependent average $\left< \mathcal{O} \right>=\left< \mathcal{O} \right>(t)$ is found from
\BEQ \label{4.1}
\frac{\D}{\D t}\left< \mathcal{O} \right> = \Tr{\mathcal{O}\, \partial_t \dicht} + \Tr{\dicht\, \partial_t \mathcal{O}}
\EEQ
In principle, all quantities $\alpha_{\vec{k}}$, $E_{\vec{k}}$, $\gamma_{\vec{k}}$ 
and $\bar{n}_{\vec{k}}$ should be considered as being time-dependent, if
the spherical constraint is taken into account. These explicit time-dependencies come from the second term on the right-hand-side in (\ref{4.1}). 
For the sake of simplicity of the presentation, we shall discard it for the moment but shall re-introduce it later. 

Therefore, for any observable $\mathcal{O}$ not depending explicitly on time, 
hence $\partial_t \mathcal{O}=0$, inserting the Lindblad equation (\ref{eq:lindblad}) into
(\ref{4.1}) gives 
\BEQ\label{eq:Oave}
\frac{\ud}{\ud t} \left< \mathcal{O} \right> = - \II \langle [\mathcal{O}, H] \rangle + \langle \bar{\mathcal{D}} (\mathcal{O})\rangle
\EEQ
with the adjoint dissipator
\BEQ
\bar{\mathcal{D}}(\mathcal{O}) 
 =  \sum_{{\vec{k}}\in\mathcal{B}}\gamma_{\vec{k}} \bigg(
  (\bar{n}_{\vec{k}}+1)  \bigg[ b_{\vec{k}}^\dagger \mathcal{O} b_{\vec{k}} - \frac{1}{2} \{b_{\vec{k}}^\dagger b_{\vec{k}}, \mathcal{O}\}\bigg] 
+ \bar{n}_{\vec{k}}  \bigg[ b_{\vec{k}} \mathcal{O} b_{\vec{k}}^\dagger - \frac{1}{2} \{b_{\vec{k}} b_{\vec{k}}^\dagger, \mathcal{O}\}\bigg] \bigg)
\label{Dbar}
\EEQ
Although the form of the $\gamma_{\vec{k}}$ was discussed in the previous section, 
we shall keep them here in a generic form. This will allow an alternative derivation of the
Lindblad equation (\ref{Lind3}). 

In order to understand how this adjoint dissipator arises, consider the second term as an example, namely 
$\mathcal{D}_2(\dicht)=\gamma \bar{n} \left( b^{\dagger}\dicht  b - \demi\left\{ b^{\dagger} b, \dicht\right\}\right)$, for a single mode. Then 
\BEA
\Tr{\mathcal{O}\mathcal{D}_2} &=& \gamma \bar{n} \Tr{ \mathcal{O} b^{\dagger} \dicht b 
- \demi \mathcal{O} b^{\dagger}b\dicht - \demi \mathcal{O}\dicht b{^\dagger} b}
\nonumber \\
&=& \gamma \bar{n} \Tr{ \dicht \left( b\mathcal{O}b^{\dagger} -\demi \mathcal{O} b^{\dagger} b - \demi b^{\dagger} b \mathcal{O}\right)} 
\:=\: \gamma \bar{n} \left\langle b\mathcal{O}b^{\dagger} -\demi \mathcal{O} b^{\dagger} b - \demi b^{\dagger} b \mathcal{O}\right\rangle
\nonumber
\EEA
which produces the second term in (\ref{Dbar}). The first term is obtained similarly. 

For the single-particle observables $\mathcal{O} \in\{ b_{\vec{k}}, b_{\vec{k}}^\dagger, \q{\vec{k}},\p{\vec{k}}\}$, 
we find from (\ref{4.1},\ref{eq:Oave},\ref{Dbar})
\begin{subequations} \label{eq:single}
\begin{align}\label{eq:b}
&\frac{\ud }{\ud t} \langle b_{\vec{k}} \rangle 
= -\bigg(\frac{\gamma_{\vec{k}}}{2}+\II E_{\vec{k}} \bigg) \langle b_{\vec{k}} \rangle  + \langle \partial_tb_{\vec{k}}\rangle
&, \;\;
& \frac{\ud }{\ud t} \langle b_{\vec{k}}^\dagger \rangle 
= - \bigg(\frac{\gamma_{\vec{k}}}{2}-\II E_{\vec{k}}\bigg)  \langle b_{\vec{k}}^\dagger \rangle+ \langle \partial_tb_{\vec{k}}^\dagger\rangle\hspace{.8cm}\\
&\frac{\ud }{\ud t} \langle q_{\vec{k}}   \rangle 
= -\frac{\gamma_{\vec{k}}}{2} \langle q_{\vec{k}} \rangle + \frac{ g}{\mathcal{S}}\Lambda_{-;{\vec{k}}}^2 \langle \pi_{-{\vec{k}}} \rangle
&, \;\;
&\frac{\ud }{\ud t} \langle \pi_{-{\vec{k}}} \rangle 
= -\frac{\gamma_{\vec{k}}}{2} \langle \pi_{-{\vec{k}}} \rangle - 2\Lambda_{+;{\vec{k}}}^2 \langle q_{{\vec{k}}} \rangle 
\label{eq:qpi}
\end{align}
\end{subequations}
where we also used the fact that $E_{\vec{k}}$, $\alpha_{\vec{k}}$ and $\gamma_{\vec{k}}$ are all even in $\vec{k}$.
In particular, the time-dependent magnetisation is expressed as  
\BEQ\label{eq:magnetization}
M = \sum_{\vec{n} \in \mathcal{L}} \langle s_{\vec{n}} \rangle  = \sqrt{\mathcal{N}\,}\, \langle q_{\vec{0}} \rangle
\EEQ
where use was made of the orthogonality of the Fourier series. 

Next, we turn to two-body correlators. 
We find, again using eqs.~(\ref{4.1},\ref{eq:Oave},\ref{Dbar}), 
\begin{subequations} \label{twoboson}
\begin{align}\label{eq:n}
\frac{\ud }{\ud t} \langle b_{\vec{k}}^\dagger b_{{\vec{k}}'} \rangle 
&= \langle \partial_t b_{\vec{k}}^\dagger b_{{\vec{k}}'} \rangle 
+  \bigg[ \II (E_{\vec{k}} - E_{{\vec{k}}'}) -\frac{\gamma_{\vec{k}} + \gamma_{{\vec{k}}'}}{2} \bigg]  \langle b_{\vec{k}}^\dagger b_{{\vec{k}}'}  \rangle 
+ \gamma_{\vec{k}} \bar{n}_{\vec{k}} \delta_{{\vec{k}},{\vec{k}}'}	\\[0.2cm]
\label{eq:offcoherences}
\frac{\ud }{\ud t} \langle b_{\vec{k}} b_{{\vec{k}}'} \rangle 
&= \langle \partial_t b_{\vec{k}} b_{{\vec{k}}'} \rangle 
+\bigg[ -\II (E_{\vec{k}} + E_{{\vec{k}}'}) -\frac{\gamma_{\vec{k}} + \gamma_{{\vec{k}}'}}{2} \bigg]    \langle b_{\vec{k}} b_{{\vec{k}}'}  \rangle
\end{align}
\end{subequations}
{}From these equations we may also compute dynamical equations for the two-point correlators
\BEQ
Q_{\vec{k}}(t)   := \langle q_{\vec{k}} q_{-{\vec{k}}} \rangle \;\; , \;\;
\Pi_{\vec{k}}(t) := \langle \pi_{\vec{k}} \pi_{-{\vec{k}}} \rangle \;\; , \;\;
\Xi_{\vec{k}}(t) := \frac{1}{2} \langle q_{\vec{k}} \pi_{\vec{k}}  + \pi_{-{\vec{k}}} q_{-{\vec{k}}} \rangle
\EEQ
The spherical constraint (\ref{eq:constraint2}) then becomes in the $\mathcal{N}\to\infty$ limit
\BEQ \label{eq:constraint3} 
\sum_{\vec{k}\in\mathcal{B}} Q_{\vec{k}}(t)=\mathcal{N}\;\;\;  \Leftrightarrow \;\;\; \int_{\mathcal{B}} \frac{\D \vec{k}}{(2\pi)^d} Q_{\vec{k}}(t) =1
\EEQ 
and we find the eqs of motion for the two-point correlators
\begin{subequations} \label{TwoPoint}
\begin{align}
\label{xk}\frac{\ud Q_{\vec{k}}}{\ud t} &= - \gamma_{\vec{k}} \left[ Q_{\vec{k}}(t) 
- \frac{1}{4}\sqrt{\frac{2 g}{\mathcal{S}}}\frac{\Lambda_{-;{\vec{k}}}}{\Lambda_{+;{\vec{k}}}} (2\bar{n}_{\vec{k}} + 1)\right]  
+  2 \frac{ g}{\mathcal{S}}\Lambda_{-;{\vec{k}}}^2 \Xi_{\vec{k}}(t) \\[0.2cm]
\label{zk}
\frac{\ud \Xi_{\vec{k}}}{\ud t} &= -\gamma_{\vec{k}} \cdot \Xi_{\vec{k}}(t)  
 + \frac{ g}{\mathcal{S}}\Lambda_{-;{\vec{k}}}^2 \cdot \Pi_{\vec{k}}(t) - 2\Lambda_{+;{\vec{k}}}^2\cdot  Q_{\vec{k}}(t)\\[0.2cm]
\label{yk}
\frac{\ud \Pi_{\vec{k}}}{\ud t} &= - \gamma_{\vec{k}} 
\left[ \Pi_{\vec{k}}(t) -\sqrt{\frac{\mathcal{S}}{2 g}}\frac{\Lambda_{+;{\vec{k}}}}{\Lambda_{-;{\vec{k}}}} (2\bar{n}_{\vec{k}} + 1) \right] 
-  4\Lambda_{+;{\vec{k}}}^2 \cdot \Xi_{\vec{k}}(t)
\end{align}
\end{subequations}
At this point we would like to stress again that the canonical commutation relation 
$\left[q_{\vec{k}},\pi_{\vec{k'}}\right] = \II \delta_{k,k'}$ is preserved due to the fact that 
$q_{\vec{k}}$ and $\pi_{\vec{k}'}$ are Schr\"odinger operators. In particular, this is connected to the trace-preserving property of the dynamics as
\begin{equation} \label{4.10}
 \partial_t\left<\left[q_{\vec{k}},\pi_{\vec{k'}}\right]\right>  
 = \tr \bigg(\left[q_{\vec{k}},\pi_{\vec{k'}}\right] \partial_t{\rho} \bigg) = \II \delta_{k,k'} \partial_t\tr \rho = 0 \ .
\end{equation}
Along with this, the commutation relation between the bosonic ladder operators is preserved since they present the same underlying algebra as
\begin{equation}
 \left[ b_{\vec{k}},b^\dagger_{\vec{k}'} \right] = -\frac{\II}{2}\bigg(\frac{\alpha_{k'}}{\alpha_{k}}\left[q_{\vec{k}},\pi_{\vec{k}'} \right] 
 - \frac{\alpha_k}{\alpha_{k'}}\left[\pi_{-\vec{k}'} ,q_{-\vec{k}}\right]\bigg) 
 = \delta_{k,k'} \ .
\end{equation}

Having completed these formal calculations, we must now take the spherical constraint 
(\ref{eq:constraint2},\ref{eq:constraint3}) into account. From (\ref{xk}), 
this becomes an integro-differential equation for
the time-dependent spherical parameter $\mathcal{S}=\mathcal{S}(t)$. 
It follows that the parameter $\alpha_{\vec{k}}=\alpha_{\vec{k}}(t)$, defined in (\ref{eq:alpha}), becomes time-dependent
as well. It describes the transformation (\ref{eq:map}) between the bosonic operators $b_{\vec{k}},b_{\vec{k}}^{\dagger}$ 
and the spins $q_{\vec{k}}$ and momenta $\pi_{\vec{k}}$.  Therefore,
one must decide whether either the pair $(b_{\vec{k}},b_{\vec{k}}^{\dagger})$ or else the pair 
$(q_{\vec{k}},\pi_{\vec{k}})$ is taken to be time-independent, 
and hence is described by the Schr\"odinger picture. 

We choose $(q_{\vec{k}},\pi_{\vec{k}})$ as time-independent operators. 
The Lindblad formalism then implies the time-independent commutator (\ref{eq:comm}). Furthermore, the equations
of motion eqs.~(\ref{eq:qpi}) and (\ref{TwoPoint}) remain valid. 
They will form the basis for our analysis of the dynamics of the {\sc qsm}. 

In consequence, in eqs.~(\ref{eq:b}) and (\ref{twoboson}) the contributions coming from the second term in (\ref{4.1}) must be worked out. For example, 
the first equation of motion in (\ref{eq:b}) now becomes, where the dot indicates the time derivative 
\BEQ
\frac{\D }{\D t}\left<b_{\vec{k}}\right> 
= -\bigg(\frac{\gamma_{\vec{k}}}{2}+\II E_{\vec{k}} \bigg) 
\langle b_{\vec{k}} \rangle -\frac{\dot{\alpha}_{\vec{k}}}{\alpha_{\vec{k}}}\langle b_{-\vec{k}}^\dagger \rangle
\EEQ
The other equations can be generalised similarly, but we shall not require them in this work. 

Before we analyse the dynamics produced by equations (\ref{TwoPoint}), we shall first consider their steady-state properties.

\subsection{Stationary solution and equilibrium properties}

The correlators in eqs.~(\ref{TwoPoint}) will relax to their stationary values, namely $\Xi_{\vec{k}}(\infty) = 0$ and 
\BEQ\label{eq:corr_equi}
Q_{\vec{k}}(\infty)   = \frac{1}{4}\sqrt{\frac{2g}{\mathcal{S}}}\frac{\Lambda_{-;{\vec{k}}}}{\Lambda_{+;{\vec{k}}}} (2\bar{n}_{\vec{k}} + 1),\qquad  
\Pi_{\vec{k}}(\infty) = \sqrt{\frac{\mathcal{S}}{2 g}}\frac{\Lambda_{+;{\vec{k}}}}{\Lambda_{-;{\vec{k}}}} (2\bar{n}_{\vec{k}} + 1)
\EEQ
These are precisely the equilibrium values expected from the {\sc saqsm} \cite{Wald15}. 
To see that more clearly, we substitute these results into the spherical constraint (\ref{eq:constraint2}) and find 
\BEQ \label{4.11}
1 = \sqrt{\frac{ g}{8\mathcal{S}}}\frac{1}{\mathcal{N}}\sum\limits_{{\vec{k}} \in \mathcal{B}} 
\frac{\Lambda_{-;{\vec{k}}}}{\Lambda_{+;{\vec{k}}}} (2\bar{n}_{\vec{k}} + 1) 
\stackrel{\mathcal{N}\nearrow \infty}{\longrightarrow} 
\sqrt{\frac{ g}{8\mathcal{S}}}\int_{\mathcal{B}}\frac{\D \vec{k}}{(2\pi)^d} \frac{\Lambda_{-;{\vec{k}}}}{\Lambda_{+;{\vec{k}}}} \coth(E_{\vec{k}}/2T)
\EEQ
This is indeed the spin-anisotropic spherical constraint at equilibrium. 
The derivation, through a canonical transformation, is given in appendix~\ref{app:cantrafo}. 
In view of the re-entrant phase diagram for a non-isotropic interaction with $\lambda\ne 1$, this is a non-trivial check of the formalism. 

We have therefore confirmed that {\em the equilibrium state of the {\sc saqsm} is a stationary solution of the Lindblad equation.}  
Details on the form of the $\gamma_{\vec{k}}$ are not required to verify this. 

In the isotropic case $\lambda = 1$, it is useful to let $\mathfrak{z}:=2(\mathcal{S}-d)$. Then (\ref{4.11}) reduces to the familiar form \cite{Oliv06}
\BEQ\label{eq:equi_equation}
\frac{\sqrt{g\,}\,}{2} \int\limits_{\mathcal{B}}\frac{\ud {\vec{k}}}{(2\pi)^d} \;  
\frac{1}{\sqrt{\mathfrak{z} +\omega_{\vec{k}}\,}\,}\coth\left(\frac{\sqrt{g\,}\,}{2T} \sqrt{\mathfrak{z}+\omega_{\vec{k}}\,}\right) = 1 \ .
\EEQ
In the following sections, we shall mainly concentrate on either the semi-classical limit $g\to 0$ 
or else on the the zero-temperature equilibrium limit $T=0$. 
In these limit cases, the spherical constraint reduces to
\BEA\label{eq:equi_expansion}
\begin{cases} 
1 - \frac{g}{12T} \simeq T \int\limits_{\mathcal{B}}\frac{\ud {\vec{k}}}{(2\pi)^d} 
\bigg[ \frac{1}{\mathfrak{z}+\omega_{\vec{k}}}  +{\rm O}\left(g^2\right) \bigg]
 & \text{ ,~ for } g\to 0 	\\[0.35cm]
\hspace{1.1truecm}1 = \sqrt{\frac{g}{4}\,}\int\limits_{\mathcal{B}}\frac{\ud {\vec{k}}}{(2\pi)^d} \; \frac{1}{\sqrt{\mathfrak{z}+\omega_{\vec{k}}\,}\,}
 & \text{ ,~ for } T=0
\end{cases} ~~
\EEA
%
\begin{figure}[tb]
\centering
\includegraphics[width=0.8\textwidth]{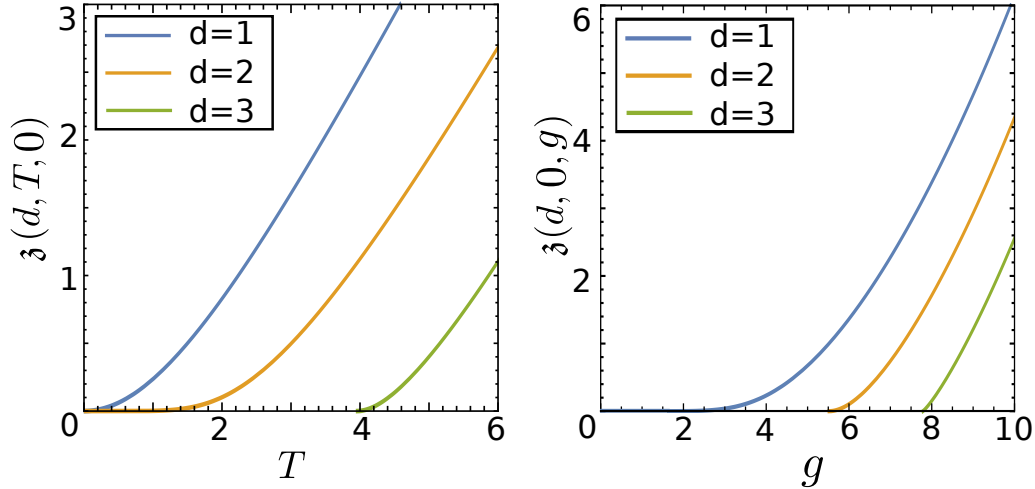}
\caption{\label{fig:crit}
Spherical parameter $\mathfrak{z}=\mathfrak{z}(d,T,g)$ as a function of $d$, the temperature $T$ and the coupling $g$. 
Left panel: classical limit $g=0$. Right panel: quantum transition at $T=0$. 
}
\end{figure}
%
The upper case in (\ref{eq:equi_expansion}) reduces to the familiar classical form of the 
equilibrium spherical constraint \cite{Berl52,Lewi52}, where the temperature
$T\mapsto T_{\rm eff}(g)=T /(1-\frac{g}{12 T})$ is replaced by  an effective temperature. 
This also shows that in the $g\to 0$ limit, one recovers the classical equilibrium state. 
The lower case in (\ref{eq:equi_expansion}) is the spherical constraint for the quantum phase transition at $T=0$
\cite{Henk84a,Vojt96,Oliv06}. For illustration, in figure~\ref{fig:crit} we  show the Lagrange multiplier $\mathfrak{z}=\mathfrak{z}(d,T,g)$. 
In the classical limit $g=0$ (left panel), the critical value $\mathfrak{z}=0$ is reached for $d\leq 2$ only for a vanishing temperature $T=0$ 
and there is no phase transition. On the other hand, for $d>2$, the line $\mathfrak{z}=0$ 
is already reached for a finite value $T=T_c(d)>0$ which defines the critical temperature. 
A qualitatively analogous behaviour is seen for the quantum phase transition at $T=0$ (right panel of figure~\ref{fig:crit}). 
While for $d=1$, the critical line $\mathfrak{z}=0$ is only reached for $g=0$, for any dimension $d>1$ one finds a finite critical value $g_c(d)>0$. 
A more detailed comparison reveals that the classical transition in $d+1$ dimensions, 
at $g=0$ and $T_c>0$ and the quantum transition in $d$ dimensions at $T=0$ and $g_c(d)>0$, are in the
same equilibrium universality class \cite{Henk84a,Vojt96,Oliv06,Wald15}.

\subsection{Formal solution of the non-equilibrium problem}

To complete the microscopic derivation of the Lindblad dissipator, 
we now give a phenomenological discussion of how to chose the dissipator in a physically motivated fashion 
in order to include the correct classical many-body dynamics.
To do so, we begin by writing down the formal solution of eqs.~(\ref{xk})-(\ref{yk}). This system can be re-written in a matrix form 
\BEQ
\frac{\D }{\D t} \begin{pmatrix}
                  Q_{\vec{k}}(t)\\[.25cm] \Xi_{\vec{k}}(t)\\[.25cm]\Pi_{\vec{k}}(t)
                 \end{pmatrix}
                 = - m_{\vec{k}}^\lambda(t)\begin{pmatrix}
                  Q_{\vec{k}}(t)\\[.25cm] \Xi_{\vec{k}}(t) \\[.25cm] \Pi_{\vec{k}}(t)
                 \end{pmatrix} + \vec{u}_{\vec{k}}^\lambda(t)
                 \label{eq:eom}
\EEQ
with the matrices
\BEQ\label{eq:matrices}
m_{\vec{k}}^\lambda (t) = \begin{pmatrix}
                   \gamma_{\vec{k}}(t)& -2 \frac{g}{\mathcal{S}}\Lambda_{-;{\vec{k}}}^2& 0\\[.25cm]
                   2\Lambda_{+;{\vec{k}}}^2& \gamma_{\vec{k}}(t) & -\frac{g}{\mathcal{S}}  \Lambda_{-;{\vec{k}}}^2\\[.25cm]
                   0&4\Lambda_{+;{\vec{k}}}^2&\gamma_{\vec{k}}(t)
                  \end{pmatrix} \ ,
           \       
           \vec{u}_{\vec{k}}^\lambda = \gamma_{\vec{k}}(t)(2\bar{n}_{\vec{k}}+1)
           \begin{pmatrix}
            -\sqrt{\frac{ g}{8\mathcal{S}}}\frac{\Lambda_{-;{\vec{k}}}}{\Lambda_{+;{\vec{k}}}}\\[.25cm]
            0\\[.25cm]
            \sqrt{\frac{\mathcal{S}}{2 g}}\frac{\Lambda_{+;{\vec{k}}}}{\Lambda_{-;{\vec{k}}}}
           \end{pmatrix} \ .
\EEQ
Here we suppressed the explicit time-dependence of the spherical parameter $\mathcal{S} = \mathcal{S}(t)$ 
of $\Lambda_{\pm;{\vec{k}}} = \Lambda_{\pm;{\vec{k}}}(t)$ for readability of the equation.
Some more comments are in order: 
\begin{itemize}
\item For a phenomenological discussion the damping rates $\gamma_{\vec{k}}$ 
were left unspecified. Since spin-anisotropy is a quantum-mechanical effect, this discussion must be carried out in the
isotropic case $\lambda=1$. Only at the end, we shall compare with the dissipator (\ref{Lind4}) 
derived form microscopic considerations in section~3. 
\item The time-dependence of the spherical parameter $\mathcal{S}(t)$ 
is to be found self-consistently from the formal solution and the spherical constraint 
$\sum_{\vec{k}\in\mathcal{B}}Q_{\vec{k}}(t)=\mathcal{N}$.
\item Already the isotropic case $\lambda=1$ turns out to be considerably more difficult than the classical non-equilibrium dynamics, 
so that we leave the non-isotropic case $\lambda\ne 1$ for future work. 
\end{itemize}
Concentrating from now on only on the isotropic case $\lambda=1$, we can simplify the matrices (\ref{eq:matrices}) by 
using the relations~(\ref{eq:LambdaLambdaLambda}, \ref{spec_density_5}) and find
\BEQ\label{eq:matrices-isotrop}
m_{\vec{k}}(t) = \begin{pmatrix}
                   \gamma_{\vec{k}}(t)& -2 g& 0\\[.25cm]
                   2\Lambda_{{\vec{k}}}^2& \gamma_{\vec{k}}(t) &-  g\\[.25cm]
                   0&4\Lambda_{{\vec{k}}}^2&\gamma_{\vec{k}}(t)
                  \end{pmatrix} \ ,
           \       
           \vec{u}_{\vec{k}}^\lambda = \gamma_{\vec{k}}(t)(2 \bar{n}_{\vec{k}}+1)
           \begin{pmatrix}
            \frac{1}{4}\frac{\sqrt{2g}}{\Lambda_{{\vec{k}}}}\\[.25cm]
            0\\[.25cm]
            \frac{\Lambda_{{\vec{k}}}}{\sqrt{2 g}}
           \end{pmatrix} \ .
\EEQ

\subsubsection{Choice of the damping parameters}
For $\lambda=1$, a well-defined classical limit $g \to 0$ exists, \textcolor{black}{and can be brought to coincide with the well-known purely 
relaxational model-A dynamics  of the $\mbox{\rm O}(n)$ model in the $n\to\infty$ limit \cite{Hohe77,Ronc78,Godr00b,Cala05,Henk10,Taeu14}, 
as we shall now see}. 
Then, the equation of motion for the spin correlator $Q_{\vec{k}}$ decouples and leads to
(recall $\mathfrak{z}=2(\mathcal{S}-d)$)
\begin{subequations} \label{4.17}
\BEQ \label{eq:Q-cl}
\frac{\D }{\D t} Q_{\vec{k}}(t) = -\gamma_{\vec{k}}(t) Q_{\vec{k}}(t) + \gamma_{\vec{k}}(t)\frac{T}{\mathfrak{z}(t)+\omega_{\vec{k}}} \ .
\EEQ
We stress that this equation of motion is qualitatively different from the classical Kramers
equation (see \cite{Wald16} for more details) since thermal fluctuations occur not just in the equation of motion 
of the momenta but already in the equation for the spins. 
The second term of the r.h.s. of eq. (\ref{eq:Q-cl}) comes from the assumed Lindblad
dissipator and generates a coherent quantum dynamics. For an initial state at infinite temperature $Q_{\vec{k}}(0)=1$. 
Then the formal solution of (\ref{eq:Q-cl}) reads 
\BEQ
Q_{\vec{k}}(t) = \e^{-\int_0^t \!\D\tau\: \gamma_{\vec{k}}(\tau)}  
\left(1+ T \int_0^t \!\D t'\: \frac{\gamma_{\vec{k}}(t')}{\mathfrak{z}(t')+\omega_{\vec{k}}}\
\e^{\int_0^{t'}\!\D \tau\: \gamma_{\vec{k}}(\tau)}\right)
\EEQ
\end{subequations}
We now compare this with the known dynamics of the classical model \cite[eq. (2.18)]{Godr00b}. 
The spin-spin correlator obeys the following equation of motion, 
which can be derived from the Langevin equation of the classical spherical model
\begin{subequations} \label{4.18}
\BEQ
\frac{\D }{\D t} Q_{\vec{k}}(t) = - 2\big( \mathfrak{z}(t) +\omega_{\vec{k}} \big) Q_{\vec{k}}(t) + 2T 
\EEQ
and has the solution 
\BEQ
Q_{\vec{k}}(t) = \e^{-2t\omega_{\vec{k}} - 2\int_0^t \!\D\tau\;\mathfrak{z}(\tau)}  \left(1+ 2T \int_0^t \!\D t'\: \e^{-2t'\omega_{\vec{k}} 
- 2\int_0^{t'} \!\D\tau\;\mathfrak{z}(\tau)}\right)
\EEQ
\end{subequations}
Our requirement that the $g\to 0$ limit should reduce to the classical Langevin equation means 
that eqs.~(\ref{4.17}) and (\ref{4.18}) must be consistent. 
This is achieved if we choose 
\BEQ \label{choix_final}
\gamma_{\vec{k}}(t) = 2 \Lambda^2_{\vec{k}}(t) = 2\left( \mathfrak{z}(t)+ \omega_{\vec{k}} \right)
\EEQ
and includes an implicit fixing of the time-scale in the classical dynamics \cite{Godr00b}. 
Remarkably, the condition (\ref{choix_final}) is identical to the result eq.~(\ref{spec_density_5}) 
obtained from the microscopic derivation of the Lindblad dissipator (\ref{Lind4}), up to
a choice of the overall damping constant $\gamma_0$. In particular, this sheds a different light on the heuristic argument 
we used above in order to fix the phenomenological 
exponent $\kappa=3$.  

Therefore, we have seen that {\em the requirements of reproducing (i) the correct quantum equilibrium state and 
(ii) the full classical dynamics in the limit $g\to 0$ are enough to fix
the precise form of the Lindblad dissipator, up to an overall choice of the time scale.}

\subsubsection{Closed formal solution}
With the final choice (\ref{choix_final}), we return to the dynamics for $g\ne 0$, but keep $\lambda=1$. The formal solution of eq.~(\ref{eq:eom}) is 
\BEQ\label{eq:matrixmotion}
 \begin{pmatrix}
        Q_{\vec{k}}(t)\\[.25cm] \Xi_{\vec{k}}(t) \\[.25cm] \Pi_{\vec{k}}(t)
 \end{pmatrix} = \e^{M_{\vec{k}}(t)} 
 \begin{pmatrix}
        Q_{\vec{k}}(0)\\[.25cm] 
        \Xi_{\vec{k}}(0)\\[.25cm]
        \Pi_{\vec{k}}(0)
 \end{pmatrix}
        + \gamma\int_0^t \!\D \tau\: \e^{M_{\vec{k}}(t) -M_{\vec{k}}(\tau)} 
        (2\bar{n}_{\vec{k}}(\tau)+1)
 \begin{pmatrix}
        \sqrt{\frac{ g}{2}}\Lambda_{{\vec{k}}}\\[.25cm]
        0\\[.25cm]
        \sqrt{\frac{2}{ g}}\Lambda_{{\vec{k}}}^3
 \end{pmatrix} \ 
\EEQ
where 
\BEQ\label{eq:Mk}
M_{\vec{k}}(t) = \int_0^t \!\D\tau\: m_{\vec{k}}(\tau ) = -\big(Z(t)+t\omega_{\vec{k}}\big) \begin{pmatrix}
                                                                      \gamma&0&0\\[.1cm]
                                                                      1&\gamma&0\\[.1cm]
                                                                      0&2&\gamma
                                                                     \end{pmatrix}
 +  g t \begin{pmatrix}
      0&2&0\\[.1cm]
      0&0&1\\[.1cm]
      0&0&0
     \end{pmatrix}
\EEQ
and we defined the integrated spherical parameter
\BEQ
Z (t) := \int_0^t \!\D \tau\: \mathfrak{z}(\tau)
\EEQ
At equilibrium, thermodynamic stability requires $\mathfrak{z}=\mathfrak{z}_{\rm eq}\geq 0$, as we have seen in section~\ref{sec:model}. 

Here, we are interested in the non-equilibrium dynamics after the systems undergoes a quench from an initial 
disordered state to a state characterised by certain values of $(T,g)$. 
Since then the initial values $\langle q_{\vec{k}}\rangle(0)=\langle \pi_{\vec{k}}\rangle(0)=0$, 
the noise-averaged global magnetisation remains zero at all times, although fluctuations around
this will be present. We therefore focus on two-body correlators. 
By analogy with classical dynamics we expect that if that quench goes towards a state in the disordered phase, 
with a single ground state of the Hamiltonian $H$, a rapid relaxation, with
a finite relaxation time, should occur towards that quantum equilibrium state. For sufficiently large times, $\mathfrak{z}(t)>0$ is expected. 
On the other hand, for quenches either onto a critical point or else into the ordered phase 
(with at least two distinct, but equivalent ground states), the formal relaxation times
become infinite. Then $\mathfrak{z}(t)$ may evolve differently. For the classical spherical model, 
quenched from a fully disordered high-temperature state to a temperature
$T$, one finds for long times the leading behaviour \cite{Ronc78,Godr00b} 
\BEQ \label{eq:Zcl}
Z^{\rm cl}(t) \sim \frac{\digamma}{2} \ln t \;\; , \;\;
\digamma = \begin{cases} -\demi(4-d)  & \mbox{\rm ~~;~ if $T=T_c$ and $d<4$} \\
                         ~~0          & \mbox{\rm ~~;~ if $T=T_c$ and $d>4$} \\
                         -\frac{d}{2} & \mbox{\rm ~~;~ if $T<T_c$ }
\end{cases}
\EEQ
In contrast to the equilibrium situation, this is non-positive and by itself gives a clear indication that after a quench to 
$T\leq T_c$, the system never reaches equilibrium. 
In the next two sections, we shall work out what happens in the case of quantum dynamics. 
As we shall show, $Z(t)<0$ may occur for non-equilibrium quantum quenches, but the time-dependence
can be quite different from the classical result, in particular for quenches deep into the ordered phase.

In order to study what happens after a quench from the disordered phase, 
the system must be prepared by choosing initial two-point correlators. For a quantum equilibrium initial state, this
requires $\Xi_{\vec{k}}(0) = 0$ and $Q_{\vec{k}}(0)=:\mathcal{A}_k$ and 
$\Pi_{\vec{k}}(0)=\mathcal{C}_k$, where $\mathcal{A}_{\vec{k}}=\mathcal{A}_{\vec{k}}(T_0,g_0)$ 
and $\mathcal{C}_{\vec{k}}=\mathcal{C}_{\vec{k}}(T_0,g_0)$ are chosen to specify the initial state further. 
The quench amounts to changing $T_0\mapsto T$ and $g_0\mapsto g$ to their final
values which are kept fixed during the system's evolution.  
The two-point correlators are found from the system (\ref{eq:matrixmotion}) by evaluating the matrix exponential which finally gives 
\begingroup
\allowdisplaybreaks
\begin{subequations} \label{2Punkt}
\begin{align}
\nonumber
Q_{\vec{k}}(t) =& \: \e^{-\frac{\gamma}{g}\Delta_t}\bigg[\mathcal{A}_{\vec{k}}\cos^2\sqrt{t\Delta_t}
+\mathcal{C}_{\vec{k}}g^2 t \frac{\sin^2\sqrt{t\Delta_t}}{\Delta_t}\bigg]
\\\nonumber
&+\frac{\gamma}{2}\int_0^t \!\D \tau\: \sqrt{\Delta_\tau'\,}\Bigg(
\bigg[\frac{\cos\sqrt{t\Delta_t}\sin\sqrt{\tau\Delta_\tau}}{\sqrt{\Delta_\tau/(\tau \Delta_\tau')}}
-\frac{\sin\sqrt{t\Delta_t}\cos\sqrt{\tau\Delta_\tau}}{\sqrt{\Delta_t/(t \Delta_\tau')}}\bigg]^2\\
&+\bigg[\cos\sqrt{t\Delta_t}\cos	\sqrt{\tau\Delta_\tau}
-\frac{\sin\sqrt{t\Delta_t}\sin\sqrt{\tau\Delta_\tau}}{\sqrt{(\Delta_t/t)/(\Delta_\tau/\tau)}}\bigg]^2\Bigg)
\e^{\frac{\gamma}{g}(\Delta_\tau-\Delta_t)}\coth\frac{\sqrt{\Delta_\tau'}}{2T}
\label{eq:qq}
\\ \nonumber \\ \nonumber \\ \ \nonumber
\Pi_{\vec{k}}(t) =& \: \e^{-\frac{\gamma}{g}\Delta_t}\bigg[\mathcal{A}_{\vec{k}}g^2 t\frac{\sin^2\sqrt{t\Delta_t}}{\Delta_t}
+\mathcal{C}_{\vec{k}}\cos^2\sqrt{t\Delta_t}\bigg]
\\ \nonumber
&+\frac{\gamma}{2g^2}\int_0^t\!\D\tau\:\sqrt{\Delta_\tau'} \Bigg(\bigg[\frac{\sin\sqrt{t\Delta_t}\sin\sqrt{\tau\Delta_\tau}}{\sqrt{t/\Delta_t}}
+\frac{\cos\sqrt{t\Delta_t}\cos\sqrt{\tau\Delta_\tau}}{\sqrt{\tau/\Delta_\tau}}\bigg]^2
\\
&+\bigg[\frac{\sin\sqrt{t\Delta_t}\cos\sqrt{\tau\Delta_\tau}}{\sqrt{t/\Delta_t}}
-\frac{\cos\sqrt{t\Delta_t}\sin\sqrt{\tau\Delta_\tau}}{\sqrt{\tau/\Delta_\tau}}\bigg]^2\Bigg)
\e^{\frac{\gamma}{g}(\Delta_\tau-\Delta_t)}\coth\frac{\sqrt{\Delta_\tau'}}{2T}
\\ \nonumber \\ \nonumber \\ \ \nonumber
\Xi_{\vec{k}}(t) =& \: \e^{-\frac{\gamma}{g}\Delta_t}\bigg[\frac{\mathcal{C}_{\vec{k}}g}{\sqrt{\Delta_t}}
-\mathcal{A}_{\vec{k}}\sqrt{\Delta_t}\bigg]\sin2\sqrt{t\Delta_t}
+ \frac{\gamma}{g}\int_0^t \!\D \tau\:\e^{\frac{\gamma}{g}(\Delta_\tau-\Delta_t)}\coth\frac{\sqrt{\Delta_\tau'}}{2T}
\\ \nonumber
&\times \sqrt{\Delta_\tau}\Bigg(\sqrt{\frac{t}{\tau}}
\bigg[\frac{\cos\sqrt{t\Delta_t}\cos\sqrt{\tau\Delta_\tau}}{\sqrt[4]{(\Delta_\tau/\tau)/(\Delta_t/t))}}
+\frac{\sin\sqrt{t\Delta_t}\sin\sqrt{\tau\Delta_\tau}}{\sqrt[4]{(\Delta_t/t)/(\Delta_\tau/\tau))}}\bigg]
\bigg[\frac{\sin\sqrt{t\Delta_t}\cos\sqrt{\tau\Delta_\tau}}{\sqrt[4]{(\Delta_\tau/\tau)/(\Delta_t/t))}}
\\ \nonumber 
&+\frac{\cos\sqrt{t\Delta_t}\sin\sqrt{\tau\Delta_\tau}}{\sqrt[4]{(\Delta_t/t)/(\Delta_\tau/\tau))}}\bigg]
+\frac{\tau\Delta_\tau'}{\sqrt{\Delta_\tau}}\sqrt{\frac{t}{\tau}}
\bigg[\frac{\sin\sqrt{t\Delta_t}\cos\sqrt{\tau\Delta_\tau}}{\sqrt[4]{(\Delta_\tau/\tau)/(\Delta_t/t))}}
-\frac{\cos\sqrt{t\Delta_t}\sin\sqrt{\tau\Delta_\tau}}{\sqrt[4]{(\Delta_t/t)/(\Delta_\tau/\tau))}}\bigg] \times
\\
& \times \bigg[\frac{\cos\sqrt{t\Delta_t}\sin\sqrt{\tau\Delta_\tau}}{\sqrt[4]{(\Delta_\tau/\tau)/(\Delta_t/t))}}
-\frac{\sin\sqrt{t\Delta_t}\cos\sqrt{\tau\Delta_\tau}}{\sqrt[4]{(\Delta_t/t)/(\Delta_\tau/\tau))}}\bigg] \Bigg)
\end{align}
\end{subequations}
\endgroup

\noindent
with the definition (the $\vec{k}$-dependence is suppressed for readability)
\BEQ\label{vartheta}
\Delta_t := g (Z(t) + t\omega_{\vec{k}}) \ .
\EEQ
and the notation $\Delta_t'=\frac{\D \Delta_t}{\D t}$. 
This gives the full solution of the quantum problem and must be evaluated by using the the spherical constraint~(\ref{eq:constraint3}), 
viz. $\int_{\mathcal{B}} \frac{\D \vec{k}}{(2\pi)^d} Q_{\vec{k}}(t) =1$. 
This leads to a formidable integro-differential equation for spherical parameter $\mathfrak{z}(t)$.

\subsubsection{Remark on the relaxation towards equilibrium}
In order to arrive at the first understanding of the correlator (\ref{eq:qq}), 
let us assume that there exists a {\em finite} relaxation time $t_r$ such that the
system is stationary for times $t\geq t_r$. For such times, we can write $\mathfrak{z}=\mathfrak{z}_\infty\simeq Z(t)/t$. 
Furthermore, the integration in (\ref{eq:qq}) can be split according to
$[0,t] = [0,t_r] \cup [t_r,t]$. In the limit $t\to \infty$ we would have 
\BEQ
Q_{\vec{k}}(\infty) = \demi \frac{\sqrt{g\,}}{\sqrt{\mathfrak{z}_\infty
+\omega_{\vec{k}}}\,}\coth\bigg[\frac{\sqrt{g\,}}{2T} \sqrt{ \mathfrak{z}_\infty+\omega_{\vec{k}}\,} \bigg] 
\EEQ
and this is consistent with the equilibrium correlator eq.~(\ref{eq:corr_equi}). We can then conclude:

\noindent
\textit{If the system relaxes towards a stationary state with a positive spherical parameter 
$\mathfrak{z}_{\infty}>0$, this stationary state has to be the unique thermodynamic equilibrium.}

\section{Semi-classsical limit}

Eqs.~(\ref{2Punkt}) contain two contributions of a different physical nature. 
The first one contains the contributions from the fluctuations in the initial state, while
the second one describes the fluctuations generated by the coupling to the external bath. 
These latter contributions appear far too formidable to yield to a  direct approach. 
We therefore restrict to the study of two limiting cases. In this section, we shall present a quasi-classical limit 
designed to reduce the complexity of the interaction with
the bath considerably, so that it can be treated. In the next section, 
we consider a quench deep into the ordered phase, where the bath interactions are expected to produce
only sub-leading terms in the long-time limit. 

The spin-spin correlator, eq.~(\ref{eq:qq}), contains complicated terms depending on $\Delta_t$, 
which in turn depends on the quantum coupling $g$. 
This suggests that a semi-classical description should mean that the quantum fluctuations generated by such terms should be 
small and could be achieved by letting $\Delta_{t} \to 0$. 
Simplifying, eq.~(\ref{eq:qq}) would then reduce to  
\BEQ \label{eq:qq2}
Q_{\vec{k}}(t) \simeq \frac{\e^{-\gamma t\omega_{\vec{k}}}}{G(t)} 
+\gamma\sqrt{\frac{g}{4}\,}\int_0^t \!\D \tau\:\frac{G(\tau)}{G(t)}\,\e^{-\gamma(t-\tau)\omega_{\vec{k}}}  
\sqrt{\mathfrak{z}(\tau)+\omega_{\vec{k}}\,}\,\coth\bigg[\frac{\sqrt{ g }}{2T} \sqrt{ \mathfrak{z}(\tau)+\omega_{\vec{k}}\,} \bigg] 
\EEQ
with the definition
\BEQ \label{G_def}
G(t) := \e^{\gamma Z(t)}
\EEQ
Inserted into the spherical constraint  $\int_{\mathcal{B}} \frac{\D \vec{k}}{(2\pi)^d} Q_{\vec{k}}(t) =1$, 
this gives a still complicated integro-differential equation for $G(t)$. 
Manageable expressions can be found by expanding the thermal occupation. We introduce as a small parameter 
\BEQ
\eps =\sqrt{ \frac{ g}{T} } \ .
\EEQ
which measures the relative importance of quantum and thermal fluctuations. For $\eps\to 0$ 
\BEQ\label{series_nk}
\coth \left( \eps \sqrt{\frac{\mathfrak{z}(\tau)+\omega_{\vec{k}}}{4T}} \right) =  
\frac{1}{\eps} \sqrt{\frac{4T}{\mathfrak{z}(\tau)+\omega_{\vec{k}}}}+\frac{\eps}{3} \sqrt{\frac{\mathfrak{z}(\tau)+\omega_{\vec{k}}}{4T}}
+{\rm O}(\eps^3)
\EEQ
The first term in this expansion reproduces the classical model while the higher-order terms give successive quantum corrections. 

\subsection{Classical limit}

Stopping at the first term in the expansion~(\ref{series_nk}) and choosing $Q_{\vec{k}}(0)=1$ 
for an infinite-temperature initial state gives the classical spin-spin correlator
\BEA
Q_{\vec{k}}(t) &=& \frac{1}{G(t)}\left(\e^{-\gamma t \omega_{\vec{k}}}
+\gamma T\int\limits_0^t\!\D\tau\:  G(\tau)\e^{-\gamma(t-\tau)\omega_{\vec{k}}} \right)
\EEA
From the spherical constraint eq.~(\ref{eq:constraint3}), in the $\mathcal{N}\to\infty$ limit, one finds a Volterra integral equation for $G(t)$ 
\BEQ\label{eq:tmp_classical}
G(t) =  F(t)   +  \gamma T\int\limits_0^t\!\D\tau \:  G(\tau) F(t-\tau) = F(t) + \gamma T (F\star G)(t)  
\EEQ
where $\star$ denotes a convolution and with the integral kernel 
\BEQ\label{F_def}
F(t) = \int_{\mathcal{B}} \frac{\D \vec{k}}{(2\pi)^d}\: e^{-\gamma t \omega_{\vec{k}}} = \left(e^{- 2 \gamma t} I_0 (2 \gamma t)\right)^d
\EEQ
and $I_0(x)$ is a modified Bessel function \cite{Abra65}. 
Up to a rescaling $\demi\gamma T\mapsto T$ of temperature, this reproduces the dynamical spherical constraint of the
classical model, see \cite[eq. (2.23)]{Godr00b}. 
Of course, this was to be expected from our derivation of the Lindblad dissipator (\ref{Lind4}). 

For a deep quench to temperatures $T\ll T_c(d)$ (or $T=0$), 
the solution of (\ref{eq:tmp_classical}) trivially is $G(t)\simeq F(t)$, up to corrections to scaling. 
As we shall see in  section~6, the solution of the analogous deep {\em quantum} quench is far from being trivial. 

\subsection{Leading quantum correction}

New insight beyond the classical limit is found if one includes the first quantum correction from the expansion (\ref{series_nk}) 
in  eq.~(\ref{eq:qq2}). We then get 
\BEA \label{5.8}
Q_{\vec{k}}(t) &\simeq& \frac{\e^{-\gamma t\omega_{\vec{k}}}}{G(t)}
+\gamma \int_0^t \!\D \tau\: \bigg[T +\frac{ g}{12 T} (\mathfrak{z}(\tau) + \omega_{\vec{k}}) \bigg] \frac{G(\tau)}{G(t)}
\:\e^{-\gamma(t-\tau)\omega_{\vec{k}}}
\EEA
The spherical constraint (\ref{eq:constraint3}) becomes again a Volterra-integral equation for $G(t)$. This is seen as follows.
From the definitions (\ref{F_def}) and (\ref{G_def}) we have
\BD
\frac{\D F(t)}{\D t} = -\gamma \int_{\mathcal{B}} \frac{\D \vec{k}}{(2\pi)^d}\: \omega_{\vec{k}}\: e^{-\gamma t \omega_{\vec{k}}} \;\; , \;\;
\frac{\D G(t)}{\D t} = \gamma \mathfrak{z}(t) G(t)
\ED
Integrating (\ref{5.8}) gives
\BEA
G(t) &=& F(t) +\gamma T \left( G\star F\right)(t) + \frac{g}{12 T} \int_0^{t} \!\D\tau\: 
\left[ \frac{\D G(\tau)}{\D \tau} F(t-\tau) + G(\tau) \frac{\D F(t-\tau)}{\D \tau}\right] \nonumber \\
&=& F(t) +\gamma T \left( G\star F\right)(t) + \frac{g}{12 T} \int_0^{t} \!\D\tau\: \frac{\D}{\D\tau} \left( G(\tau) F(t-\tau) \right) \nonumber
\EEA
and using the initial values $G(0)=F(0)=1$, this can be recast as
\BD
G(t) \left( 1 - \frac{g}{12 T} \right) = F(t) \left( 1 - \frac{g}{12 T} \right)  + \gamma T \left( G\star F\right)(t)
\ED
The spherical constraint can now be written as 
\BEQ\label{eq:Qcor}
G(t) = F(t) + \gamma T^\star \int_0^t \!\D \tau\: G(\tau) F(t-\tau)
\EEQ
which is identical to the classical constraint eq.~(\ref{eq:tmp_classical}), if one introduces an effective temperature
\BEQ \label{eq:Tg}
T^{\star} = T^\star(g) = \frac{T}{1-\frac{g}{12T}} \simeq T\left(1+\frac{g}{12T} \right)
\EEQ
Remarkably, $T^{\star}(g) = T_{\rm eff}(g)$ is exactly the effective temperature found in section~4 
for the semi-classical {\em equilibrium} {\sc qsm}, see eq.~(\ref{eq:equi_expansion}). 
\begin{figure}[tb]\centering
 \includegraphics[width=.475\textwidth]{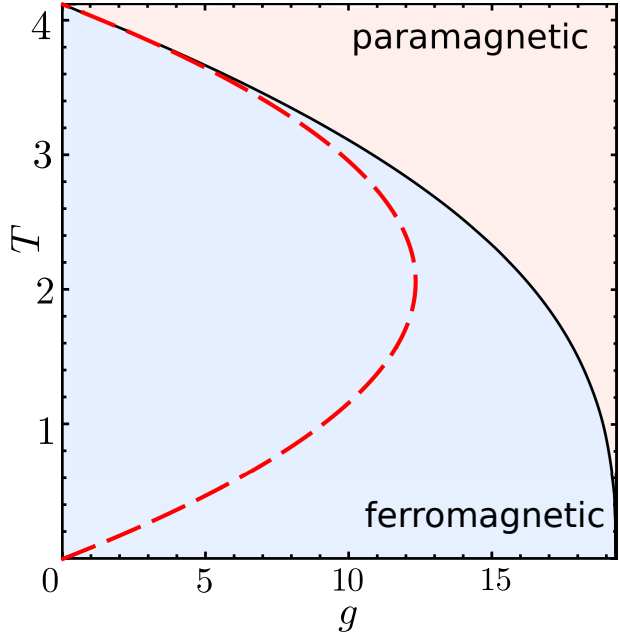}
 \caption{Phase diagram of the isotropic quantum spherical model in $d=3$ dimensions.
 The black curve is the exact critical line \cite{Oliv06} which separates the paramagnetic and ferromagnetic phases. 
 The red curve shows the critical line $T_c^{\star}(g)$ according to eq.~(\ref{eq:Tg}), to first order in $g$.}
 \label{fig:phases}
\end{figure}
In figure \ref{fig:phases} we show the phase diagram of the $3D$ isotropic {\sc qsm} ($\lambda=1)$.  
There is an ordered ferromagnetic and a disordered paramagnetic phase. 
The quantum phase transition occurs on the horizontal axis $T=0$ and the purely thermal phase transition is on the vertical axis $g=0$. 
Clearly, the effective temperature $T^{\star}(g)$ reproduces the exact critical line to first order in $g$. As expected, 
quantum fluctuations reduce the critical temperature $T_c(g)\leq T_c(0)$ with respect to the value of the classical model. 

The identity $T_{\rm eff}(g)=T^{\star}(g)$ of the effective temperatures from the equilibrium and 
dynamical analysis corroborates the correctness of our proposed Lindblad formalism. 
On the other hand, we see that the effective long-time dynamics of the {\em semi-classical} spherical model becomes purely classical, 
although the underlying microscopic dynamics is described by a Lindblad equation and explicitly preserves quantum coherence. 
Quantum effects on the dynamics will only appear in second or higher order in $g$.

%
\subsection{Equal-time spin-spin correlator}

The analysis of the Volterra equation (\ref{eq:Qcor}) is standard, \textcolor{black}{with results identical to the ones for the 
classical $\mbox{\rm O}(n)$-model with model-A dynamics, in the $n\to\infty$ limit} 
(see appendix \ref{ap:volt} for details).  

We have already seen that the formal expression for the single-time spin-spin correlator is
\BEQ \label{5.11}
Q_{{\vec{k}}}(t) =\frac{\e^{-\gamma t \omega_{\vec{k}}}}{G(t)} 
+ \frac{ g}{12 T}\left[1- \frac{\e^{-\gamma t \omega_{\vec{k}}}}{G(t)} \right] 
+ \frac{\gamma T}{G(t)}\int_0^t \!\D\tau\: G(\tau) \e^{- \gamma(t- \tau) \omega_{\vec{k}}}
\EEQ
Its long-time behaviour depends both on the dimension $d$ and on the effective temperature $T^{\star}=T^{\star}(g)$. 

\begin{enumerate}
\item \underline{$T^\star > T^\star_c$}.  This corresponds to the paramagnetic phase at equilibrium and in particular to $d<2$. 
The system relaxes within a finite time-scale $\tau_{\text{eq}}$ towards its (quantum) equilibrium state. 
For $d>2$, the critical temperature $T_c^{\star}>0$ and 
\BEQ \label{eq:Q:low1}
\gamma \tau_{\text{eq}} \stackrel{T^\star\to T^\star_c}{\simeq}  
\left[\frac{T^{\star2}_c}{T^\star-T^\star_c} \frac{|\Gamma(1-\frac{d}{2})|}{(4\pi)^{\frac{d}{2}}} \right]^{2/(d-2)} \ .
\EEQ
The limiting correlation function becomes rapidly constant in time and takes essentially an Ornstein-Zernicke form 
\BEQ \label{eq:Q:low2}
Q_{\vec{k}}(t) \rightarrow Q_{\vec{k}}(\infty) = \frac{g}{12 T} + \frac{T}{\omega_{\vec{k}} +  \xi_{\rm eq}^{-2}}
\EEQ
with the \textit{equilibrium correlation length} $\xi_{\text{eq}}^2=\gamma\tau_{\text{eq}}$. 
We also note a hard-core term, absent in the classical limit $g\to 0$ and 
which in direct space would give a contribution $\sim \frac{g}{12 T}\delta(\vec{r})$.

\item \underline{$T^\star < T^\star_c $}.  For $d>2$ dimensions, the critical point $T_c^{\star}>0$ and there is a ferromagnetic phase. 
In the scaling limit where $t\to\infty$ and $\vec{k}\to\vec{0}$
such that $\omega_{\vec{k}}t$ remains finite, we find the dynamical scaling form
\BEQ \label{eq:Q:high}
Q_{\vec{k}}(t)\simeq \frac{ g}{12 T}+ \e^{- \gamma t \vec{k^2}} (4\pi\gamma t)^{{d}/{2}}\left(1-\frac{ g}{12 T}\right) m^2 
\EEQ
Fourier-transforming to direct space gives the spin-spin correlator
\BEQ
C(t,\vec{r})\simeq \frac{g}{12 T}\:\delta(\vec{r})+   \left(1-\frac{ g}{12 T}\right) m^2 \: \e^{-\frac{\vec{r}^2}{4\gamma t}}
\EEQ
and with the short-hand $m^2 = 1 - \frac{T^{\star}}{T_c^{\star}} \simeq 1 - \frac{T}{T_c}$, sufficiently close to the critical point. 
Indeed, up to the hard-core term, and a small $g$-dependent modification of the scaling amplitude, 
this has the same long-time behaviour as the classical spherical model \cite{Ronc78,Godr00b} 
to which one reverts when taking the limit $g\to 0$.  The gaussian shape of the time-space correlator is a known property of the spherical model.

\item \underline{$T^\star = T^\star_c $}.  For quenches onto the critical line, we find the dynamical scaling form 
\BEQ \label{eq:Q:crit}
Q_{\vec{k}}(t) = \begin{cases} 
\frac{g}{12 T_c} + \frac{2 \gamma T_c}{d-2} t\:  {}_1F_1\left( 1, \frac{d}{2}; -\gamma \omega_{\vec{k}} t \right) & \mbox{\rm ~~;~ if $2<d<4$} \\
\frac{g}{12 T_c} + \frac{T_c}{\omega_{\vec{k}}} \left( 1 - e^{-\gamma \omega_{\vec{k}} t}\right) & \mbox{\rm ~~;~ if $4<d$} \\
\end{cases}
\EEQ
Apart from the hard-core term, this agrees with what is known in the classical model. 
\end{enumerate}
In particular, we recover for $T^\star \leq T^\star_c $ the dynamical exponent $z=2$, 
characteristic for diffusive dynamics of the basic degrees of freedom. 


\section{Disorder-driven dynamics after a deep quench}
%

We now turn to a different quench where quantum effects should be dominant for the long-time behaviour. 
The two-point correlators (\ref{TwoPoint}) contain contributions 
(i) from the fluctuations of the initial state and 
(ii) fluctuations which come from the exchange with the bath. 
In classical systems, the second term dominates for quenches onto the critical point, 
but only generates corrections to scaling for quenches into the two-phase coexistence region, 
where the first contribution dominates. Indeed, for classical systems the long-time scaling
behaviour in the entire two-phase region is the same as for the deep quenches to zero temperature $T=0$. 
We anticipate that a similar result should also hold true for quantum systems, quenched deep into the ordered phase with $g\ll g_c(d)$. 
At $T=0$, this is possible for dimensions $d>1$, where $g_c(d)>0$. 
Therefore, instead of eqs.~(\ref{TwoPoint}) or their formal solutions~(\ref{2Punkt}), we shall rather consider the correlators 
\begin{subequations} \label{6.1}
\begin{align}
Q_{\vec{k}}(t) =& \: \e^{-\frac{\gamma}{g}\Delta_t}
\bigg[\mathcal{A}_{\vec{k}}\cos^2\sqrt{t\Delta_t}+\mathcal{C}_{\vec{k}} g^2 t \frac{\sin^2\sqrt{t\Delta_t}}{\Delta_t}\bigg] 
\\
\Pi_{\vec{k}}(t) =& \: \e^{-\frac{\gamma}{g}\Delta_t}
\bigg[\mathcal{A}_{\vec{k}} g^2 t\frac{\sin^2\sqrt{t\Delta_t}}{\Delta_t}+\mathcal{C}_{\vec{k}} \cos^2\sqrt{t\Delta_t}\bigg]
\\ 
\Xi_{\vec{k}}(t) =& \: \e^{-\frac{\gamma}{g}\Delta_t}
\bigg[\frac{\mathcal{C}_{\vec{k}} g}{\sqrt{\Delta_t}}-\mathcal{A}_{\vec{k}} \sqrt{\Delta_t}\bigg]\sin2\sqrt{t\Delta_t}
\end{align}
\end{subequations}
along with $\Delta_t := g (Z(t) + t\omega_{\vec{k}})$. The terms neglected therein, with respect to (\ref{TwoPoint}), 
should for weak bath coupling $\gamma$ and for a quench deep into
the ordered phase only account for corrections to the leading scaling we seek. 
In our exploration of the coherent dynamics of the {\sc qsm}, we conjecture that this is so and 
we shall inquire in particular whether a dynamical behaviour distinct from the one found in the quasi-classical case can be obtained. 

As we shall see, and in contrast to the classical model, the solution of the dynamics is non-trivial.

\subsection{Spherical Constraint and Asymptotic Behaviour of the Spherical Parameter}
Accepting the reduced form (\ref{6.1}) for the two-point correlators, we concentrate on their dissipative dynamics, 
dominated by the {\em initial} disorder.  
The spherical constraint simplifies to
\BEQ
1= \int_{\mathcal{B}}\frac{\D \vec{k}}{(2\pi)^d}\:\e^{-\gamma \left(Z(t)+t \omega_{\vec{k}}\right)} 
\left( \mathcal{A}_{\vec{k}}\cos^2\sqrt{t\Delta_t} +\mathcal{C}_{\vec{k}} gt\frac{\sin^2\sqrt{t\Delta_t} }{Z(t)+t \omega_{\vec{k}}} \right)
\EEQ
and the initial conditions are characterised by the constants $\mathcal{C}_{\vec{k}}$ and $\mathcal{A}_{\vec{k}}$. 
This is still a difficult integro-differential equation without an obvious solution. 

\subsubsection{Initial conditions}
Consider a strongly disordered equilibrium initial state, situated far away from criticality. 
Then the equilibrium correlators are known \cite{Wald15}. Especially, $\Xi_{\vec{k}}(0)=0$ and
the spherical parameter $\mathfrak{z}_0 \gg 1$. We call this an \textit{infinitely disordered state}. 
Such states are characterised by an equal occupation number of all modes ${\vec{k}}$, such that
the equilibrium correlators eq.~(\ref{eq:corr_equi}) simplify to 
\begin{figure}[ht]
\centering
\includegraphics[width=.5\textwidth]{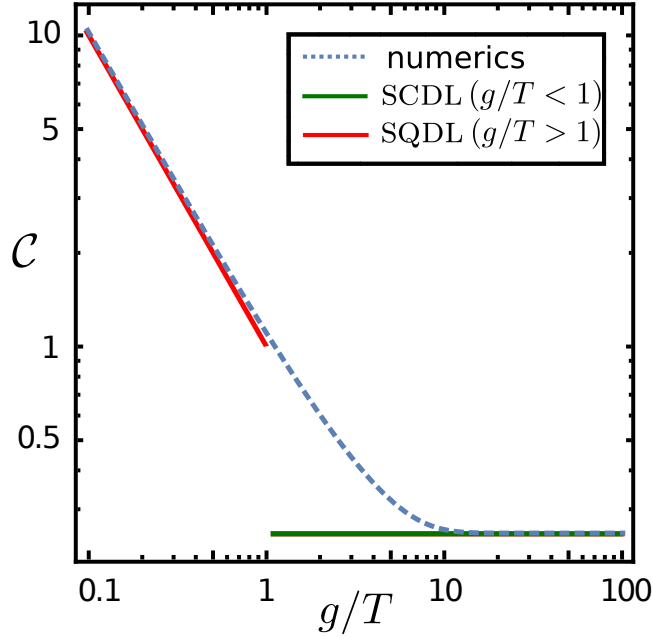}
\caption{The initial parameter $\mathcal{C}=\mathcal{C}(g_0/T_0)$ and the two limits of {\sc scdl} and {\sc sqdl}.\label{fig:C}}
\end{figure}
\BEQ \label{6.3}
Q_{\vec{k}}(0)\simeq\sqrt{\frac{g_0}{4\mathfrak{z}_0}} \coth\left(\sqrt{\frac{\mathfrak{z}_0 g_0}{4 T_0^2}}\right)\stackrel{!}{=}1
\;\; , \;\;
\Pi_{\vec{k}}(0)\simeq\sqrt{\frac{\mathfrak{z}_0}{4g_0}}\coth\left(\sqrt{\frac{\mathfrak{z}_0 g_0}{4 T_0^2}}\right)=\frac{\mathfrak{z}_0}{g_0}=:\mathcal{C}
\EEQ
where the first relation follows from the spherical constraint. 
Hence the single constant $\mathcal{C}$ characterises the infinitely disordered initial state. 
Since $\mathfrak{z}(T_0,g_0)$ is defined self-consistently by the spherical constraint, 
no explicit expression $\mathcal{C}(g_0,T_0)$ is available. Solving (\ref{6.3}) numerically, the
parameter $\mathcal{C}=\mathcal{C}(g_0/T_0)$ is traced in figure~\ref{fig:C}. 
Two limit cases can be identified, which are both obtained for $\mathfrak{z}_0\gg 1$. 
\begin{enumerate}
\item \underline{the strong classical-disorder limit} ({\sc scdl}), defined by the condition  $g_0\ll T_0^2$, 
along with $\mathfrak{z}_0\gg 1$. A first-order Taylor series of $\coth$ gives
\BEQ\label{eq:c-scdl}
\mathcal{C}\simeq \frac{T_0}{g_0}\gg 1
\EEQ
The {\sc scdl} is obtained when $\mathcal{C}$ is becoming large and positive. 

\item \underline{the strong quantum-disorder limit} ({\sc sqdl}), defined by the condition $g_0\gg T_0^2$, 
along with $\mathfrak{z}_0\gg 1$. An asymptotic expansion now gives 
\BEQ\label{eq:c-sqdl}
\mathcal{C}\simeq \frac{1}{4}
\EEQ
which is the smallest admissible value for $\mathcal{C}$ for a quantum \textit{equilibrium} initial state. 
\end{enumerate}
Clearly, more general initial conditions interpolate between the limiting cases eqs.~(\ref{eq:c-scdl},\ref{eq:c-sqdl}). 
It is conceptually significant that initial momentum correlators \textit{must} be present. 

At first  sight, one might have appealed to an analogy with classical initial disordered states 
and expected that $\mathcal{C}=0$ would be possible, but figure~\ref{fig:C} shows that 
such a state does not correspond to a quantum disordered equilibrium state. 
Choosing $\mathcal{C}=0$ means that one is considering an `artificial' initial state, inconsistent with the
laws of quantum mechanics.

We consequently parametrise our disordered initial state by
\BEQ
Q_{\vec{k}}(0) = 1, \hspace{.5cm} \Pi_{\vec{k}}(0) = \mathcal{C}
\EEQ
and then quench the system to temperature $T=0$ and a small coupling $g\ll g_c(T)$ far below the quantum critical point. 
\subsection{The spin-spin correlator}
Our first task is to solve the spherical constraint. This requires in turn to cast the spin-spin correlator into a more manageable form. 
In the deep-quench scenario just defined, the spin-spin correlator becomes 
\BEQ
Q_{\vec{k}}(t) = \e^{-\gamma(Z(t)+t\omega_{\vec{k}})}\left[\cos^2\left(\sqrt{gt(Z(t)+t\omega_{\vec{k}})}\right) 
+\frac{\mathcal{C} g t}{Z(t)+\omega_{\vec{k}}}\sin^2\left(\sqrt{gt(Z(t)+t\omega_{\vec{k}})}\right) \right]
\EEQ
Recall from the classical dynamics that $Z^{cl}(t) \simeq -\frac{d}{2}\ln t$ for $t\to \infty$ at $T=0$. 
In order to prepare for the possibility that $Z(t)<0$ also in the quantum case, 
it will turn out to be advantageous to 
rewrite the correlator in terms of a hyper-geometric function\footnote{We suppress the explicit time-dependence of $Z=Z(t)$.}
\BEA
Q_{\vec{k}}(t) &=&\demi\left[1+\frac{\mathcal{C}gt}{Z+t\omega_{\vec{k}}}
+\left(1-\frac{\mathcal{C}gt}{Z+t\omega_{\vec{k}}}\right)
\leftidx_0 F_1\left(\demi ;- g t(Z+t\omega_{\vec{k}})\right)\right]\e^{-\gamma(Z+t\omega_{\vec{k}})}
\nonumber \\
&=&\left[ 1 +\demi \left(1-\frac{\mathcal{C}gt}{Z+t\omega_{\vec{k}}}\right) 
\sum_{n=1}^\infty\frac{(-gt)^n}{\left(\demi\right)_n}\frac{(Z+t\omega_{\vec{k}})^n}{\Gamma(n+1)}\right]\e^{-\gamma(Z+t\omega_{\vec{k}})}
\EEA
where $(a)_n=\frac{\Gamma(a+n)}{\Gamma(a)}$ denotes the Pochhammer symbol. 
The evaluation of the spherical constraint becomes more simple if all dependence on $\vec{k}$ 
is brought into the exponential. This will allow to derive factorised representations which in turn
will permit to rewrite the expressions where the dimension $d$ becomes a parameter which then can be generalised and considered as real $d\in\mathbb{R}$.  
This is easily achieved as
\BEQ
Q_{\vec{k}}(t)=\left[ 1 +\demi \sum_{n=1}^\infty\left(\partial_\gamma^n+\mathcal{C}gt\partial_\gamma^{n-1}\right) 
\frac{1}{\left(\demi\right)_n}\frac{(gt)^n}{n!}\right]\e^{-\gamma(Z+t\omega_{\vec{k}})}
\label{eq:dqQ}
\EEQ
\subsection{The spherical constraint}
We recall the spherical constraint (\ref{eq:constraint3}), written as $1=\int_{\mathcal{B}}\frac{\D\vec{k}}{(2\pi)^d}Q_{\vec{k}}$, 
and define\footnote{In the short-hand 
$\mathfrak{f}=\mathfrak{f}(\gamma)$, the dependence on $Z$ and $t$ is suppressed.}
\BEQ \label{6.10} 
\mathfrak{f}(\gamma) :=\int_{\mathcal{B}}\frac{\D\vec{k}}{(2\pi)^d}\: \e^{-\gamma(Z+t\omega_{\vec{k}})} 
= \e^{-\gamma Z} \left(\e^{-2\gamma t}I_0(2\gamma t)\right)^d \; \stackrel{t\to\infty}{\simeq}\e^{-\gamma Z}(4\pi\gamma t)^{-\frac{d}{2}}
\EEQ
Thus, we can rewrite the constraint using eq.~(\ref{eq:dqQ}) as\footnote{We use throughout the notation 
$\Gamma\begin{bmatrix}  a_1 & \ldots & a_n \\  b_1 & \ldots & b_m \end{bmatrix} = \frac{\Gamma(a_1) \cdots \Gamma(a_n)}{\Gamma(b_1)\cdots \Gamma(b_m)}$.}
\BEA \label{eq:dqsc}
1 = \mathfrak{f}(\gamma)+ \sum_{n=1}^\infty \frac{(gt)^n}{2}\: \Gamma\begin{bmatrix}
                                                                   \demi&\\
                                                                   n+\demi&n+1
                                                                  \end{bmatrix}
\left(\partial_\gamma^n+\mathcal{C}gt\partial_\gamma^{n-1}\right)\mathfrak{f}(\gamma)
\EEA
It is shown in appendix~\ref{app:derivative} that in the long-time limit, the derivative can be written as
\BEQ \label{eq:derivative}
\partial_\gamma^n \mathfrak{f}(\gamma) \simeq (-1)^n \mathfrak{f}(\gamma)\sum_{k=0}^n \Gamma \begin{bmatrix}
                                                         n+1&\frac{d}{2}+k& \\
                                                         \frac{d}{2}& n-k+1&k+1
                                                        \end{bmatrix} \gamma^{-k} Z^{n-k}
\EEQ
At this point, we have achieved a first goal: $d$ merely enters as a parameter and from now, 
we can treat it as continuous by means of an analytic continuation. 
Consequently, the spherical constraint can be cast in the form
\BEQ \label{eq:constraint4}
1 = \frac{\mathfrak{f}(\gamma)}{2}\left(1+\mathfrak{s}_1+\mathfrak{s}_2\right)
\EEQ
with the two double sums (see appendix~\ref{app:DQ} for the derivation)
\BEA
\mathfrak{s}_1&:=&\sum_{n=0}^\infty\sum_{k=0}^n \Gamma\begin{bmatrix}
							    \demi&\frac{d}{2}+k& &\\
							    n+\demi&\frac{d}{2}& n-k+1&k+1
							    \end{bmatrix} 
							    \left(-\frac{gt}{\gamma}\right)^n (\gamma Z)^{n-k} \nonumber\\
		&=&\Phi_3\left(\frac{d}{2};\demi;-gt Z,-\frac{g}{\gamma} t\right) \label{eq:6.14_S1} \\ 
		&\ &\nonumber \\
\mathfrak{s}_2&:=&-\gamma\mathcal{C}gt\sum_{n=1}^\infty\sum_{k=0}^{n-1} \Gamma\begin{bmatrix}
                                                                   \demi	&	\frac{d}{2}+k	&	&\\
                                                                   n+\demi	&	\frac{d}{2}	& n-k	&	k+1
                                                                  \end{bmatrix}\frac{1}{n}
\left(-\frac{gt}{\gamma}\right)^n (\gamma Z)^{n-1-k} \nonumber \\
&=& 2\mathcal{C} g^2 t^2 \int_0^1\!\D w \;\Phi_3\left(\frac{d}{2};\frac{3}{2};-\frac{gt}{\gamma}w,-gtZw\right) \label{eq:6.15_S2}
\EEA
that can be expressed in terms of the Humbert function $\Phi_3$ \cite{Humbert20a,Humbert20b}. 
This function is a confluent of one of Appell's generalisations $F_3$ \cite{Appell26} 
of Gauss' hyper-geometric function to two independent variables \cite{Sriv85}. 
The analysis of the spherical constraint requires the asymptotics of 
these functions when the absolute values of both arguments become simultaneously large. 
Since no information on these appears to be known in the mathematical literature, we shall derive it, 
as is outlined in appendix~\ref{app:DQ}. Indeed, very similar methods can be applied to different, 
but related confluents of the Appell function $F_3$ and will be presented elsewhere \cite{Wald17}. 
For our purposes, we simply state the main result: both sums can be expressed \textit{exactly} as Laplace convolutions
\BEA
\mathfrak{s}_1&=&\Gamma\begin{bmatrix}
                        \frac{1}{2}\\
                        \frac{1}{2}-\epsilon&\epsilon
                       \end{bmatrix}
\sqrt{t\,}\int_0^t\!\D v\:\frac{\leftidx{_1}F_1\left(\frac{d}{2};\demi-\epsilon;-\frac{g}{\gamma}v\right)}{v^{\demi-\epsilon}}
\frac{\leftidx{_0}F_1\left(\epsilon;-gZ(t-v)\right)}{(t-v)^{1-\epsilon}}
\label{eq:6.16S1} \\
\mathfrak{s}_2&=&\mathcal{C} g^2 t^{{3}/{2}}\Gamma\begin{bmatrix}
                        \frac{1}{2}\\
                        \frac{3}{2}-\epsilon&\epsilon
                       \end{bmatrix}
\int_0^1\!\D w \int_0^t \!\D v\: \frac{\leftidx{_1}F_1\left(\frac{d}{2};\frac{3}{2}-\epsilon;-\frac{g}{\gamma} w v\right)}{v^{\epsilon-\demi}}
\frac{\leftidx{_0}F_1\left(\epsilon;-gZ w(t-v)\right)}{(t-v)^{1-\epsilon}}\hspace{.5cm} 
\label{eq:6.17S2}
\EEA
(where $0<\epsilon<\demi$ in $\mathfrak{s}_1$ and $0<\epsilon<\frac{3}{2}$ in $\mathfrak{s}_2$). 
In appendix~\ref{app:DQ}, we first show how these integrals can be de-convoluted and then how their asymptotic limit for 
$t\to \infty$ can be found, using Tauberian theorems \cite{Fell71}. 
We then arrive at the following expression for the spherical constraint
\BEA\nonumber
1&\simeq&\frac{\mathfrak{f}(\gamma)}{2}\left\{1+\left[ 1 + \mathcal{C}\frac{gt}{Z}\left(\e^{\gamma Z} - 1\right) \right]
\left(\frac{\gamma}{g t}\right)^{\frac{d}{2}}
\frac{\leftidx{_0}{F}_1\left(\frac{1-d}{2};-g t Z\right)}{\Gamma\left(\frac{1-d}{2}\right)/ \sqrt{\pi}} \right. \\
& & \left. +\gamma \mathcal{C}g t \bigg[\frac{\leftidx{_1}{F}_1\left(1;2-\frac{d}{2};\gamma Z\right)}{\frac{d}{2}-1}
+\left(\frac{gt}{\gamma}\right)^{1-\frac{d}{2}}
\frac{\leftidx{_1}{F}_2\left( 1-\frac{d}{2};2-\frac{d}{2},\frac{3-d}{2};-gtZ\right)\e^{\gamma Z}}{\left(1-\frac{d}{2}\right)
\Gamma\left(\frac{3-d}{2}\right)/\sqrt{\pi}}
\bigg]\right\} ~~~
\label{eq:hypersc}
\EEA
and recall $\mathfrak{f}(\gamma)$ from (\ref{6.10}). 
This representation, which depends on the initial condition through the parameter $\mathcal{C}$ and contains the dimension $d$ as a continuous parameter, 
will be the basis of our analysis of the physics contained in quantum spherical constraint.

We must solve this equation for $Z=Z(t)$, in the asymptotic limit $t\to\infty$, and for fixed parameters $\gamma$, $g$ and $\mathcal{C}$ and for a given 
dimension $d>1$. The most simple case is given by the initial condition $\mathcal{C}=0$ 
and serves as an illustration on how to solve the spherical constraint. We then have
\BEA
2 e^{\gamma Z} \left(4\pi\gamma t\right)^{d/2} &=& 
1+ \left(\frac{\gamma}{g t}\right)^{{d}/{2}}
\frac{\leftidx{_0}{F}_1\left(\frac{1-d}{2};-g t Z\right)}{\Gamma\left(\frac{1-d}{2}\right)/ \sqrt{\pi}} 
\nonumber \\[.5cm]
&=& 1 + \gamma^{d/2} \left( \pi gt\right)^{1/2} \left(\frac{|Z|}{gt}\right)^{(d+1)/4} I_{-(d+1)/2}\left( 2\sqrt{gt|Z|\,}\,\right)
\EEA
\begin{figure}[tb]
 \centering
 \includegraphics[width=.5\textwidth]{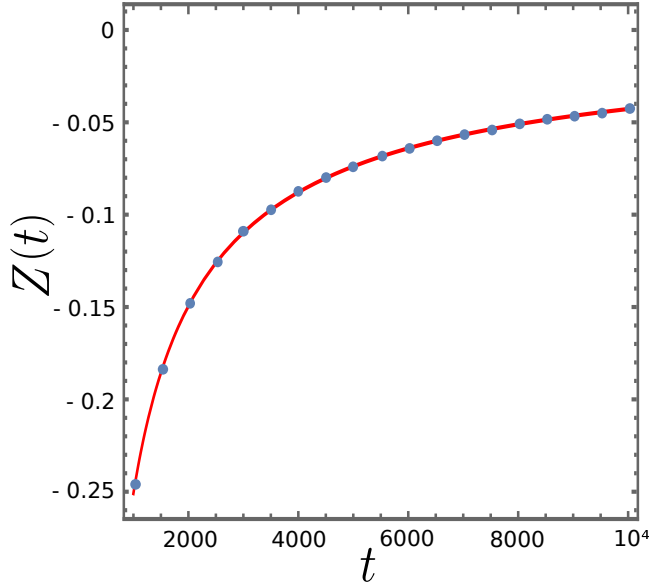}
 \caption{Time-dependence of the integrated Lagrange multiplier $Z(t)$, in $d=2$ dimensions and for the parameters $g=0.2$, 
 $\gamma=0.1$ and $\mathcal{C} = 0$. The full curve is
 the asymptotic form  eq.~(\ref{eq:Z-as}) and the dots come from solving numerically (\ref{eq:hypersc}).}
 \label{fig:numerics-Z}
\end{figure}
where we anticipated that the solution is negative $Z=-|Z(t)|<0$ and $I_{\nu}$ is a modified Bessel function \cite{Abra65}. 
To illustrate this point, we display in fig.~\ref{fig:numerics-Z} 
a typical example of the numerical solution $Z=Z(t)$ of (\ref{eq:hypersc}) with $\mathcal{C}=0$. 
Indeed, the solution is negative and we also observe that $Z(t)\to 0$ for $t\to \infty$. 
The asymptotic form of $I_{\nu}$ then leads to the following simplified form
\BD
2 \left(4\pi g\right)^{d/2} = \left( \frac{\gamma^2 |Z|}{gt}\right)^{d/4} e^{2\sqrt{gt|Z|\,}\,}
\ED
which has the solution
\BEQ\label{eq:Z-as}
|Z(t)| = \frac{d^2}{16 g t}\, W^2\left(\frac{\pi}{d} 16^{\frac{1+d}{d}} g  t^2\right) \simeq \frac{d^2}{16 g} \frac{\ln^2 t}{t}
\EEQ
where $W=W_0$ denotes the principal branch of the Lambert-W function 
\cite{Corl96}.\footnote{Asymptotically, $W(x)\simeq \ln x - \ln\ln x +{\rm o}(1)$ for $x\to\infty$.} 
The agreement with the numerical solution is illustrated in fig.~\ref{fig:numerics-Z}. 
Clearly, this solution applies to all values of $d$ and is distinct from the classical result (\ref{eq:Zcl}). The logarithmic factor indicates corrections
to a simple power-law scaling. We also notice that it is
independent of the coupling $\gamma$ between the system and the bath. 

Any equilibrium initial state must have $\mathcal{C}\geq \frac{1}{4}$. Clearly, eq.~(\ref{eq:hypersc}) with $\mathcal{C}\ne 0$ 
is still too complicated for an explicit solution. However, it turns out that a case distinction between the
dimensions $1<d<2$, $d=2$ and $d>2$ leads to more manageable forms. The details of the calculations are given in appendices~\ref{sec:constraint2d} 
for $d=2$ and~\ref{sec:constraint_d} for $d\ne2$. 
Here, we quote the results. 

\noindent {\bf A.} For $d>2$, $Z=-|Z(t)|<0$ turns out to be negative, in such a way that $t|Z|$ becomes large for large $t$. We have the equation 
\begin{subequations} \label{eq:6.19}
\BEQ
2\,\e^{\gamma Z} (4\pi\gamma t)^{d/2}\simeq 
1+\demi\gamma^{{d}/{2}}\left(1+\mathcal{C}\frac{gt}{|Z|}\right)
\left(\frac{|Z|}{g t}\right)^{\frac{d}{4}}\e^{2\sqrt{g t |Z|}} \;\; \hspace{1truecm}
\mbox{\rm ~~;~ for $d> 2$} 
\label{eq:sc-dq-exact}
\EEQ

\noindent {\bf B.} For $d=2$, we find again $Z=-|Z(t)|<0$ for large times, but now such that $t|Z|\to \vph$ tends to a constant. 
This constant is given by the transcendent equation
\BEQ \label{eq:constraint6_phi} 
\frac{4\pi}{\mathcal{C} g^2} = \varphi\; \leftidx{_2}F_3\left(1,1;\frac{3}{2},2,2;g \varphi \right) \;\; \hspace{6.0truecm}
\mbox{\rm ~~;~ for $d=2$} 
\EEQ

\noindent {\bf C.} Finally, for $1<d<2$, the integrated Lagrange multiplier 
$Z=Z(t)>0$ becomes positive for large enough times and it increases with increasing $t$ beyond any bound. Its value is determined from
\BEA
\lefteqn{2  \left(4\pi\gamma t\right)^{d/2} = \frac{\mathcal{C}gt}{Z} e^{-\gamma Z} }\\
&+& \frac{d\mathcal{C}\gamma^{d/2}}{2}\left(\frac{Z}{gt}\right)^{d/4-1}  
\left[ \frac{3(d+2)(4-d)}{64}\frac{\cos\left(2\sqrt{gtZ\,}\,+\frac{\pi d}{4}\right)}{Z} 
- \frac{\sin\left(2\sqrt{gtZ\,}\,+\frac{\pi d}{4}\right)}{\sqrt{gtZ\,}\,} \right] ~~~~~
\nonumber \\
& & \;\; \hspace{10.7truecm} \mbox{\rm ~~;~ for $1<d<2$} \nonumber 
\EEA
\end{subequations}
However, we also find an intermediate regime, with large but not enormous times, 
where $Z(t)<0$ is still negative. In that regime the effective behaviour  is analogous to the
one found above for $d>2$. 

Summarising, for large times, the  leading asymptotics of the solutions of eqs.~(\ref{eq:6.19}) become 
\BEQ \label{eq:Z_sol}
 |Z(t)| \simeq 
 \begin{cases}
               \frac{(d-2)^2}{4g}\frac{\ln^2 t}{t}           &, ~~  \ d> 2\\[.25cm]
               \varphi\ t^{-1}                               &, ~~  \ d=2 \\[.25cm]
               \left(1-\frac{d}{2}\right) \gamma^{-1} \ln t  &, ~~  \ \frac{4}{3}<d<2 
 \end{cases}
\EEQ
where $\vph$ is given by (\ref{eq:constraint6_phi}). Recall that $Z(t)$ is negative for $d\geq 2$ and positive for $1<d<2$. 
More precisely, for $\frac{4}{3}<d<2$, the large-time behaviour is given by
\BEQ
Z(t) \simeq \left(1-\frac{d}{2}\right) \gamma^{-1} \ln \gamma t 
+ \mathscr{B}(d)\cos\left(2\sqrt{gtZ\,}\,+\frac{\pi d}{4}\right)\frac{t^{1-3d/4}}{\ln^{2-d/4} \gamma t}
\EEQ
and where $\mathscr{B}(d)$ is a known dimension-dependent amplitude. 
Hence the oscillatory term can no longer be treated as a  mere correction for $d<\frac{4}{3}$.\footnote{The occurrence of such a second
`critical dimension' which a qualitative change in the systems' behaviour is a 
little reminiscent of the classical reaction-diffusion process reactions $2A\to\emptyset$ and $A\to 3A$, 
which has the critical dimensions $d_c=2$ and $d_c'\simeq \frac{4}{3}$ \cite{Card96b,Card98}.}  

The intermediate regime seen for dimensions $1<d<2$  for large, but not enormous times where $Z(t)<0$, 
is effectively described by $|Z(t)|\approx \frac{(d-2)^2}{4g}\frac{\ln^2 t}{t}$. 

\begin{figure}[tb]
 \centering
 \includegraphics[width=\textwidth]{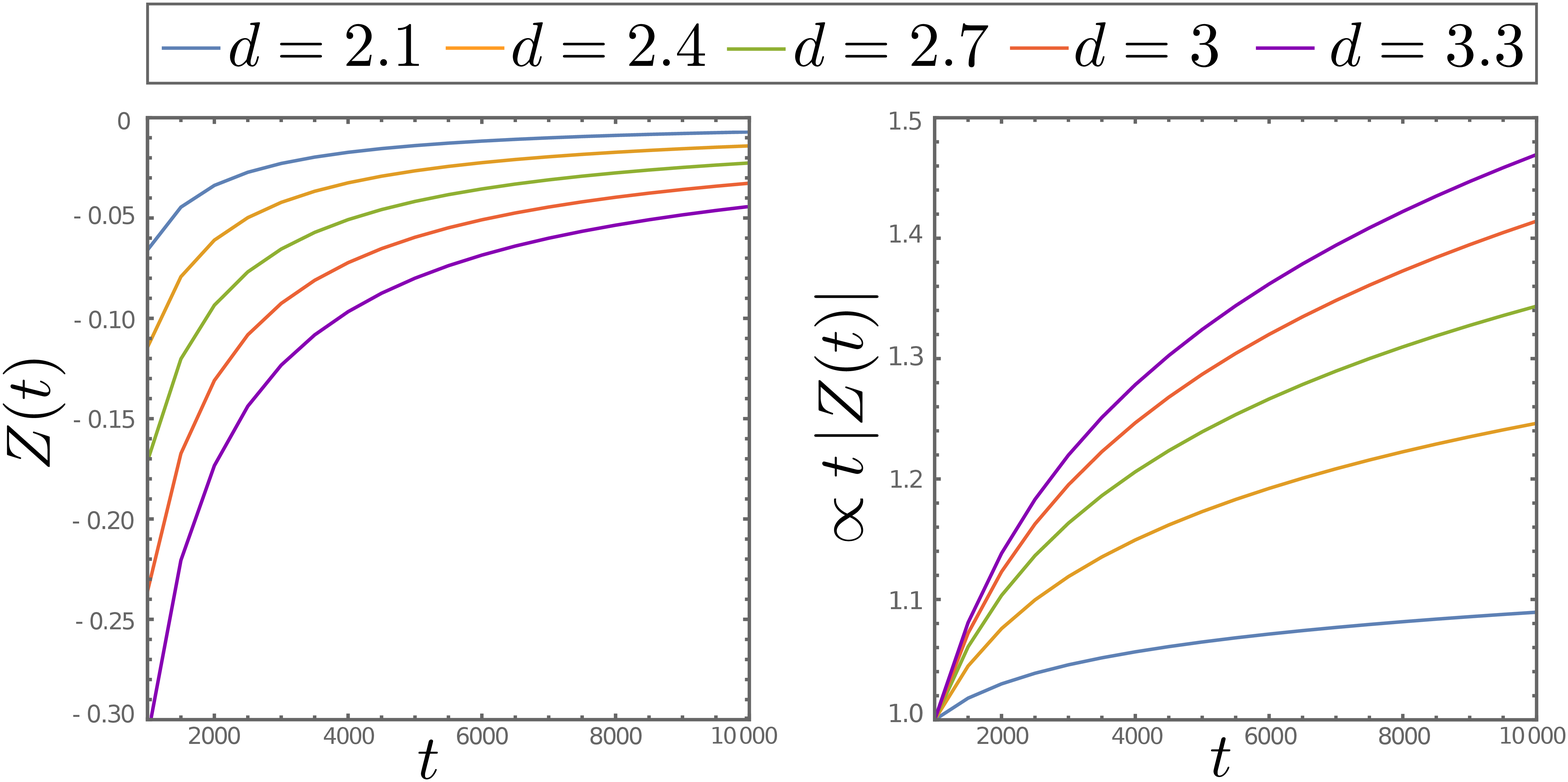}
 \caption[figZ]{\underline{Left panel}: Integrated Lagrange multiplier $Z(t)$ as a function of time $t$  
 obtained by solving eq.~(\ref{eq:hypersc}) numerically, 
 for $d =[2.1,2.4,2.7,3,3.3]$, from top to bottom, and for the parameters $\gamma = 1$, $g = 0.2$, $\mathcal{C} = 1$. \\
 \underline{Right panel}: Integrated Lagrange multiplier $t|Z(t)|$, normalised to unity at $t=1000$, 
 as a function of time and for $d =[2.1,2.4,2.7,3,3.3]$, from bottom to to top, 
 and the same parameters.} 
 \label{fig:behavior}
\end{figure}

In fig.~\ref{fig:behavior}, we illustrate the solution for $d>2$. 
%

Several comments are in order: 
\begin{enumerate}
\item Although the toy initial condition $\mathcal{C}=0$ does indeed reproduce one instance of the long-time behaviour  
found from the physically more sensible
equilibrium initial states with $\mathcal{C}\geq \frac{1}{4}$, it does not capture the full complexity of possible behaviours. 
\item For equilibrium initial states, in $d=2$ dimensions there is a qualitative change in the long-time behaviour of the solution $Z(t)$. 
\item For $d<2$ where the {\sc saqsm} undergoes a quantum phase transition at $T=0$ but where the thermal critical temperature $T_c(d)=0$ vanishes, 
the behaviour of $Z(t)$ is analogous to the one of the classical solution (\ref{eq:Zcl}), although with the opposite sign. 
The Lagrange multiplier $\mathfrak{z}(t)\sim t^{-1}$ has a simple algebraic behaviour. 

Very large times are required to see this regime. In  addition, we find an intermediate regime of large, 
but not enormous times, where the system behaves effectively as for
dimensions $d>2$, up to an amplitude. 
\item For $d>2$ where the system also has a finite critical temperature $T_c(d)>0$, strong logarithmic corrections modify the leading
scaling behaviour, which is distinct from the classical one. 
\item The case $d=2$ is intermediate between the two, with a simple power-law scaling behaviour $|Z(t)|\sim t^{-1}$. 
\item Surprisingly, the influence of the coupling of the coupling $\gamma$ with the bath is also dimension-dependent. 
For $d\geq 2$ dimensions, $\gamma$ disappears from the
leading long-time  behaviour of $Z(t)$, while it is present for $d<2$. 
Therefore, for $d\geq 2$ dimensions, as well as in the intermediate regime for $d<2$, the limit
$\gamma\to 0$ can be formally taken.
\end{enumerate}

The physical meaning of these properties will be understood by analysing the behaviour of the two-point correlators.

\subsection{Correlation function and relevant length scales}

\begin{figure}[t]
 \centering
 \includegraphics[width=.6\textwidth]{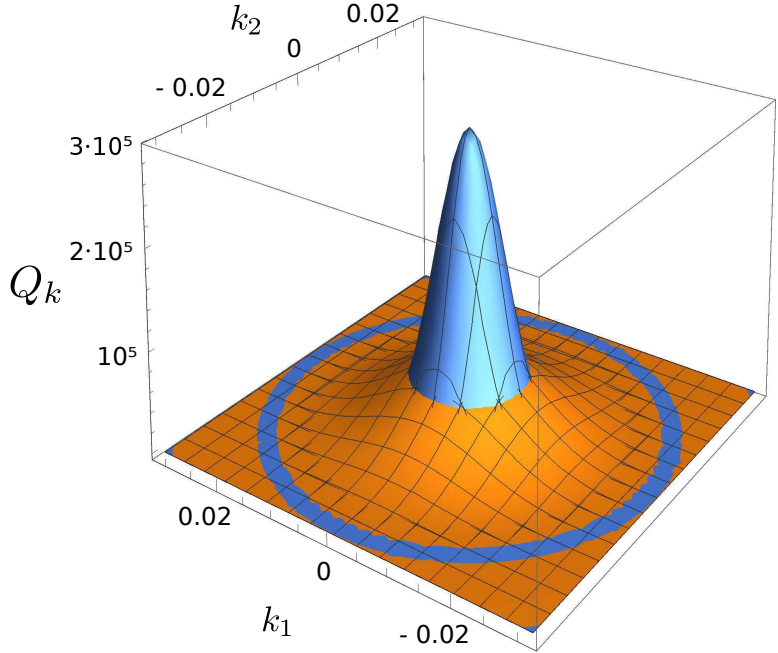}
 \caption[\ Structure factor $Q_k$ in $d=2$ dimensions.]{Structure factor $Q_k$ in $d=2$ dimensions shown in the first Brillouin zone 
 for the parameter values $c=1;g=0.1;\gamma =0.1;t=500\ (\text{orange});\ 1000 \ (\text{blue})$. We observe that the function is 
 sharply peaked around the centre of the Brillouin zone with the peak sharpening with time increasing.}
 \label{fig:Q2d}
\end{figure}

For the deep-quench dynamics the spin-spin correlation function in Fourier space reads
\begin{equation}
Q_{\vec{k}}(t) =\demi\left[1+\frac{\mathcal{C}gt}{Z+t\omega_{\vec{k}}}
+\left(1-\frac{\mathcal{C}gt}{Z+t\omega_{\vec{k}}}\right)\leftidx_0 F_1\left(\demi,- g t(Z+t\omega_{\vec{k}})\right)\right]
\e^{-\gamma(Z+t\omega_{\vec{k}})} 
\end{equation}
We are now interested in transforming this expression back to real space and studying the large-distance behaviour of the correlation. 
This is routinely revealed by a small-$|\vec{k}|$ expansion and in fig~\ref{fig:Q2d}
we see on the $2D$ example that such an expansion is more than reasonable in the asymptotic limit $t\to \infty$. 
We consequently write $\omega_{\vec{k}} \approx |\vec{k}|^2 = k^2$ and observe that $Q_{\vec{k}}$ solely depends on $k=|\vec{k}|$.
This leads to the following simplified expression for the $d$-dimensional inverse Fourier transform 
\begin{equation}
f(R) \propto R^{1-\frac{d}{2}} \int_0^\infty \!\D k \: k^{\frac{d}{2}}  J_{\frac{d}{2}-1}(k R)\hat{f}(k)
\end{equation}

\begin{figure}\centering
 \includegraphics[width=.8\textwidth]{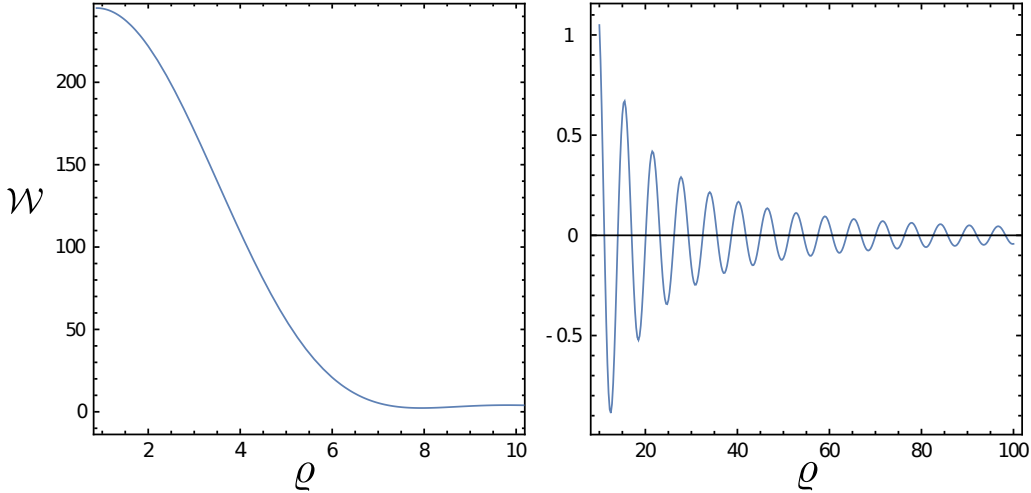}
 \caption[\ Illustration of the scaling function $\mathcal{W}$ in $d=2$.]{Illustration of the scaling function $\mathcal{W}$ 
 in $d=2$ dimensions for different ranges of the scaling variable $\varrho$.}
 \label{fig:scaling2d}
\end{figure}

\noindent{\bf A.} We start the investigation with the case $d=2$. In fig~\ref{fig:Q2d} 
we show a typical structure factor $Q_k$ in 2D for different times
and observe that the distribution is peaked around the zero momentum mode $k=0$ and the peak sharpens
for larger times. One can argue that the main contribution is given by the interval $[0,k^*]$ where $k^*$ 
is the mode where the argument of the hyper-geometric function changes signs and $\leftidx_0F_1$ 
reduces from an exponential contribution to a geometric
function at this point.  We can thus write
%
\begin{equation}
C(R) \propto \int_0^{k^*}\hspace{-.25cm} \D k\; \demi\left[1+\frac{\mathcal{C}gt}{Z+tk^2}
+\left(1-\frac{\mathcal{C}gt}{Z+tk^2}\right)\leftidx_0 F_1\left(\demi,- g t(Z+tk^2)\right)\right]
\e^{-\gamma(Z+tk^2)} k J_{0}(kR)
\end{equation}
By introducing the {\it scaling variable} $\varrho : = \sqrt{\varphi} \frac{R}{t}$ where $\varphi$ is the solution to 
eq.~(\ref{eq:constraint6_phi}), we find in a straightforward fashion using the variable transform
$\mu = \frac{|Z|-tk^2}{|Z|}$ the scaling form
\begin{equation}
 C(R) \propto \mathcal{C} g  \int_0^1 \D\mu\; \frac{\leftidx_0 F_1\left(\demi; g \varphi \mu \right)-1}{\mu} 
 J_0\left(\varrho\sqrt{1-\mu}\right)
 =: \mathcal{C} g \mathcal{W}(\varrho)
\end{equation}
This shows explicitly the dynamical scaling behaviour of the spin-spin correlator with the dynamical exponent $z=1$. 
In fig.~\ref{fig:scaling2d} we show the behaviour of the scaling function $\mathcal{W}$ for different ranges of $\varrho$. For small $\varrho$ the scaling 
function decays in a Gaussian fashion (left hand side) while it shows decaying oscillations for larger values.
%
%
It is instructive to compare with the dynamical scaling seen in the classical spherical model, 
quenched to temperature $T\ll T_c(d)$. For a purely relaxational dynamics without
any conservation law (model A), dynamical scaling is found \cite{Ronc78, Godr00b}, 
whereas in the case of a conserved order-parameter (model B), the existence to two logarithmically
distinct length scales was established long ago \cite{Coni94}. This logarithmic breaking of scale-invariance for conserved dynamics was later shown to be
a peculiarity of the spherical model, see e.g. \cite{Maz06}. The quantum dynamics we are considering here actually
has an infinite number of prescribed conservation laws, namely all canonical commutators between the spherical spins $s_n$ and their conjugate moment $p_n$.
Our finding that at least for $d=2$ a standard dynamical
scaling is found clearly suggests that the {\sc qsm} should not be considered to be as special as its classical counterpart. 
Any breaking of dynamical
scaling which we may find for different values cannot be as readily dismissed as a specific model property  
but could rather be a typical feature for more general models.

\begin{figure}[ht!]
 \centering
 \includegraphics[width=.5\textwidth]{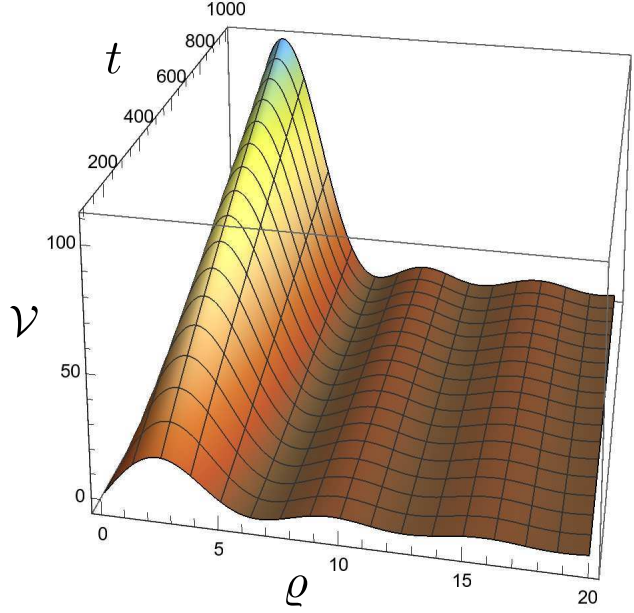}
 \caption[\ Functional dependence of the correlation function for $d = 3$.]{Functional dependence of the correlation function for $d = 3$ and $g=0.1$. 
 We do not find a single scaling function but rather find a dependence on the variable $\varrho$ and the time $t$.}
 \label{fig:multi}
\end{figure}

\vspace{.25cm}
\noindent{\bf B.} In the case $d>2$ the treatment is similar to the case $d=2$ 
since the argument of the hyper-geometric function presents once again a change of signs.
However we have to respect that $\varphi$ is no longer a constant but diverges logarithmically 
as it is shown in eq~(\ref{eq:Z_sol}). This leads to a modified multi-scaling behaviour
\begin{align}
C(R) &\propto \mathcal{C}g
R^{2-d}\underbrace{\int_0^1 \D\mu\;
\frac{\leftidx_0 F_1\left(\demi; \frac{(d-2)^2}{4} \mu \ln^2 t  \right)-1}{\mu} \left(\varrho\sqrt{1-\mu}\right)^{\frac{d}{2}-1}
J_{\frac{d}{2}-1}\left(\varrho\sqrt{1-\mu}\right)}_{=:\mathcal{V}(\varrho,t)} \\  \nonumber
\end{align}
since $\varrho \simeq \frac{\frac{d}{2}-1}{\sqrt{g}} R \ln( t) /t$ which is illustrated in fig~\ref{fig:multi}. 
The explicit logarithmic terms do break simple scale-invariance and point towards the existence of several length scales, 
which are distinguished by logarithmic factors.
We observe a behaviour in terms of $\rho$ which is is qualitatively not too different from the case $d=2$. 
However, the functional dependence on $\rho$ changes strongly when the time is 
increased which is a manifest  of breaking of simple scaling behaviour.

Phenomenologically, this looks analogous to the well-known behaviour of the classical spherical model with conserved order-parameter 
(model B) \cite{Coni94} but here
we obtain this breaking of dynamical scaling by a mere change of the dimension $d$. 
Such a feature has never been seen before, to the best of our knowledge.

\vspace{.25cm}
\noindent{\bf C.}
In $\frac{4}{3}<d<2$ spatial dimensions the situation is different since the spherical parameter is $Z>0$ 
positive and there is no intrinsic cut-off for the integral. We investigate 
first the structure factor $Q_{\vec{k}}$ by pointing out that the contribution $\sim \mathcal{C}g t$ is leading for large times.
\begin{equation}
 Q_{\vec{k}} \simeq \demi \frac{\mathcal{C}g t}{Z+t k^2}\left[1-\leftidx_0F_1\left(\demi; -gt(Z+tk^2)\right)\right] \e^{-\gamma (Z+tk^2)}
\end{equation}
We show in appendix~\ref{app:rewriteQ} that this expression can be rewritten as
\begin{equation}\label{eq:Q-re}
Q_{\vec{k}} \simeq \mathcal{C}g^2 t^2\:  \leftidx_1F_2\left(1;\frac{3}{2},2; -gt(Z+tk^2)\right) \e^{-\gamma (Z+tk^2)} \ .
\end{equation}
Hence, its Fourier transform is readily cast in the form\footnote{One simply uses the change of variables $\mu = t^2 k^2$.}
\begin{equation}
 C(R) \propto \frac{\mathcal{C} g^2}{R^{\frac{d}{2} -1}}\int_0^\infty \D \mu \; \mu^{\frac{d-2}{4}} 
 J_{\frac{d}{2}+1}\left( \sqrt{\mu} R/t \right)\leftidx_1F_2\left( 1;\frac{3}{2},2;-g(t Z + \mu) \right)\e^{-\gamma \mu/t} \ .
\end{equation}
We observe that the integral is exponentially cut off and thus only small $\mu$ values contribute. 
Thus, we can omit the $\mu$ contribution in $\leftidx_1F_2$ since $t Z \to \infty$ and the integral can be evaluated explicitly
\cite[eq~(2.12.9.3)]{Prudnikov2}
\begin{equation}
C(R) \propto \frac{\mathcal{C}g }{(2\gamma)^{d/2}} 
\frac{\sin^2\left( \sqrt{g t Z}\right)}{Z}
\exp\left[ {-\frac{R^2}{4\gamma t}}\right]
\end{equation}
revealing the dynamical exponent $z=2$.

\textcolor{black}{On the other hand, we can study a very large fixed time for which the exponential cutoff does not matter
any more ($t\to \infty$  such that $R/t = \text{cste}$.) In this scenario, one can introduce a cutoff $\mathscr{C}$ to regularise
the integral and find
\begin{align}
 C(R) &\propto \frac{\mathcal{C} g^2}{R^{\frac{d}{2} -1}}\frac{\sin^2\left( \sqrt{g t Z}\right)}{Z}\int_0^{\mathscr{C}}\D \mu \; \mu^{\frac{d-2}{4}} 
 J_{\frac{d}{2}+1}\left( \sqrt{\mu} R/t \right) \nonumber \\
 &=\frac{\mathcal{C} g^2}{R^{-2}}\frac{\sin^2\left( \sqrt{g t Z}\right)}{Z} 
 \left(\frac{\mathscr{C}}{2t}\right)^{\frac{d+2}{2}} \: {}_1F_2\left(\frac{d}{2}+1;\frac{d}{2}+2,\frac{d}{2}+2;-\frac{R^2 \mathscr{C} }{4 t^2}\right)
\end{align}
This implies a dynamical exponent $z=1$ and we thus conclude that depending on the particular limit, the dynamical exponent 
varies between $z=1$ and $z=2$, an effect that will become more apparent in the following analysis of the relevant length scales.
}

Having completed the analysis of the spin correlation function we mention that 
the momentum correlation function can be obtained from the spin correlator by simply exchanging 
$\mathcal{A}_{\vec{k}}$ and $\mathcal{C}_{\vec{k}}$ in eq~(\ref{6.1}).
We thus expect a qualitatively analogous behaviour. The off-coherence $\Xi_{\vec{k}}(t)$ will be considered below in section 6.7.

Having studied the real-space correlation function, we now investigate the \textit{relevant length scale} given by
\begin{equation}
L^2(t) \sim - \frac{\partial^2_k Q_k}{Q_k}\bigg|_{k=0}
\end{equation}
This is readily evaluated to
\begin{align}
\nonumber
L^2\sim \frac{2 t}{Z}\bigg(\frac{\mathcal{C} g t \left[1+\gamma Z - (1+\gamma Z)\leftidx_0 F_1\left(\demi,-gtZ \right) 
-2 g t\: Z\leftidx_0 F_1\left(\frac{3}{2},-gtZ \right)\right]}{
\mathcal{C} gt \left[ 1-\leftidx_0 F_1 \left(\demi,-gt Z\right) \right] + Z\left[ 1+\leftidx_0F_1\left(\demi,-gtZ\right) \right]}\\
+\frac{\gamma Z^2 \left[1+\leftidx_0 F_1\left(\demi;-gt Z\right)+ 2\frac{g}{\gamma} t\: \leftidx_0 F_1\left(\frac{3}{2},-gtZ\right)\right]}{
\mathcal{C} gt \left[ 1-\leftidx_0 F_1 \left(\demi,-g t Z\right) \right] + Z\left[ 1+\leftidx_0F_1\left(\demi,-gtZ\right) \right]}
\bigg)
\label{eq:L2}
\end{align}
For a vanishing quantum coupling $g\to 0$, the relevant length scale reduces to a purely diffusive behaviour introduced by the heat bath
\begin{align}
L^2_\gamma\sim 2 \gamma t  \ .
\label{eq:L3}
\end{align}
The length scale allows to read of the dynamical exponent $z$ according to $ L^2 \sim t^{2/z}$ and we deduce from eq.~(\ref{eq:L2}) $z = 2$, 
as expected for the classical dynamics \cite{Godr00b,Henk10}. A first impression on the different behaviour in the quantum case comes from the
toy initial condition ${\cal C}=0$. Simplifying eq.~(\ref{eq:L2}), we find, for large enough times 
\BEQ
L^2 \simeq 4 g t^2 \frac{\tanh 2 \sqrt{g t |Z|\,}\,}{ 2 \sqrt{g t |Z|\,}\,} \sim \frac{t^2}{\ln t}
\EEQ
hence a logarithmic correction to a dynamical exponent $z = 1$, typical of quantum dynamics. 

In order to evaluate accurately the intrinsic length-scale, taking into account the quantum effects, we have to distinguish, once more the cases 

\vspace{.5cm}

\noindent{\bf A.} \underline{$d=2$}:  Here $Z$ is negative and we can rewrite the hyper-geometric functions as hyperbolic functions. 
Moreover the correlation function obeys a clean scaling behaviour and we find
\begin{equation}
 L^2 \simeq 2 t \left( \gamma  +  \sqrt{\frac{gt}{|Z|}}\;\right) = 2\gamma t +2\sqrt{\frac{g}{\varphi}} t^2 
\end{equation}
indicating a crossover from \textit{diffusive} to \textit{ballistic transport}. The dynamical critical exponent crosses from 
$z=2$ to $z=1$ as we expect for a true quantum 
dynamics\footnote{An exception from the fast ballistic transport are many-body localised systems where information 
spreads much slower\cite{Pal10,Basko16}. Such slow transport has as well
been observed in translation-invariant $1D$ quantum lattice models \cite{Mich17}.} \cite{Cala06,Dutt10,Dutt15,Delf17}.

\textcolor{black}{More commonly, quenches exactly {\em onto} the quantum critical point are studied. In these situations, one finds with increasing times
a crossover from ballistic to diffusive transport, see e.g. \cite{Dutt15}. Here, on the contrary we study a quantum quench from a totally disordered
system deep into the quantum ordered region. Heuristically, the system should order locally and should from `bubbles' which are locally in one of
the equivalent quantum ground states, and whose size should increase with time. As long as these bubbles remain small enough, they should spread like single
quantum particles for which one expects an effective diffusive behaviour. At later times, when the different `bubbles' will interact which each other, 
many-body  quantum properties should dominate and lead to ballistic transport.}

\vspace{.5cm}

\noindent{\bf B.} \underline{$d>2$}: This case can be treated analogously to the case {\bf A} since $Z$ is still negative. 
Nevertheless, we do not have a clean scaling behaviour and logarithmic 
corrections are present in the long-time limit. The length scale reads
\begin{equation}
 L^2 \simeq 2 t \left( \gamma  +  \sqrt{\frac{gt}{|Z|}}\;\right) = 2\gamma t + 2 g(d-2) \frac{t^2}{\ln t}
\end{equation}
and up to logarithmic corrections, we observe the same diffusive to ballistic crossover as for $d=2$ with $z=1$.
\vspace{.5cm}

\begin{figure}[t]
 \centering
 \includegraphics[width=.9\textwidth]{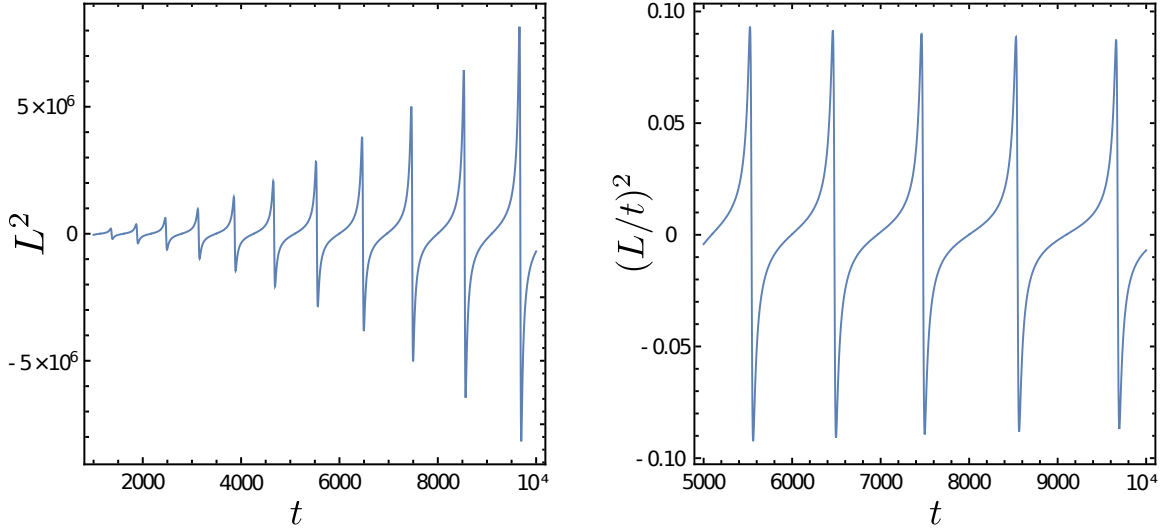}
 \caption[\ Effective characteristic length for $d=1.5$]{\underline{left panel}: effective characteristic length $L^2$ 
 for $d=1.5$, $\gamma = 1$, $g = 0.1$ and $\mathcal{C}= 1$. \underline{right panel}: $(L/t)^2$ for the same parameters as in the left panel.}
 \label{fig:upperboundz}
\end{figure}

\noindent{\bf C.} \underline{$\frac{4}{3}<d<2$}: In this case, 
the spherical parameter is positive and the hyper-geometric functions reduce to trigonometric contributions. The length scale then reduces to
\begin{equation}\label{eq:Llow}
 L^2 \simeq 2 \gamma t - 2\sqrt{\frac{g t }{Z}} \frac{ \mathcal{C}g t^2  \sin2\sqrt{gtZ} }{\mathcal{C} g t\sin^2\sqrt{gtZ} + Z \cos^2\sqrt{gtZ}} 
\end{equation}
and can be recast up to a removable singularity as
\begin{equation}
 L^2 \simeq 2 \gamma t - 4\mathcal{C}g t^2\sqrt{\frac{g t }{Z}} \frac{\tan\sqrt{gtZ} }{\mathcal{C} g t\tan^2\sqrt{gtZ} + Z} 
\end{equation}
This length scale shows an oscillatory behaviour which is shown in the left panel of fig~\ref{fig:upperboundz} 
to which we shall come back later. For now we want to focus on the right panel 
where we show $L^2/t^2$ as a function of time. We see that the peaks are rather constant and $\left|L/t\right|$ 
remains bound for all times what indicates that the dynamical exponent should be $z\geq1$. The specific value of $z$ will depend strongly 
on the specific time window.

\begin{figure}[t]
 \centering
 \includegraphics[width=.5\textwidth]{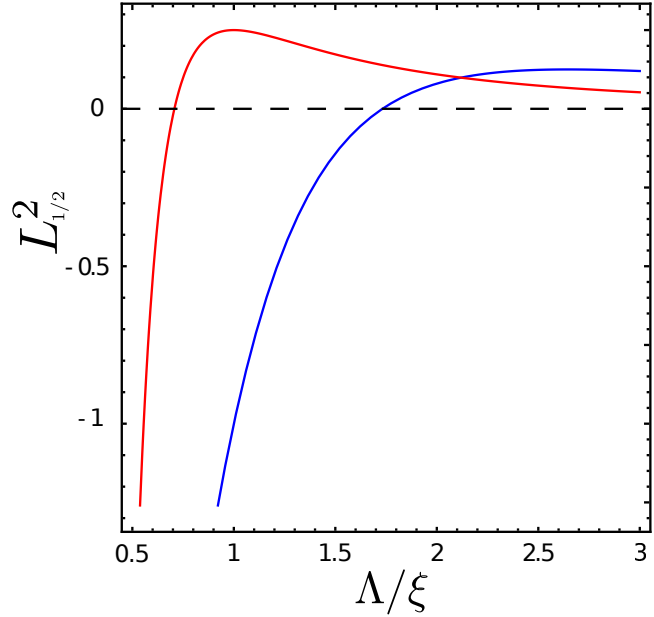}
 \caption[\ Effective squared length scale for damped oscillating correlation functions.]{
 Effective squared length scale $L_i^2(t)$, $i=1,2$ as a function of $\Lambda/\xi$ 
 for a modulated exponential correlator (blue) and a modulated Gaussian correlator (red) introduced in eq~(\ref{eq:toycorrelator}).}
 \label{fig:scales}
\end{figure}

Furthermore, we observe a strongly kinked oscillatory behaviour that even renders $L^2$ negative. 
This can be better understood by referring to simple correlation functions as
\begin{equation}\label{eq:toycorrelator}
 C_{1} = \e^{-R/\xi} \cos(R/\Lambda), \hspace{1cm} C_{2} = \e^{-(R/\xi)^2} \cos(R/\Lambda)
\end{equation}
For simplicity, we refer to $d=1$ here, since dimensionality is not changing the key aspect and 
it is straightforward to generalise the calculation. The characteristic length scale $L_i^2$ with $i=1,2$ associated with 
the correlation function $C_i$ is readily obtained from the scales second moment 
\begin{equation}\label{eq:oscillatingscale}
L_1^2 \simeq 2\Lambda^2\frac{(\Lambda/\xi)^2 - 3}{(1+(\Lambda/\xi)^2 )^2} \;\; , \;\;
L_2^2\simeq \frac{\Lambda^2}{4}\frac{1}{(\Lambda/\xi)^2} \left( 2-\frac{1}{(\Lambda/\xi)^2}\right)\ .
\end{equation}
While the overall time-dependence of this effective length scale can still be used to extract the dynamical exponent from
the scaling relation $L_i^2(t) \sim t^{2/z}$, the sign of the amplitude does depend on the ratio $\Lambda/\xi$. 
This change of sign, according to eq~(\ref{eq:oscillatingscale}), is illustrated in fig~\ref{fig:scales}.

Hence the change of signs in the effective squared length $L^2(t)$ can be attributed to oscillating correlators, and the competition between the two 
distinct length scales $\xi$ and $\Lambda$. While $L(t)$ itself can no longer be interpreted as a length scale, it should still be possible
to read off the value of the dynamical exponent.  
The oscillatory nature of eq.~(\ref{eq:Llow}) indicates consequently
a competition between at least two different length scales in the system.

\subsection{Dynamic susceptibility}

By means of eq~(\ref{eq:Z_sol}) we can calculate the dynamic susceptibility which is essentially proportional to $Q_0$
\begin{equation}
 \chi \sim Q_0 = \demi\left[1+\frac{\mathcal{C}gt}{Z}+\left(1-\frac{\mathcal{C}gt}{Z}\right)\leftidx_0 F_1\left(\demi,- g t Z\right)\right]
\e^{-\gamma Z} \ .
\end{equation}
We find for the leading contribution for large times
\begin{align}
 \chi(t) \sim \begin{cases}
              \mathcal{C}g/\varphi\sinh^2(\sqrt{g \varphi})\; t^2, &d=2\\[.5cm]
              \frac{\mathcal{C} g^2}{(d-2)^2} \;t^{d}/\ln^2 t, & d>2\\[.5cm]
               \frac{	\mathcal{C}g \gamma}{2-d} \frac{t^{2-\frac{d}{2}}}{\ln t} \sin^2\left( \sqrt{g/\gamma (1-d/2)  t \ln t }\right), &\frac{4}{3}<d<2
              \end{cases}
\end{align}
In general, for systems with simple scaling, one expects $\chi(t) \sim L(t)^d \sim t^{d/z}$, 
or said in  words, the susceptibility is proportional to the volume explored up to time $t$ \cite{Godr00b}. 
In $d=2$ dimensions, this expectation, is fully confirmed by our exact solution, since 
\begin{equation}
 \chi_{2D}(t) \sim t^2 \sim L^2
\end{equation}
and in  particular, we see once more that indeed $z=1$, in contrast to classical dynamics.
For dimensions, $d>2$, this scaling expectation for $\chi(t)$ is again confirmed, 
but only up to logarithmic corrections. In addition, the effective length scale
$L_{\rm eff}(t) \sim t \left( \ln t \right)^{-2/d}$ is different from the length scale extracted above from the second moment.

Finally, for $d<2$, not only does the exponent of the leading time-dependence deviate from the expected value 
(to say nothing on the logarithmic correction), but furthermore, a strong time-dependent modulation of $\chi(t)$ is found.

We can understand this as a further justification of the strong competition between different length scales 
as we already discussed in the previous section.

\subsection{Off-coherences}

We now want to study the off-coherence term and which reads
\begin{equation} 
\Xi_{\vec{k}}(t) = \: \e^{-\frac{\gamma}{g}\Delta_t}
\bigg[\frac{\mathcal{C} g}{\sqrt{\Delta_t}}- \sqrt{\Delta_t}\bigg]\sin2\sqrt{t\Delta_t}
\end{equation}
with $\Delta_t = g (Z(t)+t\omega_{\vec{k}})$.

\vspace{.5cm}

\noindent {\bf A.} For $d=2$ we know that $Z<0$ and thus $\Delta_t$ changes signs from negative to positive for 
after a time $t^*$ for fixed $k\neq 0$. Consequently, all $\Xi_{\vec{k}}\to0$ for $\vec{k} \neq 0$ due to the exponential damping.
For the zero mode we find
\begin{equation}
 \Xi_0 \simeq \mathcal{C}\sqrt{\frac{g}{\varphi}} \sinh(2\sqrt{g \varphi}) \sqrt{t} \; \; \; \stackrel{t\to\infty}{\to}\; \; \; \infty 
\end{equation}
and see a diverging off-coherence. This is a strong indicator that the system will not relax towards 
its thermal equilibrium but will rather stay in a non-equilibrium state for all times.
It seems possible that this is a hint that the model should undergo physical ageing. Of course, a definite assertion would require
a test of the three defining properties of physical ageing  
(slow dynamics, breaking of time-translation invariance, dynamical scaling) \cite{Henk10} and this requires at least an analysis of two-time correlators. 
We hope to return to an analysis of physical ageing in the {\sc qsm} elsewhere.

\vspace{.5cm}

\noindent {\bf B.} For $d>2$ the situation is, up to logarithmic corrections, similar to $d=2$. We find immediately
\begin{equation}
 \Xi_0 \simeq \frac{\mathcal{C}g}{d-2}\frac{t^{d-\frac{3}{2}}}{\ln t}
 \end{equation}
while all non-zero mode off-coherences vanish in the asymptotic limit. 

\vspace{.5cm}

\noindent {\bf C.} In $\frac{4}{3}<d<2$ the behaviour is qualitatively different. 
While it remains true, that all non-zero mode off-coherences vanish, we find for the zero mode
\begin{equation}
 \Xi_0 \sim t^{-\left(1-\frac{d}{2}\right)} \ln t
\end{equation}
which decays to zero and thus at least indicates that a relaxation into thermal equilibrium might be possible. 
Moreover, we observe that in this scenario the solution $Z(t)$ depends on the bath quantity $\gamma$ while for $d\geq 2$ the bath scales
entirely out. All these observations point towards the fact that \textit{the actual coupling to the reservoir 
becomes less important in higher-dimensional open quantum dynamics}.
%
%
\section{Conclusions}
We studied the {\sc qsm} as a simple exactly solvable model in order to 
explore exact quantum dynamics and compare classical to quantum dynamical properties. 
We used certain consistency criteria, in order to construct the precise form of the Lindblad master equation, namely 
(i) the quantum equilibrium is a stationary state of the chosen dynamics and 
(ii) the classical Langevin dynamics is included in the limit $g\to 0$. 
This guarantees that the equilibrium state is a stationary solution and that the canonical commutator relations are obeyed. 
As in equilibrium, for the {\sc qsm} the full $N$-body problem reduces to solving to a single integro-differential equation, for the time-dependent
spherical parameter. 
The full solution of this equation is still an open and difficult problem.

We have focussed in this work on two special cases. First, we considered weakly quantum dynamics and calculated the 
leading quantum corrections to the classical dynamics. It turns out that the effective quantum dynamics is classical and quantum effects
only renormalise the temperature and produce a hard-core effect in the spin-spin correlator. \textcolor{black}{Therefore, the heuristic expectation that
the thermal noise should wash out the quantum properties of the long-time dynamics is indeed confirmed and the dynamics is equivalent to the purely
relaxational classical model-A dynamics.} 
This confirmation serves as a useful consistency check of the formalism we set up to describe the open quantum dynamics of the spherical model.

Second, we studied the true quantum dynamics driven by the initial disorder for a quantum quench across the critical point 
and deep into the ordered phase. \textcolor{black}{In this regime, not explored before to the best of our knowledge, 
the model's long-time behaviour is distinct from any heuristic expectation. We found that the conserved canonical quantum commutators lead to profound
modifications of the dynamics, with respect to its classical limit.} In order to carry out this analysis, 
we explored new mathematical methods that are related to asymptotic expansions of confluent hyper-geometric functions 
in two variables. 
It turns out that the long-time behaviour of the integrated spherical parameter $Z(t)$ is extremely complex to deduce and 
that it depends on the spatial dimension in a non-trivial fashion. We have found 
\begin{equation} 
 |Z(t)| \simeq 
 \begin{cases}
               \frac{(d-2)^2}{4g}\frac{\ln^2 t}{t}           &, ~~  \ d> 2\\[.25cm]
               \varphi\ t^{-1}                               &, ~~  \ d=2 \\[.25cm]
               \left(1-\frac{d}{2}\right) \gamma^{-1} \ln t  &, ~~  \ \frac{4}{3}<d<2 
 \end{cases}
\end{equation}
This behaviour is qualitatively different from the classical case where simply $|Z^{\rm cl}(t)|\sim \ln t$.

Due to this strong dependence on the dimensionality of the system, we observed prominent differences in the scaling behaviour. 
In $d=2$ dimensions we find a regular scaling with a unique 
characteristic length scale. Thus, the {\sc qsm} is able to reliably predict general qualitative properties. 
In $d\neq 2$ dimensions, we find strong logarithmic corrections which destroy a 
simple scaling behaviour, through the presence of several time-dependent length scales with differ by power of $\ln t$. 
One might be tempted to view these corrections as a peculiarity of the 
{\sc sm}, as found long ago for the classical spherical model with a conserved order parameter \cite{Coni94,Maz06} 
and interpreted in a multi-scaling scenario.
However, since we find a clean scaling in $d=2$ dimensions, we believe that the logarithmic corrections through several logarithmically different
length scales should not be too readily dismissed as a peculiarity of the {\sc qsm}. These results are asymptotically independent of the
damping rate $\gamma$ such that the limit $\gamma\to 0$ towards closed quantum systems may be taken. The dynamical exponent turns out to be $z=1$, 
indicative of ballistic motion, as seen before 
in the quantum dynamics of models with fermionic degrees of freedom. 
\textcolor{black}{While ballistic motion is common for quantum systems near their
quantum critical point \cite{Calabrese07,Calabrese16,Dutt15,Eisl11,Znid15} and actually is expected to occur for very general reasons \cite{Delf17}, 
here we find it for quenches deeply into the two-phase ordered region, and so far unstudied with field-theoretical
methods.} 
For dimensions $d<2$ we find again logarithmic corrections to scaling, but of a different kind, and in addition strong time-dependent modulations 
of the spin-spin correlator $C(R)$ in terms of the distance $R$. Here, the damping constant $\gamma$ does appear in the scaling amplitudes. 

These features are confirmed and strengthened through an analysis of the leading long-time behaviour of the characteristic length scale $L(t)^2$ 
and of the time-dependent susceptibility $\chi(t)$. They show simple power-law scaling for $d=2$, with associated logarithmic corrections whenever $d\ne 2$
and thereby confirm the existence of several logarithmically different length scales.   

This work should be seen as a first, tentative, exploration of non-equilibrium quantum dynamics, far from a critical point, 
of an interacting many-body system such as the {\sc qsm}. 
Several essential assumptions and hypotheses were admitted throughout in our exploration. First of these, are the intrinsic Born approximation and
the Markov property which underlie the Lindblad approach. Second, the main new results come from our study of the deep quantum quenches into the 
ordered phase. Our results crucially depend on the conjecture that the integral term in eq~(\ref{eq:qq}) is irrelevant, hence will
give rise only to finite-time corrections to scaling. Testing this conjecture remains a difficult open problem. 
Another aspect which should be further analysed is the precise nature of the relaxation process. Is the quantum relaxation in the {\sc qsm} in some
way reminiscent to the {\em physical ageing} seen in the classical analogues~? Although we have found some preliminary indications which might point 
into this direction, a full testing of this will require to analyse the behaviour of two-time
correlators, via the quantum regression theorem \cite{Breu02,Scha14}, or even to include an external field and look at two-time response functions. 
We hope to return to this elsewhere. It would also be important to compare our results with what can be found from different approaches, notably Keldysh
field-theory \cite{Sieb15,Sieb16} or the generalised hydrodynamics of strongly interacting non-equilibrium
quantum systems \cite{Bertini16,Castro16,Doyon17,Caux17,Piroli17,Dubail16}. 

An attractive feature of the {\sc qsm} is that the r\^ole of the dimension $d$ can be analysed explicitly. 
Our result suggest, to the extend that the {\sc qsm} is a reliable guide for collective quantum dynamical behaviour, 
that $2D$ quenched quantum systems should show simple dynamical scaling, with an easily achieved data-collapse, whereas in
$3D$ quenched quantum systems it should only be possible to find a data-collapse in small time-dependent windows with effective time-dependent exponents. 
To what extent such an expectation is borne out in more general quantum models
remains an important challenge for the future.

\noindent 
{\bf Acknowledgements:} 
It is a pleasure to thank R. Betzholz, J.-Y. Fortin, D. Karevski and G. Morigi for useful discussions. 
SW is grateful to the `Statistical Physics Group' at
University of S\~ao Paulo, Brazil and the Group `Rechnergest\"utzte Physik der Werkstoffe' at ETH Z\"urich, Switzerland,
for their warm hospitality
and to UFA-DFH for financial support through grant CT-42-14-II.  
GTL thanks the financial support of the S\~ao Paulo Research Foundation under grant number 2016/08721-7.
  

\newpage
\appendix

\section{Equilibrium quantum spherical constraint}
\label{app:cantrafo}
We present the exact derivation of the quantum spherical constraint in the equilibrium {\sc saqsm}, 
by diagonalising the hamiltonian via canonical transformations. 
Consider the following hamiltonian, with bosonic operators $a_n$  such that $[a_n,a_m^{\dagger}]=\delta_{n,m}$ 
\BEQ \label{A1}
{H} = \sum_{n,m\in \mathcal{L}} \left[ {a}_{n}^{\dagger} A_{nm}{a}_{m}
-\demi \left( {a}_{n} B_{nm} {a}_{m} + \text{h.c.} \right) \right]
+\sum_{n\in \mathcal{L}} C_n \left({a}_{n}+{a}_{n}^{\dagger}\right)
\EEQ
which for a specific choice of the matrices $A,B$ reduces to the hamiltonian (\ref{eq:H}) of the {\sc saqsm}. 
In addition, the vector $\vec{C}$ allows to consider the
effects of an external field. We shall present an exact derivation of the equilibrium spherical constraint, which should also arise from the
stationary state ($t\rightarrow \infty$ limit) of the dynamics. Many aspects of the treatment are analogous 
to the one of free fermion hamiltonians, see e.g. \cite{Lieb61,Henk99}. 
For the sake of notational simplicity, we only treat the $1D$ case explicitly, the generalisation to any $d>1$ being obvious. 

Define the harmonic oscillator ladder operators \cite{Wald15}
\BEQ
{s}_{n} = \left(\frac{ g }{8\mathcal{S}}\right)^{1/4}\left({a}_{n}+{a}^\dagger_n\right)
\;\; , \;\; 
{p}_{n} = -\II \left(\frac{  \mathcal{S}}{2 g}\right)^{1/4}\left({a}_{n}-{a}^\dagger_n\right)
\EEQ
and the spherical constraint is then
\BEA \nonumber
  \mathcal{N} \sqrt{\frac{8\mathcal{S}}{ g}} &=& 
  \sum_{n\in\mathcal{L}}\bigg( \big<{a}_n{a}_n\big>+ \big<{a}^\dagger_n {a}^\dagger_n\big> 
  +2\big<{a}^\dagger_n {a}_n\big> +1 \bigg) \\
  &=& \big|\big<\vec{{a}}\big>\big|^2+ \big|\big<\vec{{a}}^\dagger\big>\big|^2 
  +2\big<\vec{{a}}^\dagger \cdot \vec{{a}}\big> + \mathcal{N} \ ,
\EEA
where we have introduced the vector $\vec{{a}} = ({a}_1,{a}_2, \ldots , {a}_{\mathcal{N}-1}, {a}_\mathcal{N} )$ 
and its element-wise adjoint.
We now apply the canonical transformation, used for the diagonalisation  in \cite{Wald15}
\BEQ
\vec{{a}} = \vec{r} + v^t \vec{{b}} - w^t \vec{{b}}^\dagger
\EEQ
to the spherical constraint and find
\BEA\nonumber
 \mathcal{N} \sqrt{\frac{8\mathcal{S}}{g}} &=& 4 \left| \vec{r} \right|^2 + 4 \vec{r} \cdot (v-w)^t\left<\vec{{b}}+\vec{{b}}^\dagger\right>\\\nonumber
 & &+ \sum_{lmn}\left( v_{ml}v_{nl}-2v_{ml}w_{nl}+w_{ml}w_{nl}\right) \left<{b}_m{b}_n\right>\\\nonumber
 & &+ \sum_{lmn}\left( w_{ml}w_{nl}-2w_{ml}v_{nl}+v_{ml}v_{nl}\right) \left<{b}^\dagger_m{b}^\dagger_n\right>\\\nonumber
 & &+ \sum_{lmn}\left(2 v_{ml}v_{nl}-v_{ml}w_{nl}+w_{ml}v_{nl}\right) \left<{b}^\dagger_m{b}_n\right>\\
 & &+ \sum_{lmn}\left(2 w_{ml}w_{nl}-v_{ml}w_{nl}+w_{ml}v_{nl}\right) \left<{b}_n{b}^\dagger_m\right> + \mathcal{N}
\label{eq:sc-trafo}
 \EEA
Following \cite{Wald15}, we define the matrix 
\BEA
\underline{\Psi} : = (v - w )^t
\EEA
with
the eigenvectors $\vec{\Psi}_n$ of $(A-B)(A+B)$ as column entries, see (\ref{A1}).
Analogously, we define
\BEA
\underline{\Phi} : = (A+B)\underline{\Psi} 
\EEA
(for a full analysis of the diagonalisation of $H$ via canonical transformations, see \cite[app. A]{Wald15}). 

Since $\left[{b}_n,{b}_m \right] = \left[{b}^\dagger_n,{b}^\dagger_m \right] =  0$,
we can exchange the indices $m$ and $n$ in line $2$ and $3$ of eq.~(\ref{eq:sc-trafo}) to find the same prefactor for $\left<{b}_m{b}_n\right>$ and
$\left<{b}^\dagger_m{b}^\dagger_n\right>$. In the fifth line we use the commutation relation to achieve
a normal order and estimate the prefactor of $\left<{b}^\dagger_m{b}_n\right>$ from this and the fourth line. We find
\BEA \nonumber
 2\mathcal{N}\sqrt{\frac{\mathcal{S}}{g}} &=& \mathcal{N} 
 + 4 \left| \vec{r} \right|^2 + 4 \vec{r} \cdot \underline{\Psi}\left<\vec{{b}}+\vec{{b}}^\dagger\right>
 + \sum_{n}\left(\left|\vec{\Psi}_n\right|^2-\vec{\Psi}_n\cdot\vec{\Phi}_n\right)\\
 \nonumber
 &+& \sum_{mn}\vec{\Psi}_m\cdot \vec{\Psi}_n \left( \left<\wht{b}_m{b}_n\right>
 +\left<{b}^\dagger_m{b}^\dagger_n\right>+2\left<{b}^\dagger_m{b}_n\right>\right) 
\EEA
Using the property $\vec{\Phi}_n \cdot \vec{\Psi}_n = 1$ \cite{Wald15}, we can rewrite the spherical constraint as
\BEA
 \frac{\mathcal{N}}{2}\sqrt{\frac{\mathcal{S}}{g}}=  \left| \vec{r} \right|^2 +  \vec{r} 
 \cdot \underline{\Psi}\left<\vec{{b}}+\vec{{b}}^\dagger\right>
 + \sum_{mn}\frac{\vec{\Psi}_m  .\vec{\Psi}_n}{4} \left( \left<{b}_m {b}_n\right>
 +\left<{b}^\dagger_m {b}^\dagger_n\right>+2\left<{b}^\dagger_m {b}_n\right> + \delta_{nm}\right)
\EEA
Finally, we use the orthogonality of the eigenvectors of Toeplitz matrices to find
\BEA 
 \frac{\mathcal{N}}{2}\sqrt{\frac{\mathcal{S}}{g}} 
 =  \left| \vec{r} \right|^2 +  \vec{r} \cdot \underline{\Psi}\left<\vec{{b}}+\vec{{b}}^\dagger\right>
 + \sum_{n}\frac{\left|\vec{\Psi}_n\right|^2}{4} 
 \left( \left<{b}_n {b}_n\right> +\left< {b}^\dagger_n {b}^\dagger_n\right>+2\left< {b}^\dagger_n {b}_n\right> + 1\right)
\EEA
For systems without an external magnetic field $\vec{r}=\vec{0}$ which we shall admit from now on. 
The absolute value of the eigenvectors was found in \cite{Wald15} to be
\BEQ
\left|\vec{\Psi}_n\right|^2 = \frac{\Lambda_k^-}{\Lambda_k^+} = 
\sqrt{\frac{\mathcal{S}-\frac{1-\lambda}{2}\cos k}{\mathcal{S}-\frac{1+\lambda}{2}\cos k}}
\EEQ
With this result we can write the final result, in zero external field
\BEA \label{eq:constraint}
 \sqrt{\frac{8}{g}}\sqrt{\mathcal{S}} = 
 \int_{\mathcal{B}} \frac{\D k}{2\pi}
 \frac{\Lambda_{-,k}}{\Lambda_{+,k}} 
 \left( \left<{b}_k {b}_k\right> +\left< {b}^\dagger_k {b}^\dagger_k\right>+2\left< {b}^\dagger_k {b}_k\right> + 1\right)
\EEA
which is easily generalised to $d$ dimensions.

In equilibrium, the off-diagonal averages  $\left<{b}_k {b}_k\right> = \left< {b}^\dagger_k {b}^\dagger_k\right> \to 0$ 
decay to zero and the number operator $\left< {b}^\dagger_k {b}_k\right>$ 
is given by the thermal occupation of the corresponding mode
\BEA \label{eq:constraint-eq}
 \sqrt{\frac{8}{g}\,}\, {\mathcal{S}}^{1/2} =  \int_{\mathcal{B}} \frac{\D \vec{k}}{(2\pi)^d} 
 \frac{\Lambda_{-,k}}{\Lambda_{+,k}}  \left( 2 n_k + 1\right)
\EEA
which is equivalent to eq.~(\ref{4.11}) in the main text. 
  
\section{Analysis of the Volterra equation \label{ap:volt}}

Solving the linear Volterra equation (\ref{eq:Qcor}), at an effective temperature $T^{\star}$, is standard, 
e.g. \cite{Ronc78,Cugl95,Godr00b,Henk15}. Define the Laplace transform 
\BEQ
\lap{f}(p) = \int_0^\infty \!\D t\: f(t) \e^{-pt}
\EEQ
such that the Laplace-transformed equation (\ref{eq:Qcor}) reads simply
\BEQ \label{eq:Qcorr-L}
\lap{G}(p) = \frac{\lap{F}(p)}{1-\gamma T^{\star}\: \lap{F}(p)}
\EEQ
Tauberian theorems \cite[ch. XIII]{Fell71} permit to extract the long-time behaviour of $G(t)$ from the behaviour of $\lap{G}(p)$ for $p\to 0$. 
We require $\lap{F}(p)=\lap{F}_{\rm uni}(p)+\lap{F}_{\rm reg}(p)$, for $p$ small, 
conveniently decomposed into an universal and a regular part, which have been derived countless times before
\BEQ
\lap{F}_{\text{uni}} (p) \stackrel{p\to 0}{\approx} \frac{\Gamma\left(1-\frac{d}{2}\right)}{\gamma(4\pi)^{\frac{d}{2}}} 
\left(\frac{p}{\gamma}\right)^{\frac{d}{2}-1}
\;\; , \;\;
\lap{F}_{\text{reg}}(p) = \frac{1}{\gamma}\left( A_1 -A_2 \frac{p}{\gamma} + A_3 \left(\frac{p}{\gamma} \right)^2 \mp \ldots \right)
\EEQ
where the last expansion can only be carried to the point where the coefficients 
\BEQ
A_n = \int_{\mathcal{B}}\frac{\D \vec{k}}{(2\pi)^d} \frac{1}{\omega_{\vec{k}}^n} \ .
\EEQ
exist ($\mathcal{B}=[-\pi,\pi]^d$ is the Brillouin zone). 
For example, even $A_1$ does not exist for $d\leq 2$ and $A_2$ only exists for $d>4$. We conclude that
\BEA
\lap{F}(p) \stackrel{p\to 0}{\approx}
\frac{1}{\gamma} \begin{cases}
 \Gamma\left(1-\frac{d}{2}\right)(4\pi)^{-\frac{d}{2}} \left(p/\gamma\right)^{\frac{d}{2}-1}\                                      &\text{ ,~~ if} \ 0<d<2\\ 
 A_1 -\left|\Gamma\left(1-\frac{d}{2}\right)\right|(4\pi)^{-\frac{d}{2}} \left(p/\gamma\right)^{\frac{d}{2}-1}\                    &\text{ ,~~ if} \ 2<d<4\\ 
 A_1 - A_2\  p/\gamma\ -\left|\Gamma\left(1-\frac{d}{2}\right)\right|(4\pi)^{-\frac{d}{2}} \left(p/\gamma\right)^{\frac{d}{2}-1}\  &\text{ ,~~ if} \ 4<d<6
             \end{cases}
             \label{eq:behaviorF}
\EEA
In the last, we included the regular term which dominates for $d<6$. 
Inserting into (\ref{eq:Qcorr-L}) gives $\lap{G}(p)$ which in turn must inserted into the generic
expression (\ref{5.11}) for the spin-spin correlator, which we repeat here  for convenience 
\BEQ \label{B6}
Q_{{\vec{k}}}(t) =\frac{\e^{-\gamma t \omega_{\vec{k}}}}{G(t)} + \frac{ g}{12 T}\left[1- \frac{\e^{-\gamma t \omega_{\vec{k}}}}{G(t)} 
\right] + \gamma T\frac{1}{G(t)}\int_0^t \!\D\tau\: G(\tau) \e^{- \gamma(t- \tau) \omega_{\vec{k}}}
\EEQ
We shall now study the three cases from (\ref{eq:behaviorF}) separately.

\subsection{$0<d<2$}

In this case, $\lap{F}(p)$ is a monotonous and surjective function on the interval $(0,\infty)$, 
hence the equation $1-\gamma T^{\star} \lap{F}(p)=0$ always has a solution at $p=p_0$.
Hence $\lap{G}(p)$ has a simple pole at some $p_0=t_{\text{eq}}^{-1}$, for all $T^{\star}>0$. 
The leading long-time behaviour of $G(t)$ is exponential, with the explicit relaxation time 
\BEQ
G(t) \sim \e^{t/t_{\text{eq}}} \;\; , \;\; 
t_{\text{eq}} = \gamma^{-1}\left[T^\star \Gamma\left(1-\frac{d}{2}\right) (4\pi)^{-d/2} \right]^{-\frac{2}{d-2}}
\EEQ
Inserting this into (\ref{B6}) leads straightforwardly to (\ref{eq:Q:low2}). 

\subsection{$2<d<4$} 

Since for dimensions $d>2$ the coefficient $A_1$ is finite, its value can be used to define a critical temperature 
\BEQ
T^\star_c = \frac{1}{A_1}
\EEQ
Then three distinct situations can arise:  
(i) The case $T^\star > T^\star_c$ is treated analogously to the case $d<2$. 
Here, the relaxation time is modified, because the phase transition does occurs at finite 
temperature, according to 
\BEQ
t_{\text{eq}} = \gamma^{-1}\left[\frac{T^\star - T^\star_c}{T^\star T^\star_c } 
|\Gamma\left(1-\frac{d}{2}\right)| (4\pi)^{-d/2} \right]^{-\frac{2}{d-2}}
\EEQ
but the correlator retains the form (\ref{eq:Q:low2}).

(ii) For $T^\star<T^\star_c$ we have to analyse eq. (\ref{eq:Qcorr-L}) carefully. Define the short-hand 
\BEQ
m^2 = 1-T^\star/T^\star_c
\EEQ
and expand $\lap{G}(p)$ to lowest non-trivial order in $p$ to find
\BEA
\lap{G}(p) &=& \frac{1}{\gamma}\frac{A_1 -(4\pi)^{-d/2}\left|\Gamma\left( 1-\frac{d}{2} \right)\right| 
\left(p/\gamma\right)^{d/2-1}}{m^2 +T^\star (4\pi)^{-d/2} \left|\Gamma\left(1-\frac{d}{2}\right) \right|(p/\gamma)^{d/2-1}}\nonumber \\
&\stackrel{p\to0}{\simeq}&\frac{1}{\gamma}\left[\frac{A_1}{m^2} - \frac{(4\pi)^{-d/2}}{m^4}\left| \Gamma\left(1-\frac{d}{2}\right) \right| 
\left(\frac{p}{\gamma}\right)^{\frac{d}{2}-1}\right] + \ldots 
\EEA
A Tauberian theorem \cite{Fell71} then gives the long-time behaviour of $G(t)$ 
by a formal inverse Laplace transform ($\delta(t)$ is the Dirac distribution)
\BEQ \label{B12}
G(t) \simeq \frac{1}{m^2 \gamma T_c^{\star}}\:\delta(t) +\frac{(4\pi \gamma t)^{-d/2}}{m^4},\ \text{for} \ t\rightarrow \infty \ \text{and} \ 2<d<4
\EEQ
The singular term therein, of course, does not appear in the long-time limit, 
but is required to evaluate the correlator. Following \cite{Henk15}, we insert into (\ref{B6}) and obtain
\BEA
Q_{\vec{k}}(t) &=& e^{-\gamma \omega_{\vec{k}} t} m^4 (4\pi \gamma t)^{d/2} \left( 1 - \frac{g}{12 T} \right) + \frac{g}{12 T} 
\nonumber \\
& & 
+ e^{-\gamma \omega_{\vec{k}} t} m^4 (4\pi \gamma t)^{d/2} \frac{\gamma T}{m^2 \gamma T_c^{\star}}
+\gamma T t^{d/2} \mathscr{L}^{-1} \left( \Gamma(1-d/2) p^{1-d/2} \frac{1}{p+\gamma \omega_{\vec{k}}} \right)(t)
\nonumber \\
&=& e^{-\gamma \omega_{\vec{k}} t} m^4 (4\pi \gamma t)^{d/2} \left( 1 - \frac{g}{12 T} +\frac{1}{m^2}\frac{T}{T_c^{\star}} \right) + \frac{g}{12 T} 
\nonumber \\
& & + \gamma T\: t \frac{1}{1-d/2}\: {}_1F_1\left( 1, 2-\frac{d}{2}; -\gamma \omega_{\vec{k}} t\right) 
\nonumber \\
&=& e^{-\gamma \omega_{\vec{k}} t} m^2 (4\pi \gamma t)^{d/2} \left( 1 - \frac{g}{12 T} \right) + \frac{g}{12 T} \\ \nonumber
& & + \frac{\gamma T }{1-d/2} t\: e^{-\gamma \omega_{\vec{k}} t}\: {}_1F_1\left( 1-\frac{d}{2}, 2-\frac{d}{2}; \gamma \omega_{\vec{k}} t\right)
\EEA
Herein, in the first two lines the terms proportional to $T$ come from the integral in (\ref{B6}). 
The first of those in the contribution from the singular term in (\ref{B12})
and the other is cast into an inverse Laplace transformation. In the next step, 
this inverse transformation is found using \cite[eq. (2.1.2.1)]{Prudnikov5} and the coefficient of the
other term is simplified using the definitions of $m^2$ and of $T^{\star}$. Finally, 
we used the identity \cite[eq. (13.1.27)]{Abra65}. We are interested in the limit $\vec{k}\to\vec{0}$,
$t\to\infty$ such that $\omega_{\vec{k}}t$ remains finite. Then the last term is sub-dominant and we arrive at (\ref{eq:Q:high}).  

(iii) For $T^\star=T^\star_c = 1/A_1$, the leading terms in small-$p$ expansion are 
\BEQ
\lap{G}(p)=\frac{1}{\gamma}\left(\frac{1}{T_c^\star}\right)^2 
\frac{(4\pi)^{d/2}}{\left|\Gamma(1-\frac{d}{2})\right|}\left(\frac{p}{\gamma}\right)^{1-d/2} -\frac{1}{\gamma T_c^{\star}} +\mbox{\rm o}(p)
\EEQ
hence 
\BEQ
G(t) = G_d t^{d/2-2} - \frac{1}{\gamma T_c^{\star}}\: \delta(t)
\EEQ
where $G_d$ is a known constant whose value will not be required. 
Inserting into (\ref{B6}) and taking into account the contribution of the singular term in the integral gives
\BEA
Q_{\vec{k}}(t) &=& \frac{e^{-\gamma \omega_{\vec{k}} t} t^{2-d/2}}{G_d} 
\underbrace{\left( 1 - \frac{g}{12 T_c} - \frac{T_c}{T_c^{\star}}\right)}_{=0} + \frac{g}{12 T_c} 
\nonumber \\
& & +\gamma T_c t^{2-d/2} \Gamma(d/2-1)\mathscr{L}^{-1}\left( p^{1-d/2}\frac{1}{p+\gamma \omega_{\vec{k}}}\right)(t) 
\nonumber \\
&=& \frac{g}{12 T_c} +\frac{\gamma T}{d/2-1}\;{}_1F_1\left(1,\frac{d}{2};-\gamma \omega_{\vec{k}} t\right) t
\EEA
Herein, the first term vanishes because of the definition of $T^{\star}$ 
and we re-used \cite[eq. (2.1.2.1)]{Prudnikov5}. This gives the first eq.~(\ref{eq:Q:crit}). 

\subsection{$d>4$}

The discussion is analogous to the previous ones. At $T^{\star}=T_c^{\star}$, expansion gives for small $p$ gives 
$\lap{G}(p)\simeq \frac{1}{{T_c^{\star}}^2 A_2}\frac{1}{p} - \frac{1}{\gamma T_c^{\star}}$, hence
\BEQ
G(t) \simeq -\frac{1}{\gamma T_c^{\star}}\:\delta(t) +\frac{1}{{(T_c^{\star}})^2 A_2}
\EEQ
Inserting this into (\ref{B6}) leads to
\BEQ
Q_{\vec{k}}(t) = ({T_c^{\star}})^2 A_2 e^{-\gamma \omega_{\vec{k}} t} 
\underbrace{\left( 1 - \frac{g}{12 T_c} +\frac{T_c}{T_c^{\star}}\right)}_{=0} + \frac{g}{12 T_c} +
\frac{T_c}{\omega_{\vec{k}}} \left( 1 - e^{-\gamma \omega_{\vec{k}} t} \right)
\EEQ
where we used again the definition of $T^{\star}$ and have thus found the second eq.~(\ref{eq:Q:crit}). 
Finally, below criticality, we must expand up to the first universal term.  We obtain
for $p$ small (as it stands, this holds for $d<6$, but extensions are obvious)
\BEA
\lap{G}(p) \simeq
\frac{1}{\gamma} \frac{A_1 -A_2 \frac{p}{\gamma} -|{\cal F}|_1 
\left(\frac{p}{\gamma}\right)^{d/2-1}}{m^2+T^\star A_2 \frac{p}{\gamma}+\gamma T^{\star} |{\cal F}_1\left(\frac{p}{\gamma}\right)^{d/2-1} }  
\simeq \frac{1}{\gamma}\left(\frac{A_1}{m^2} - \frac{A_2}{m^4} \frac{p}{\gamma}\right) 
-\frac{|{\cal F}|_1}{m^4} \left(\frac{p}{\gamma}\right)^{d/2-1} 
\EEA
which gives for large times
\BEQ
G(t) \simeq \frac{1}{m^2 \gamma T_c^{\star}}\: \delta(t) - \frac{A_2}{m^4 \gamma^2}\:\delta'(t) +\frac{(4\pi\gamma t)^{-d/2}}{m^4}
\EEQ
and from which one readily arrives again at eq.~(\ref{eq:Q:high}). 

We remark that the small-$p$ expansions must be carried up to including 
(i)  eventual constant terms and (ii) the leading universal contribution. The first contribution is required for the
correct evaluation of the correlator (unless one prefers to derive sum rules instead, 
as carried out in \cite{Godr00b}) and the second contribution gives the leading time-dependence. 

We did not  discuss the case $d=4$ explicitly, although this can be done without much extra difficulty \cite{Hase06,Ebbi08,Henk15}.
Below criticality, there is no dimension-dependent singularity
and one may simply set $d=4$ in the final result (\ref{eq:Q:high}) and at criticality, additional logarithmic singularities will appear.

\section{Proof of an identity}
\label{app:derivative}
We prove the asymptotic identity eq.~(\ref{eq:derivative}). 

\noindent
\textbf{Lemma}: {\it The function $\mathfrak{f}(\gamma) = e^{-\gamma Z} (4\pi \gamma t)^{-d/2}$ obeys for all $d>0$ and all $Z,t$ the identity} 
\BEQ\label{Eq:theorem}
\partial_\gamma^n \mathfrak{f}(\gamma) = (-1)^n \mathfrak{f}(\gamma)\sum_{k=0}^n \Gamma\begin{bmatrix}
                                                         n+1&\frac{d}{2}+k& \\
                                                         \frac{d}{2}& n-k+1&k+1
                                                        \end{bmatrix} \gamma^{-k} Z^{n-k}
\EEQ

\noindent
\textbf{Proof}: This proceeds via mathematical induction, with the habitual two steps. 

\noindent 
$\bullet$ \underline{Basis $n=1$}: it suffices to calculate the first derivative and compare with (\ref{Eq:theorem}). 
We find straightforwardly, in both cases 
\BD
\partial_\gamma \mathfrak{f}(\gamma) = -\mathfrak{f}(\gamma) \left[Z +\frac{d}{2\gamma} \right]
\ED

\noindent $\bullet$ \underline{Step $n\to n +1$}: We write
\BD
\partial_\gamma^{n+1}\mathfrak{f}(\gamma) = \partial_\gamma\partial_\gamma^n \mathfrak{f}(\gamma)
\ED
and use the expression (\ref{Eq:theorem}) to find 
\BD
\partial_\gamma^{n+1}\mathfrak{f}(\gamma) = (-1)^{n+1}\mathfrak{f}(\gamma) \sum_{k=0}^n
\Gamma\begin{bmatrix}
       n+1& \frac{d}{2}+k&\\
       \frac{d}{2}&k+1&n-k+1
      \end{bmatrix}
\gamma^{-k} Z^{n-k} \left\{Z+\frac{d}{2\gamma}+\frac{k}{\gamma}\right\}
\ED
Shifting the index $n$ to $n+1$ produces 
\BEA
\partial_\gamma^{n+1}\mathfrak{f}(\gamma) &=& (-1)^{n+1}\mathfrak{f}(\gamma) \sum_{k=0}^{n+1}\bigg\{
\Gamma\begin{bmatrix}
       n+2& \frac{d}{2}+k&\\
       \frac{d}{2}&k+1&n-k+2
      \end{bmatrix}
\gamma^{-k} Z^{n+1-k} \times\nonumber \\
& & \times \left[1-\frac{k}{n+1}\right]\left[1+\frac{d}{2Z\gamma}+\frac{k}{Z\gamma}\right]\bigg\} \nonumber
\EEA
Herein, the first line is already the sought expression for the $(n+1)^{\rm st}$ derivative.
It only remains to show that the residual terms
\BEQ \label{appE_D2}
\sum_{k=0}^{n+1}
\Gamma\begin{bmatrix}
       n+2& \frac{d}{2}+k&\\
       \frac{d}{2}&k+1&n-k+2
      \end{bmatrix}
\gamma^{-k} Z^{n+1-k}\bigg\{
\frac{d}{2Z\gamma}+\frac{k}{Z\gamma}-\frac{k}{n+1}\left[1+\frac{d}{2Z\gamma}+\frac{k}{Z\gamma}\right]\bigg\}
\EEQ
cancel. For simplicity we omit non-zero multiplicative factors and consider\footnote{In (\ref{appE_D2}), 
bring the curly bracket to the common denominator, which does not depend on k and
hence can be dropped.}
\BEA\nonumber
& &\sum_{k=0}^{n+1}\Gamma\begin{bmatrix}
       \frac{d}{2}+k&\\
       k+1&n-k+2
      \end{bmatrix}
(\gamma Z)^{-k} \left[\left(\frac{d}{2}+ k\right) (n+1-k)+ k \gamma Z\right]\\
&=&\sum_{k=0}^{n}\Gamma\begin{bmatrix}
       \frac{d}{2}+k+1&\\
       k+1&n-k+1
      \end{bmatrix}(\gamma Z)^{-k}
-\sum_{k=1}^{n}\Gamma\begin{bmatrix}
       \frac{d}{2}+k&\\
       k&n-k+2
      \end{bmatrix}(\gamma Z)^{-k-1} = 0\hspace{1cm} \nonumber
\EEA
which completes the proof. \hfill $\qed$


\section{Asymptotic analysis of some double series}
\label{app:DQ}
In the main text, we introduced two double series 
\BEA
\mathfrak{s}_1 &:=& \sum_{n=0}^\infty\sum_{k=0}^n \Gamma\begin{bmatrix}
							    \demi&\frac{d}{2}+k& &\\
							    n+\demi&\frac{d}{2}& n-k+1&k+1
							    \end{bmatrix} 
							    \left(-\frac{gt}{\gamma}\right)^n (\gamma Z)^{n-k} \label{eq:C1_S1}\\
\mathfrak{s}_2 &:=& -\gamma\mathcal{C}gt\sum_{n=1}^\infty\sum_{k=0}^{n-1} \Gamma\begin{bmatrix}
                                                                   \demi	&	\frac{d}{2}+k	&	&\\
                                                                   n+\demi	&	\frac{d}{2}	& n-k	&	k+1
                                                                  \end{bmatrix}\frac{1}{n}
\left(-\frac{gt}{\gamma}\right)^n (\gamma Z)^{n-1-k} \label{eq:C2_S2}
\EEA
and we require their asymptotic behaviour for $t\gg 1$ large, where $Z$ is either being kept fixed or varies slowly with $t$.  

\noindent 
{\bf 1.} We start our analysis with the treatment of $\mathfrak{s}_1$. 
Begin with (\ref{eq:C1_S1}) and exchange the order of summation, followed by a shift in the second summation variable. This results in
\begin{subequations} \label{C3}
\BEA 
\mathfrak{s}_1 &=& \sum_{k=0}^\infty\sum_{n=k}^{\infty} \Gamma\begin{bmatrix}
							    \demi&\frac{d}{2}+k& &\\
							    n+\demi&\frac{d}{2}& n-k+1&k+1
							    \end{bmatrix} 
							    \left(-\frac{gt}{\gamma}\right)^n (\gamma Z)^{n-k} 
\nonumber \\
			   &=& \sum_{k=0}^\infty \sum_{n=0}^\infty\Gamma\begin{bmatrix}
                                                         \demi&\frac{d}{2}+k& &\\
                                                        n+k+\frac{1}{2}&\frac{d}{2}& n+1&k+1
                                                        \end{bmatrix}
\left(-\frac{gt}{\gamma}\right)^{n+k}(\gamma Z)^{n} \label{C3a} \\
&=& \frac{\Gamma(\demi)}{\Gamma(\frac{d}{2})} \sum_{k=0}^{\infty} \sum_{n=0}^{\infty}  
\frac{\Gamma(k+\frac{d}{2})}{\Gamma(k+n+\demi)}  \frac{(-gt/\gamma)^k}{k!} \frac{(-gtZ)^n}{n!} \label{C3b}
\EEA
\end{subequations}
Recalling the definition of the Humbert function \cite{Humbert20a,Humbert20b}
\begin{equation}
\Phi_3\left(\beta;\gamma;x,y\right) = \sum_{m=0}^{\infty}\sum_{n=0}^{\infty} \frac{(\beta)_m}{(\gamma)_{m+n}} \frac{x^m}{m!} \frac{y^n}{n!}
\label{eq:humdef}
\end{equation}
we can identify $\mathfrak{s}_1=\Phi_3\left(\frac{d}{2};\demi;-gtZ, -\frac{gt}{\gamma}\right)$, as stated in (\ref{eq:6.14_S1}) in the main text. 
Sums such as (\ref{C3}) would be easy to evaluate if they would factorise, but in fact they are coupled by the factor  
$\Gamma\big( n+k+\frac{1}{2} \big)$ in the denominator. 
In order to achieve a factorisation, we use the following identity,  which  involves Euler's Beta function, 
with an arbitrary constant $0<\epsilon<\demi$
\BEA
\frac{1}{\Gamma\left(n+k+\demi\right)}
=\frac{B(n+\epsilon,\demi+k-\epsilon)}{\Gamma(\epsilon+n)\Gamma(\demi-\epsilon+k)}
=\frac{t^{\demi-n-k}}{\Gamma(\epsilon+n)\Gamma(\demi-\epsilon+k)}
\int_0^t \!\D x\:  x^{k-\epsilon -\demi} (t-x)^{n+\epsilon-1} ~
\EEA
which is obtained from eqs.~(6.2.1) and (6.2.2) in \cite{Abra65}. 
Now, insert this identity into (\ref{C3a}) such that the sums over $n$ and $k$ decouple. We then find 
\BEQ \label{eq:C5}
\mathfrak{s}_1 = \Gamma\begin{bmatrix}
			\demi&\\
			\demi - \epsilon,&\epsilon
                        \end{bmatrix} \sqrt{t} \left( \mathfrak{u}_1 \star \mathfrak{v}_1\right) (t)
\EEQ
with $0<\epsilon<\demi$ and the functions
\BEQ \label{eq:C6_S1}
\mathfrak{u}_1(x) = x^{-\frac{1}{2}-\epsilon}\leftidx{_1}F_1\left(\frac{d}{2};\demi-\epsilon;-\frac{g}{\gamma}x\right) \;\; , \;\;
\mathfrak{v}_1(x) = x^{\epsilon-1}\leftidx{_0}F_1\left(\epsilon;-gZ x\right)
\EEQ
Inserting the functions from (\ref{eq:C6_S1}) then gives the exact representation of $\mathfrak{s}_1$ 
as a Laplace convolution, stated in (\ref{eq:6.16S1}) in the main text,   
where the Laplace transform is defined as
\BEQ
\lap{\mathfrak{h}}(p) := \mathscr{L}(\mathfrak{h})(p) = \int_0^{\infty} \!\D x \:  \mathfrak{h}(x) \e^{-p x} \ ,
\EEQ
The {\it Laplace convolution theorem} states 
$(\mathfrak{u}_1\star\mathfrak{v}_1)(t)=\mathscr{L}^{-1}\left( \lap{\mathfrak{u}}_1(p) \lap{\mathfrak{v}}_1(p)\right)(t)$.  

In addition, combining the representation (\ref{eq:6.16S1},\ref{eq:C5}) 
with the Laplace convolution theorem gives access to the large-$t$ asymptotics of $\mathfrak{s}_1$, 
via a Tauberian theorem \cite{Fell71}: 
{\it find the small-$p$ behaviour for $\lap{\mathfrak{u}}_1(p)$ and 
$\lap{\mathfrak{v}}_1(p)$ and then carry out the inverse Laplace transform.} Therefore, we use eq.~(3.38.1.1) from \cite{Prudnikov4} and find
\BEQ
\lap{\mathfrak{u}}_1(p) = \Gamma\left(\demi -\epsilon\right)p^{\epsilon-\demi}\left( 1+\frac{g}{\gamma p} \right)^{-{d}/{2}},\hspace{1cm}
\lap{\mathfrak{v}}_1(p)=\Gamma(\epsilon)p^{-\epsilon} \e^{-\frac{gZ}{p}}
\EEQ
The small-$p$ expansion of the product $\lap{\mathfrak{u}}_1(p)\lap{\mathfrak{v}}_1(p)$ yields\footnote{Here we 
explicitly treat the quantum case 
$g\neq 0$. Admitting $g=0$ leads to a different small-$p$ expansion that results in the well-known classical 
zero-temperature quench dynamics \cite{Godr00b}}
\BEA
\mathfrak{s}_1&\stackrel{p\searrow0}{\simeq}& \sqrt{\pi t} \left(\frac{\gamma}{g}\right)^{\frac{d}{2}}
\mathscr{L}^{-1}\bigg( p^{\frac{d-1}{2}} \e^{-\frac{g Z}{p}}\bigg)(t)
\EEA
and the inverse Laplace transform can be extracted from eq. (2.2.2.1) in \cite{Prudnikov5}
\BEQ \label{eq:app:s1}
\mathfrak{s}_1 \simeq \sqrt{\pi}\left(\frac{\gamma}{g t }\right)^{\frac{d}{2}} 
\frac{\leftidx{_0}{F}_1\left(\frac{1-d}{2};-g t Z\right)}{\Gamma\left(\frac{1-d}{2}\right)}
\EEQ

\noindent
{\bf 2.} For $\mathfrak{s}_2$ our approach is analogous. Starting from (\ref{eq:C2_S2}), 
we shift variables and exchange the order of summation to arrive at 
\BEA
\mathfrak{s}_2 
&=& -\gamma\mathcal{C}gt\sum_{n=1}^\infty\sum_{k=1}^{n} \Gamma\begin{bmatrix}
                                                                   \demi	&	\frac{d}{2}-1+k	&	&\\
                                                                   n+\demi	&	\frac{d}{2}	& n-k+1	&	k
                                                                  \end{bmatrix}\frac{1}{n}
\left(-\frac{gt}{\gamma}\right)^n (\gamma Z)^{n-k} \nonumber \\
&=& -\gamma\mathcal{C}gt\sum_{k=1}^\infty\sum_{n=k}^{\infty} \Gamma\begin{bmatrix}
                                                                   \demi	&	\frac{d}{2}-1+k	&	&\\
                                                                   n+\demi	&	\frac{d}{2}	& n-k+1	&	k
                                                                  \end{bmatrix}\frac{1}{n}
\left(-\frac{gt}{\gamma}\right)^n (\gamma Z)^{n-k} \nonumber \\
&=& -\gamma\mathcal{C}gt\sum_{k=0}^\infty\sum_{n=0}^{\infty} \Gamma\begin{bmatrix}
                                                                   \demi	&	\frac{d}{2}+k	&	&\\
                                                                   n+1	&	\frac{d}{2}	& n+k+\frac{3}{2}	&	k+1
                                                                  \end{bmatrix}\frac{1}{n+k+1}
\left(-\frac{gt}{\gamma}\right)^{n+k} (\gamma Z)^{n} \nonumber \\
&=& \gamma\mathcal{C}g^2 t^2 \int_0^{\infty}\!\D v\: \e^{-v} \sum_{k=0}^\infty\sum_{n=0}^{\infty} \Gamma\begin{bmatrix}
                                                                   \demi	        &	\frac{d}{2}+k		               \\
                                                                   ~\frac{d}{2}	    & n+k+\frac{3}{2}
                                                                  \end{bmatrix}
\frac{\left(-gt \e^{-v}/\gamma\right)^{k}}{k!} \frac{(-gtZ \e^{-v})^{n}}{n!} 
\EEA
With the definition (\ref{eq:humdef}) of the Humbert function $\Phi_3$, we can also identify \\
$\mathfrak{s}_2=2\mathcal{C}g^2 t^2 \int_0^1\!\D w\: \Phi_3\left(\frac{d}{2};\frac{3}{2};-\frac{gt}{\gamma}w, -gtZ w\right)$, 
as stated in (\ref{eq:6.15_S2}) in the main text.  

The two sums can be decoupled via the identity, with $0<\epsilon<\frac{3}{2}$ 
\BEA
\frac{1}{\Gamma\left(n+k+\frac{3}{2}\right)}
=\frac{t^{\frac{1}{2}-n-k}}{\Gamma(\epsilon+n)\Gamma(\frac{3}{2}-\epsilon+k)}
\int_0^t \!\D x\:  x^{k-\epsilon -\frac{3}{2}} (t-x)^{n+\epsilon-1} ~
\EEA
such that we finally recast the double sum into an integrated  convolution 
\BEA
\mathfrak{s}_2 =  \mathcal{C} g^2 t^{\frac{3}{2}}\Gamma\begin{bmatrix}
                                                              \demi&\\
                                                              \frac{3}{2}-\epsilon,&\epsilon
                                                             \end{bmatrix}
\cleanint{0}{1}\!\D w\: (\mathfrak{u}_2 \star \mathfrak{v}_2)(t)
\EEA
with 
\BEA
\mathfrak{u}_2(x) = x^{\frac{1}{2}-\epsilon}\leftidx{_1}F_1\left(\frac{d}{2},\frac{3}{2}-\epsilon,-\frac{g}{\gamma} w x\right) \;\; , \;\; 
\mathfrak{v}_2(x) = x^{\epsilon-1}\leftidx{_0}F_1\left(\epsilon,-gZ w x\right)
\EEA
as stated in (\ref{eq:6.17S2}) in the main text. 
Finally, the asymptotics for $t\to \infty$ is found as before from a Tauberian theorem. 
The Laplace transforms of the above functions read \cite[(3.38.1.1)]{Prudnikov4}
\BEA
\lap{\mathfrak{u}}_2(p) = \Gamma\left( \frac{3}{2} - \epsilon \right)
p^{\epsilon-\frac{3}{2}}\left(1+\frac{g w}{\gamma p}\right)^{-\frac{d}{2}} \;\; , \;\;
\lap{\mathfrak{v}}_2(p) = \Gamma(\epsilon)p^{-\epsilon} \e^{-gZ\frac{w}{p}}
\EEA
Inserting leads to the expression
\BEA
\mathfrak{s}_2 &=& \mathcal{C} g^2\sqrt{\pi t^3}
\mathscr{L}^{-1}\bigg( p^{\frac{d-3}{2}} \int_0^1\!\D w\: \left( p +\frac{g}{\gamma} w \right)^{-\frac{d}{2}}\e^{-gwZ/p}\bigg)(t)
\EEA
The $w$-integration can be expressed exactly as an incomplete Gamma function \cite{Abra65} 
\BEA
\mathfrak{s}_2 =\gamma \mathcal{C} g\sqrt{\pi t^3} (\gamma Z)^{\frac{d}{2}-1}\e^{\gamma Z}
\mathscr{L}^{-1}\bigg[\Gamma\left(1-\frac{d}{2},\gamma Z\right)\frac{1}{\sqrt{p}}
-\frac{1}{\sqrt{p}}\Gamma\left(1-\frac{d}{2},Z\frac{g}{p}+\gamma Z\right) \bigg](t)
\EEA
Since we now want to study this expression in the $p\to 0$ limit, it is adequate to use an asymptotic expansion 
for the last term, which we extract from eq.~\cite[(6.5.30)]{Abra65}
\BEQ
\Gamma(a,x+y) \stackrel{x\to\infty}{\simeq}\Gamma(a,x) - \e^{-x}x^{a-1}\left( 1-\e^{-y} \right) 
\EEQ
In order to evaluate the inverse Laplace transform, we consult eqs.~(2.2.2.1), (3.10.2.2) and (2.1.1.3) in \cite{Prudnikov5} and find
\BEA\nonumber
\mathfrak{s}_2 \simeq \gamma \mathcal{C}g t 
\bigg\{ \frac{\leftidx{_1}F_1\left(1;2-\frac{d}{2};\gamma Z\right)}{\frac{d}{2}-1}&+&
\sqrt{\pi}\left(\frac{\gamma}{gt}\right)^{\frac{d}{2}-1}\bigg[
\frac{\leftidx{_1}F_2\left( 1-\frac{d}{2};2-\frac{d}{2},\frac{3-d}{2};-gtZ\right)}{\left(1-\frac{d}{2}\right)\Gamma\left(\frac{3-d}{2}\right)}
\,\e^{\gamma Z}\\&+&
\frac{\e^{\gamma Z}-1}{gtZ}\frac{\leftidx{_0}F_1\left(\frac{1-d}{2};-gtZ\right)}{\Gamma\left(1-\frac{d}{2}\right)}\bigg]
\bigg\}
\label{eq:app:s2}
\EEA
Finally, combining eqs.~(\ref{eq:app:s1},\ref{eq:app:s2}) and inserting into the constraint (\ref{eq:constraint4}), 
we arrive at the asymptotic form (\ref{eq:hypersc}) of the spherical constraint. 

Similar methods can be applied to find the asymptotics of several confluents of Appell's hyper-geometric function $F_3$ \cite{Appell26,Sriv85}, 
when both arguments become large. This will be presented elsewhere \cite{Wald17}.


\section{Spherical constraint in two spatial dimensions}
\label{sec:constraint2d}
The constraint (\ref{eq:constraint4}) requires a special analysis in two spatial dimensions, 
due to apparent divergences in eq.~(\ref{eq:app:s2}) for $d\to 2$. 
We carry this out by writing $d=2(1+\eps)$ and studying the limit $\eps \to 0$. We want to show that 
eq.~(\ref{eq:app:s2}) is indeed well-defined in the $d\to 2$ limit and to find this limit.  

The critical sum is $\mathfrak{s}_2$, which may be rewritten as
\BEQ
\mathfrak{s}_2\simeq\mathcal{C} g t \gamma\bigg[\e^{\gamma Z}\left(\frac{1}{\eps} 
-  \frac{\leftidx{_1}F_2\left(-\eps;1,\demi;-g t Z\right)}{\eps} \right)
+ \frac{\e^{\gamma Z} -1}{g t Z} \sqrt{\pi\,}\,  \frac{\leftidx{_0}F_1\left(-\demi;-g t Z\right)}{\Gamma\left(-\demi\right)}\bigg]
\EEQ
The limit where $\eps$ goes to zero can be taken using the formula (derived below)
\BEQ \label{E.2}
\lim_{\eps\to 0}\left(\frac{1}{\eps} -  \frac{\leftidx{_1}F_2\left(-\eps;1,\demi; x\right)}{\eps} \right) 
= 2 x\;\leftidx{_2}F_3\left(1,1;\frac{3}{2},2,2;x\right)
\EEQ
and renders the  sum $\mathfrak{s}_2$ into the form 
\BEQ
\mathfrak{s}_2\simeq\mathcal{C} g t \gamma\bigg[-\e^{\gamma Z}2g t Z\;
\leftidx{_2}F_3\left(1,1;\frac{3}{2},2,2;-g t Z\right)-2 \frac{\e^{\gamma Z} -1}{g t Z} 
\: \leftidx{_0}F_1\left(-\demi;-g t Z\right)\bigg]
\EEQ
%
\begin{figure}[tb]\centering
 \includegraphics[width=.7\textwidth]{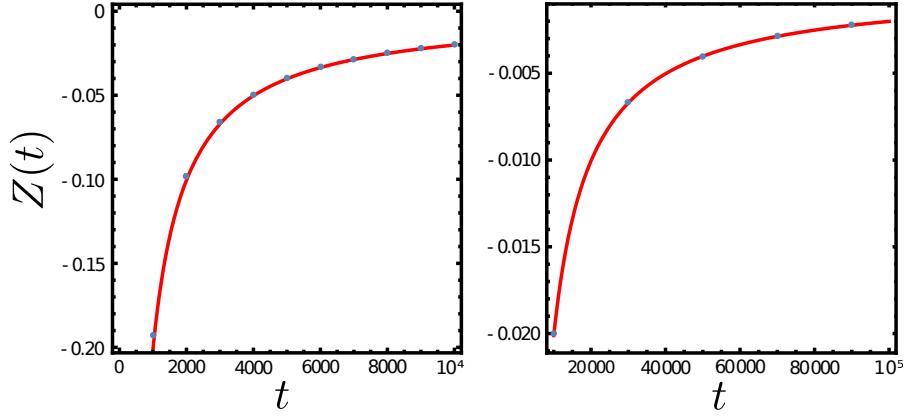}
 \caption[figE]{Solution $Z(t)$ of the spherical constraint for $d=2$, $g=0.1$, $\gamma = 1$ and $\mathcal{C}=1/4$. 
 The full curve is from eq.~(\ref{E5}) and the dots are numerical data. 
 The left and right panel display different intervals for $t$. \label{fig:AppE}}
\end{figure}
%
Recalling (\ref{eq:app:s1}), we can now study the constraint (\ref{eq:constraint4}) in $2D$. 
Solving the spherical constraint numerically, see figure~\ref{fig:AppE}, 
we remark that the observations $Z=-|Z|<0$ and $Z\to 0$ still hold true in the long-time limit $t\to \infty$. 
However, an asymptotic expansion for $t|Z| \to \infty$ fails. Therefore, we must consider 
$t |Z| =: \varphi \to \text{cste.}$ and proceed to determine this constant. 
Asymptotically, the constraint (\ref{eq:constraint4}) reads
\BEQ
8\pi \gamma t \simeq 
1 - \sqrt{\frac{1}{4t}} \frac{\gamma}{g}\: \leftidx{_0}F_1\left(-\demi; g \varphi\right)
+\mathcal{C} g \gamma t \left[ 2g\varphi\: \leftidx{_2}F_3\left(1,1;\frac{3}{2},2,2;g \varphi \right)
-\frac{2\gamma}{gt}\: \leftidx{_0}F_1\left( -\demi;g \varphi \right)\right]
\EEQ
where we replaced $\e^{\gamma Z} \mapsto 1$. We also observe that the first and the last term on the right-hand site are sub-dominant. 
For the constant $\varphi$ we thus find the transcendental equation
\BEQ \label{E5}
\frac{4\pi}{\mathcal{C} g^2} = \varphi\; \leftidx{_2}F_3\left(1,1;\frac{3}{2},2,2;g \varphi \right)
\EEQ
which is eq.~(\ref{eq:constraint6_phi}) in the main text. It always has an unique solution since the image of the right-hand side is $\mathbb{R^+}$ 
and the function is monotonous. The spherical parameter then reads $Z \simeq -\varphi/t$.

In the limit of an extreme {\sc scdl} with $\mathcal{C}\to\infty$, see fig.~\ref{fig:C}, we have simply $\varphi\mathcal{C}=4\pi g^{-2}$. 
The opposite limit of an
extreme {\sc sqdl} with $\mathcal{C}=\frac{1}{4}$ gives an upper bound for the admissible values of $\varphi$. 

\underline{{\bf Proof }of eq.~(\ref{E.2}):} insert the expansion $\Gamma(n-\eps)/\Gamma(-\eps)\simeq -\Gamma(n)\eps+\mbox{\rm O}(\eps^2)$ into
\BEA
\lefteqn{ \frac{1}{\eps}\left[ 1 - {}_1F_2\left(-\eps;1,\demi;x\right)\right] 
= -\frac{1}{\eps} \sum_{n=1}^{\infty} \frac{x^n}{n!\, (1)_n (\demi)_n}\frac{\Gamma(n-\eps)}{\Gamma(-\eps)} } \nonumber \\
&\simeq & \sum_{n=0}^{\infty} \frac{x^{n+1}}{n!} \frac{n!\, (1)_n}{(2)_n (2)_n (\frac{3}{2})_n \demi} +\mbox{\rm O}(\eps) 
\:=\: 2x\: {}_2F_3\left(1,1;2,2,\frac{3}{2};x\right) +\mbox{\rm O}(\eps)\nonumber
\EEA
by also using $(a)_{n+1} = (a+1)_n \frac{\Gamma(a+1)}{\Gamma(a)}$. \hfill $\qed$

\newpage


\section{Analysis of the spherical constraint for $d\ne 2$}
\label{sec:constraint_d}

We present the asymptotic analysis of the spherical constraint (\ref{eq:hypersc}) in generic dimensions $d\ne 2$. 

\subsection{$d>2$}

\begin{figure}[tb]
\centering
\includegraphics[width=\textwidth]{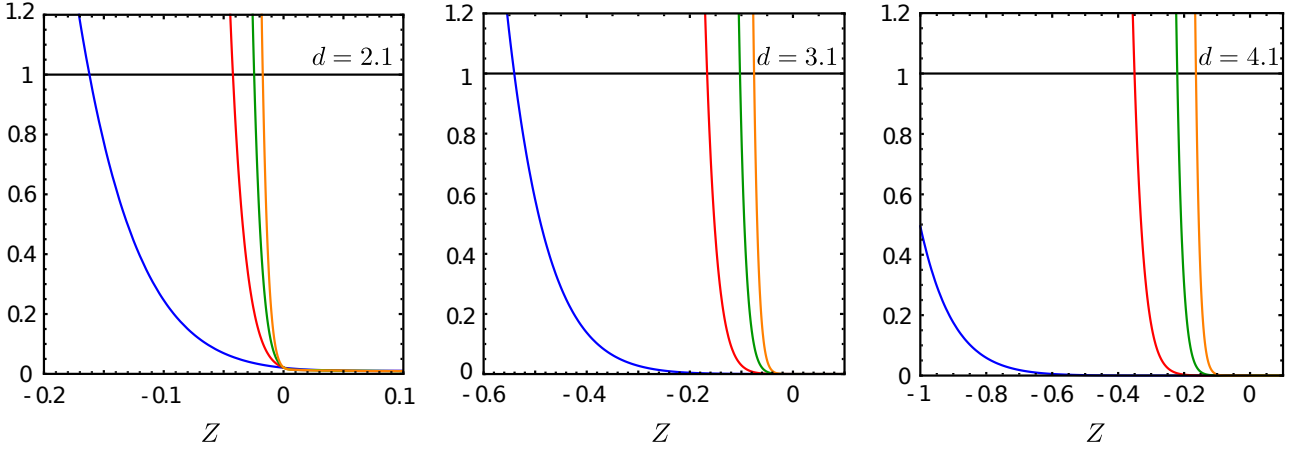}
\caption[.]{Solving the constraint (\ref{eq:hypersc}) as a function of $Z$: 
the {\sc lhs} is shown in black and the {\sc rhs} is shown for different times 
$t=[1000,4000,7000,10000]$ corresponding to the blue, red, green and orange lines, from left to right. 
The other parameters are $\mathcal{C} = 1$,  $\gamma = 0.1$ and $g = 0.1$. 
Different frames correspond to different dimensions: left panel $d=2.1$, middle panel $d=3.1$, right panel $d=4.1$.} 
\label{fig:constraint}
\end{figure}

In order to define the goals of an asymptotic analysis, we first consider the qualitative behaviour of the numerical solution $Z=Z(t)$, 
illustrated in fig.~\ref{fig:constraint}. 
Therein, both the left-hand side ({\sc lhs}) and the right-hand-side ({\sc rhs}) are displayed as a function of $Z$, 
for certain values of $t$, and for typical values of $\mathcal{C}$, $g$ and $\gamma$. 
The solution $Z=Z(t)$ is given by the intersections of the black and one of the coloured lines, respectively. 
For large times and for dimensions $d>2$, the numerical examples suggest  the following properties,  
which we shall need for our further analysis: 
\begin{enumerate}
 \item The solution to the spherical constraint is \textit{unique} and \textit{negative}, 
       which is clear from fig.~\ref{fig:constraint}.\footnote{We have checked numerically that $Z<0$ 
 f     or times  up to  $t\approx 10^{51}$.}  
 \item In the asymptotic limit where $t\to \infty$, the solution tends to $Z\to 0^{-}$. This is apparent in fig.~\ref{fig:constraint} and 
       further shown in the left panel of fig.~\ref{fig:behavior} in the main text. 
 \item the decay of $Z$ is slower than ${\rm O}(t^{-1})$, such that $t|Z(t)|$  still increases with $t$,
       as further illustrated in the right panel of fig.~\ref{fig:behavior}.
\end{enumerate}
In fig.~\ref{fig:behavior} in the main text, the time-dependence of $Z(t)$ is further illustrated for the generic spatial dimensions $d\ne 2$. 
The qualitative shape of these curves does not depend much on the specific values of the other parameters. 
Therefore, these examples suggest that the sought long-time behaviour can be obtained by studying the
asymptotics for $t|Z(t)|\to \infty$ in (\ref{eq:hypersc}), at least when $d\ne 2$. 
A more detailed study further suggests that this growth is more slow than any power-law.  
%
%

Therefore, we need the following expansions of the various hyper-geometric functions in (\ref{eq:hypersc}) for
$t|Z(t)|\gg 1$ and $|Z(t)|\ll 1$. This is achieved by the asymptotic identities, see \cite{Abra65} and especially \cite[(07.22.06.0005.01)]{Wolfram} 
\begin{subequations}
\BEA
\leftidx{_1}{F}_2\left( 1-\frac{d}{2};2-\frac{d}{2},\frac{3-d}{2};gt|Z|\right)&\simeq&
-\left(1-\frac{d}{2}\right)\frac{\Gamma(\frac{3}{2}-\frac{d}{2})}{\Gamma(-\demi)} 
       \frac{\e^{2\sqrt{g t |Z|}}}{\left(gt|Z|\right)^{1-{d}/{4}}}\\[.5cm]
\leftidx{_0}{F}_1\left(\frac{1-d}{2};g t |Z|\right)&\simeq& -\frac{\Gamma(\demi-\frac{d}{2})}{\Gamma(-\demi)} 
\left(g t |Z|\right)^{{d}/{4}}\e^{2\sqrt{g t |Z|}}\\[.5cm]
\leftidx{_1}{F}_1\left(1;2-\frac{d}{2};\gamma Z\right)&\simeq& 1 +\frac{\gamma Z}{2-\frac{d}{2}}
\EEA
\end{subequations}
which simplify the constraint (\ref{eq:hypersc}) to the following form
\BEA
\hspace{-0.3truecm}\e^{\gamma Z} (4\pi\gamma t)^{d/2} \simeq 
\demi +\left(1+\mathcal{C} g t \gamma\left[ 1 +\frac{1}{\gamma |Z|} \right]\right)
\left(\frac{\gamma^2|Z|}{g t}\right)^{\frac{d}{4}}\frac{\e^{2\sqrt{g t |Z|}}}{4}
+\frac{\gamma \mathcal{C}gt}{d-2}\bigg[ 1 +\frac{4}{d-4}\gamma |Z| \bigg]
\EEA
Herein, the last term on the right-hand site is sub-dominant. 
We can therefore neglect it and arrive at the following final form of the constraint 
\BEQ
2\,\e^{\gamma Z} (4\pi\gamma t)^{d/2}\simeq \Bigg[
1+\frac{\gamma^{\frac{d}{2}}}{2}\left(1+\mathcal{C}\frac{gt}{|Z|}\right)
\left(\frac{|Z|}{g t}\right)^{\frac{d}{4}}\e^{2\sqrt{g t |Z|}} \Bigg]
\label{eq:Contrainte}
\EEQ
which is eq.~(\ref{eq:sc-dq-exact}) in the main text. 
As before in the toy case where $\mathcal{C}=0$ and analysed in the main text, the constraint can be solved explicitly in terms of $W$-functions, 
but some care is needed to select the correct real-valued branch \cite{Corl96}, which is either $W_0$ or $W_{-1}$, 
such that positive values for $|Z(t)|$ are produced. We find
\BEQ \label{eq:F4}
t |Z(t)| \simeq \frac{(d-4)^2}{16 g}
\begin{cases}
      W_{-1}^2\left(\frac{2g}{d-4}\left[\frac{(8\pi)^d}{\mathcal{C}^2} \right]^{\frac{1}{d-4}} t^{2\frac{d-2}{d-4}}\right) &, ~~ d<4\\ \\
      \left(\frac{2}{d-4}\right)^2 \ln^2\left( \frac{(8\pi t)^2}{\mathcal{C}} \right)                                      &, ~~ d=4\\ \\
      W_{0}^2\left(\frac{2g}{d-4}\left[\frac{(8\pi)^d}{\mathcal{C}^2} \right]^{\frac{1}{d-4}} t^{2\frac{d-2}{d-4}}\right)  &, ~~ d>4
\end{cases}
\EEQ
The leading behaviour is found from the known asymptotics of the 
$W$-function\footnote{One uses $W_{-1}(x)\simeq \ln(-x) -\ln(-\ln(-x)) + {\rm o}(1)$ for $x\to 0^{-}$ \cite{Corl96}.} to be 
\BEQ \label{eq:F5}
|Z(t)| \simeq \frac{(d-2)^2}{4g} \frac{\ln^2 t}{t} 
\EEQ
for all dimensions $d>2$. This asymptotic result does neither depend explicitly on the initial condition $\mathcal{C}$ 
nor on the coupling $\gamma$ to the bath.

\subsection{$1<d<2$} 

\begin{figure}[tb]
\centering
\includegraphics[width=\textwidth]{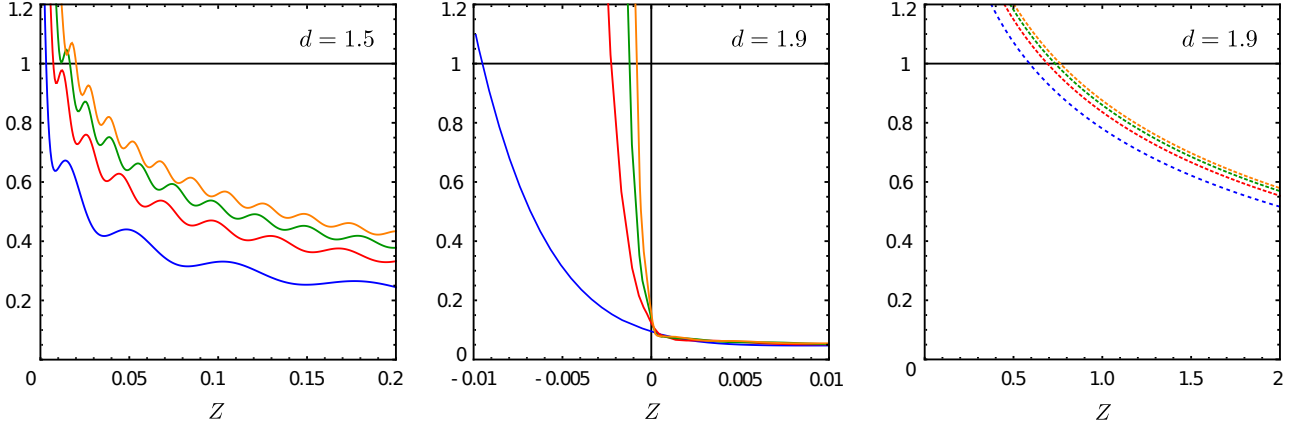}
\caption[cd]{Solving the constraint (\ref{eq:hypersc}) as a function of $Z$: 
the {\sc lhs} is shown in black and the {\sc rhs} is shown for different times 
$t=[10000,40000,70000,100000]$ corresponding to the blue, red, green and orange lines, from left to right, in the left and middle panels. 
In the right panel, the {\sc rhs} with $t=[1,4,7,10]\cdot 10^{40}$ corresponds to the blue, red, green and orange dashed lines, from left to right. 
The other parameters are $\mathcal{C} = 1$,  $\gamma = 0.1$ and $g = 0.1$. 
Different frames correspond to different dimensions: left panel $d=1.5$, middle and right panels $d=1.9$.
\label{fig:constraint_d}} 
\end{figure}

Again, we try to identify the correct mathematical setting by looking at some numerical solutions of the constraint (\ref{eq:hypersc}). 
We illustrate in fig.~\ref{fig:constraint_d} 
some typical behaviour, for several values of $d$. Clearly, the left panel shows  that for $d<2$ 
the qualitative behaviour is different from what was seen for $d>2$.  
We observe as generic features
\begin{enumerate}
 \item For large enough times, the solution to the spherical constraint becomes \textit{positive}.  
 \item In the asymptotic limit $t\to \infty$, the solution $Z(t)$ grows beyond all bounds, but its growth is very slow compared to $t$.  
 \item Strong oscillations are  superposed onto this growth, the frequency of  whom apparently increase with $t$, while the amplitude decreases. 
 \item There is a regime of large intermediate times, where the solution $Z=-|Z(t)|<0$ 
 is negative and qualitatively behaves as seen above for dimensions $d>2$. 
 This is illustrated in the middle panel of fig.~\ref{fig:constraint_d}, which is very similar to fig.~\ref{fig:constraint}. 
 In the right panel, it is further 
 shown that for truly enormous times the final true asymptotic regime with $Z>0$ is reached. 
\end{enumerate}
%
Therefore, for intermediate times, we can take over the analysis for $d>2$ and recover eq.~(\ref{eq:F5}) 
as an effective description.\footnote{Eq.~(\ref{eq:F4}) with $d<4$ applies.} 
One can estimate the order of the time-scale $t_{\times}$ where this cross-over happens by setting $Z=0$ in the constraint (\ref{eq:hypersc}). 
For $d=2-\vep$ dimensions,  
we find $t_{\times}\approx \frac{\gamma}{g}\,\e^{8\pi/\mathcal{C}g}$ which for the chosen parameters can become very large indeed. 

In order to find the true final asymptotics for really large values of $t$, 
we must re-analyse (\ref{eq:hypersc}) in the limit where $t\to\infty$ and $Z\gg 1$. We then require the following asymptotic expansions,
see \cite{Abra65} and \cite[(07.22.06.0011.01)]{Wolfram}
\begin{subequations} \label{eq:F6}
\BEA
{}_1F_1\left(1;2-\frac{d}{2};\gamma Z\right) &=& \left(1-\frac{d}{2}\right)\e^{\gamma Z} \left(\gamma Z\right)^{d/2-1} 
\left[ \Gamma\left(1-\frac{d}{2}\right) - \Gamma\left(1-\frac{d}{2},\gamma Z\right)\right]\nonumber \\
&\simeq & \left(1-\frac{d}{2}\right)\e^{\gamma Z} \left(\gamma Z\right)^{d/2-1} \Gamma\left(1-\frac{d}{2}\right) 
- \left(1-\frac{d}{2}\right) \left(\gamma Z\right)^{-1}
\EEA
\BEA
\frac{{}_0F_1\left(\frac{1-d}{2};-gtZ\right)}{\Gamma(1-d/2)} &=&  \left( gtZ\right)^{(d+1)/4} J_{-(d+1)/4}\left( 2 \sqrt{gtZ\,}\,\right) \nonumber \\
&\simeq & \frac{\left( gtZ\right)^{(d+1)/4}}{\pi^{1/2}} \left[ 
\cos\left(2\sqrt{gtZ\,}\,+\frac{\pi d}{4}\right) \left(1 -\frac{d(d+2)[(d+1)^2-9]}{512}\frac{1}{gtZ}\right)\right. \nonumber \\
& & \left. - \sin\left(2\sqrt{gtZ\,}\,+\frac{\pi d}{4}\right) \frac{d(d+2)}{16}\frac{1}{\sqrt{gtZ\,}\,}\right]
\EEA
\begin{align}
\frac{{}_1F_2\left(1-\frac{d}{2};2-\frac{d}{2},\frac{3}{2}-\frac{d}{2};-gtZ\right)}{\Gamma(\frac{3}{2}-\frac{d}{2})(1-\frac{d}{2})\pi^{-1/2}} &\simeq
\Gamma\left(1-\frac{d}{2}\right) \left( gtZ\right)^{d/2-1} \nonumber \\
 +\cos\left(2\sqrt{gtZ\,}\,+\frac{\pi d}{4}\right)&\left( gtZ\right)^{d/4-1}\left[ -1+\frac{d(d+2)(d^2-14d+56)}{512}\frac{1}{gtZ} \right] \nonumber \\
 &+\sin\left(2\sqrt{gtZ\,}\,+\frac{\pi d}{4}\right)\left( gtZ\right)^{d/4-3/2} \frac{d(d-6)}{16} \nonumber \\
\end{align}
\end{subequations}
and where $J_{\nu}$ is a Bessel function and $\Gamma(a,x)$ an incomplete Gamma function \cite{Abra65}. 
Inserting these expansions into (\ref{eq:hypersc}), several leading terms cancel. The constraint takes the form
\BEA
\lefteqn{2\left(4\pi\gamma t\right)^{d/2} = \frac{\mathcal{C}gt}{Z}\: \e^{-\gamma Z}} \nonumber \\
&+&\frac{d\mathcal{C}\gamma^{d/2}}{2}\left(\frac{Z}{gt}\right)^{d/4-1} 
\left[ \frac{3(d+2)(4-d)}{64}\frac{\cos\left(2\sqrt{gtZ\,}\,+\frac{\pi d}{4}\right)}{Z} 
- \frac{\sin\left(2\sqrt{gtZ\,}\,+\frac{\pi d}{4}\right)}{\sqrt{gtZ\,}\,} \right] ~~~~~
\label{eq:F7}
\EEA
which is eq.~(\ref{eq:6.19}) in the main text. 
In order to solve this equation, consider first only the first term on the right-hand side. 
If one assumes that asymptotically $\e^{\gamma Z} \sim t^{\alpha}$, matching the left-hand side
with the right-hand side gives $\alpha=1-\frac{d}{2}$. 
Then, the second term on the right-hand side is of the order $t^{1-d/4+\alpha}$, up to logarithmic or oscillating factors. 
If $\alpha<d/4$, this second term merely generates a correction. This is so for $d>\frac{4}{3}$. 
Similarly, the third term is of the order $t^{1/2-d/4+\alpha}$, hence it only generates a finite-time correction for $d>1$. 

Hence, for $\frac{4}{3}<d<2$, it is enough to concentrate on the first term on the right-hand-side in (\ref{eq:F7}). 
Analogously to previous cases, the constraint is solved via the Lambert-W function
\BEQ
\gamma Z = W\left( \frac{\mathcal{C}g}{2^{d+1}\pi^{d/2}} \left(\gamma t\right)^{1-d/2}\right) 
\simeq \left(1-\frac{d}{2}\right) \ln\gamma t + {\rm O}(\ln\ln t)
\EEQ
For a better approximation, one can re-inject this solution into the second and third terms on the right-hand-side of (\ref{eq:F7}). 
Then one obtains an oscillatory correction,  
of the form quoted in the main text. 


\section{Structure factor for $4/3<d<2$}
\label{app:rewriteQ}
We derive the identity eq.~(\ref{eq:Q-re}). 
We neglect all prefactors, focus on the functional dependence and treat the expression
\begin{equation}
 Q_{\vec{k}} \propto \frac{1-\leftidx_0F_1(1/2;-x)}{x}
\end{equation}
where $x= gt (Z + tk^ 2)$. Clearly, replacing the hyper-geometric function by its series representation gives 
\begin{equation}
 \frac{1-\leftidx_0F_1(1/2;-x)}{x} = \sum_{n=1}^\infty \frac{1}{(1/2)_n}\frac{(-x)^{n-1}}{n!} 
 = \sum_{n=0}^\infty \frac{1}{(1/2)_{n+1}}\frac{(-x)^{n}}{(n+1)!} \label{eq:Q-rewrite}
\end{equation}
We now rewrite
\begin{equation}
 \frac{1}{(1/2)_{n+1}}\frac{1}{(n+1)!} = \Gamma\begin{bmatrix}
                                          n+1 &3/2&2&1/2& \\
                                          1& n+3/2  & n+2 &2& 3/2
                                         \end{bmatrix}\frac{1}{n!} = 2\:  \frac{(1)_n}{(3/2)_n (2)_n}\frac{1}{n!}
\end{equation}
and can consequently recast eq.~(\ref{eq:Q-rewrite}) as
\begin{align}
 \frac{1-\leftidx_0F_1(1/2;-x)}{x} = 2 \sum_{n=0}^\infty \frac{(1)_n}{(3/2)_n (2)_n}\frac{(-x)^n}{n!} = 2\; \leftidx_1F_2(1;3/2,2;-x)
\end{align}
which is the representation used in the main text.

\newpage


\end{document}